\documentclass[twocolumn,tighten,times]{aastex631}
\usepackage{amsmath}
\accepted{ApJ}

\hypersetup{linkcolor=blue,citecolor=blue,filecolor=blue,urlcolor=magenta}
\newcommand{\um}{$\mu$m}
\def\columndensity{\ensuremath{N_{\mathrm{HI+H_{2}}}}}
\def\cmtwo{\ensuremath{{\rm{cm}}^{-2}}}

\definecolor{aborlaff}{RGB}{200,  34, 34}
\definecolor{rbeck}{RGB}{34,  200, 34}

\definecolor{kostas}{rgb}{1.00, 0.00, 0.0}

\shorttitle{Extragalactic magnetism with SOFIA -- IV: Overview and first results on the polarization fraction}
\shortauthors{Lopez-Rodriguez et al.}
\begin{document}

\title{Extragalactic magnetism with SOFIA (SALSA Legacy Program) - IV: \\
Program overview and first results on the polarization fraction\footnote{SALSA provides a software repository at \url{https://github.com/galmagfields/hawc}, and publicly available data at \url{http://galmagfields.com/}}}

\correspondingauthor{Lopez-Rodriguez, E.}
\email{elopezrodriguez@stanford.edu}

\author[0000-0001-5357-6538]{Enrique Lopez-Rodriguez}
\affil{Kavli Institute for Particle Astrophysics \& Cosmology (KIPAC), Stanford University, Stanford, CA 94305, USA}

\author[0000-0001-8906-7866]{Sui Ann Mao}
\affil{Max-Planck-Institut f\"ur Radioastronomie, Auf dem H\"ugel 69, 53121 Bonn, Germany}

\author{Rainer Beck}
\affil{Max-Planck-Institut f\"ur Radioastronomie, Auf dem H\"ugel 69, 53121 Bonn, Germany}

\author[0000-0003-3249-4431]{Alejandro S. Borlaff}
\affil{NASA Ames Research Center, Moffett Field, CA 94035, USA}

\author[0000-0002-4324-0034]{Evangelia Ntormousi}
\affil{Scuola Notmale Superiore di Pisa, Piazza dei Cavalieri 7, 56126 Pisa, Italy} 
\affil{Institute of Astrophysics, Foundation for Research and Technology-Hellas, Vasilika Vouton, GR-70013 Heraklion, Greece}

\author[0000-0002-8831-2038]{Konstantinos Tassis}
\affil{Department of Physics \& ITCP, University of Crete, GR-70013, Heraklion, Greece} 
\affil{Institute of Astrophysics, Foundation for Research and Technology-Hellas, Vasilika Vouton, GR-70013 Heraklion, Greece}

\author[0000-0002-5782-9093]{Daniel~A.~Dale}
\affil{Department of Physics \& Astronomy, University of Wyoming, Laramie, WY, USA} 

\author[0000-0001-6326-7069]{Julia Roman-Duval}
\affil{Space Telescope Science Institute, 3700 San Martin Drive, Baltimore, MD21218, USA}

\author[0000-0002-4210-3513]{Kandaswamy Subramanian}
\affil{Inter-University Centre for Astronomy and Astrophysics, Post Bag 4, Ganeshkhind, Pune 411007, India}

\author[0000-0002-4059-9850]{Sergio Martin-Alvarez}
\affil{Institute of Astronomy and Kavli Institute for Cosmology, University of Cambridge, Madingley Road, Cambridge CB3 0HA, UK}

\author{Pamela M. Marcum}
\affil{NASA Ames Research Center, Moffett Field, CA 94035, USA}

\author[0000-0002-7633-3376]{Susan E. Clark}
\affil{Kavli Institute for Particle Astrophysics \& Cosmology (KIPAC), Stanford University, Stanford, CA 94305, USA}

\author[0000-0001-8362-4094]{William T. Reach}
\affil{Universities Space Research Association, NASA Ames Research Center, Moffett Field, CA 94035, USA}

\author{Doyal A. Harper}
\affil{Department of Astronomy and Astrophysics University of Chicago, Chicago, IL 60637, USA}

\author{Ellen G. Zweibel}
\affil{Department of Astronomy, University of Wisconsin–Madison, 475 North Charter Street, Madison, WI 53706, USA}
\affil{Department of Physics, University of Wisconsin–Madison, 1150 University Avenue, Madison, WI 53706, USA}

%%%%%%%%%%%%%%%%
%%%%% ABSTRACT %%%%%
%%%%%%%%%%%%%%%%

\begin{abstract} 
We present the first data release of the Survey on extragALactic magnetiSm with SOFIA (SALSA Legacy Program) with a set of 14 nearby ($<20$ Mpc) galaxies with resolved imaging polarimetric observations using HAWC+ from $53$ to $214$~\um\ at a resolution of $5-18\arcsec$~($90$~pc $-$ $1$~kpc). We introduce the definitions and background on extragalactic magnetism, and present the scientific motivation and sample selection of the program. Here, we focus on the general trends in the emissive polarization fraction. Far-infrared polarimetric observations trace the thermal polarized emission of magnetically aligned dust grains across the galaxy disks with polarization fractions of $P=0-15$\% in the cold, $T_{\rm d} = [19,48]$~K, and dense, $\log_{10}(\columndensity[\cmtwo]) = [19.96,22.91]$, interstellar medium. The spiral galaxies show a median $\langle P_{154\mu m} \rangle = 3.3\pm0.9 $\% across the disks. We report the first polarized spectrum of starburst galaxies showing a minimum within $89-154$~\um. The falling $53-154$~\um~polarized spectrum may be due to a decrease in the dust grain alignment efficiency produced by variations in dust temperatures along the line-of-sight in the galactic outflow. We find that the starburst galaxies and the star-forming regions within normal galaxies have the lowest polarization fractions. We find that 50\% (7 out of 14) of the galaxies require a broken power-law in the $P-\columndensity$ and $P-T_{\rm d}$ relations with three different trends. Group 1 has a relative increase of anisotropic random B-fields produced by compression or shear of B-fields in the galactic outflows, starburst rings, and inner-bar of galaxies; and Groups 2 and 3 have a relative increase of isotropic random B-fields driven by star-forming regions in the spiral arms, and/or an increase of dust grain alignment efficiency caused by shock-driven regions or evolutionary stages of a galaxy. 
\end{abstract}

%%%%%%%%%%%%%%%%%%%
%%%%% INTRODUCTION %%%%%
%%%%%%%%%%%%%%%%%%%

\section{Introduction} \label{sec:INT}

Recent far-infrared (FIR; $50-220$~\um) and sub-mm ($850$~\um) imaging polarimetric observations have opened a new window of exploration for extragalactic magnetism. The orientation of the magnetic fields (B-fields) has been measured in the dense and cold regions of the interstellar medium (ISM) of nearby galaxies by means of magnetically aligned dust grains \citep{ELR2018,Jones2019,ELR2020,Jones2020,ELR2021a,ELR2021c,ELR2021,Borlaff2021,Pattle2021b}. These results have been obtained using the High-Angular Wideband Camera Plus \citep[HAWC+;][]{Vaillancourt2007,Dowell2010,Harper2018} on board the 2.7-m Stratospheric Observatory For Infrared Astronomy (SOFIA) and POL-2 on the James Clerk Maxwell Telescope (JCMT). 

This manuscript presents the Survey on extragALactic magnetiSm with SOFIA (SALSA) Legacy Program (PI: Lopez-Rodriguez, E. \& Mao, S. A.). SALSA aims to construct a comprehensive empirical picture of the B-field structure in the multi-phase ISM as a function of the gas dynamics of several types of galaxies from hundred pc- to kpc-scales. This SOFIA Legacy program relies on a multi-wavelength approach to trace the strength and morphology of the B-fields using FIR and radio polarimetric observations. These observations are combined with neutral and molecular emission lines observations, tracing the gas dynamics in the disk of galaxies. Star formation tracers are also used to study the effect of star-forming regions on the galaxy's B-field. Early results \citep{Borlaff2021,ELR2021a,ELR2021c,ELR2021} of this program have successfully demonstrated this multi-wavelength approach to characterize the B-field in the multi-phase ISM of several galaxy types (i.e., mergers, spirals, and starburst galaxies). Here, we present a brief background on extragalactic magnetism, describe the definitions of the B-fields in galaxies, and raise open questions regarding the role of B-fields in galaxy evolution (Section \ref{sec:Back}). Section \ref{sec:LegOver} presents the scientific goals, observing strategy, and galaxy sample of the program. Section \ref{sec:DR1} presents the first results of the data release focused on the polarization fraction. We discuss the trends of the polarization fraction as a function of the relative contribution of the ordered and random B-fields, the polarized spectrum of starburst galaxies, the effect of galaxy inclination on the polarization fraction, and a comparison with \textit{Planck} results on the Milky Way in Section \ref{sec:DIS}. Our conclusions are presented in Section \ref{sec:CON}.

%%%%%%%%%%%%%%%%%%%
%%%%% BACKGROUND %%%%%
%%%%%%%%%%%%%%%%%%%

\section{Background} \label{sec:Back}

%%%%%%%%%%%%%%%%%%%%%%
\subsection{Radio observational signatures of magnetic fields in galaxies}\label{subsec:RadioObs}
%%%%%%%%%%%%%%%%%%%%%%

Most of our knowledge about B-fields in galaxies has been established using radio polarimetric observations in the $3-20$~cm wavelength range \citep[e.g.,][and updates, for a review]{BW2013}. This wavelength range traces the B-fields via synchrotron emission and Faraday rotation measurements arising from the warm and diffuse phase of the ISM. These remarkable efforts have demonstrated the ubiquity of B-fields from galactic to intergalactic scales and have opened the question of the origin and evolution of the B-fields in galaxies across cosmic time.

Several decades of observations have shown that all spiral galaxies have large-scale ordered B-fields with an average strength of $\sim 5\pm2~\mu$G \citep{Beck2019}. The total B-field strength, measured by synchrotron total intensity and assuming equipartition between total B-field and total cosmic rays electron density, is estimated to be $\sim 17\pm14~\mu$G  \citep{Fletcher2010,Beck2019}. In face-on spiral galaxies, the most prominent B-field morphology is a kpc-scale spiral pattern where the polarized emission is mostly spatially coincident with the interarm regions and is strongest at the inner edges of arms \citep[e.g.][]{Han1999,Fletcher2011,Beck2007,Beck2015}. This result may be explained by the fact that B-fields in the arms are more turbulent when compared to those in the interarm regions. In edge-on spiral galaxies, the B-field morphology is observed to have a component parallel to the midplane of the disk, and another component with an X-shape structure extending several kpcs above and below the disk \citep{Hummel1988,Hummel1991,Krause2020}. Larger radio halos containing B-fields and cosmic rays are found in galaxies with higher star formation rate surface densities \citep{Wiegert2015}. The physical origin of the X-shape B-field in the halos of spiral galaxies is still unclear.

Some dwarf and irregular galaxies have been observed to have ordered B-fields with similar strengths to those in spiral galaxies, and on scales comparable to the galaxy size \citep{Chyzy2000,Chyzy2003,Kepley2010}. These ordered B-fields have even been observed in galaxies without spiral arms \citep[e.g., NGC~4736,][]{Chyzy2008}. Ordered spiral B-field patterns have also been observed in the Large Magellanic Cloud \citep[LMC;][]{Klein1993,Mao2012} and Small Magellanic Cloud \citep[SMC;][]{Mao2008,Livingston2022}. 

Starburst galaxies have the strongest measured B-field strengths of $\ge50~\mu$G \citep{Adebahr2013,LB2013,Adebahr2017} with ordered B-fields along the plane of the galaxy and an X-shape away from the disk \citep{Heesen2011,Adebahr2013,Adebahr2017}. The starburst ring in the central $1$~kpc of NGC~1097 has the strongest B-field strength, $\ge60~\mu$G, in a barred galaxy \citep{Tabatabaei2018} with an ordered spiral B-field morphology toward the center of the galaxy \citep{Beck1999,Beck2005}.

In interacting galaxies, large-scale B-fields have been found to a) be partially ordered and tracing tidal tails, and b) have stretched spiral arms with stronger total B-field strengths than in non-interacting spiral galaxies \citep{Chyzy2004,Drzazga2011,Basu2017}. Interacting galaxies in the Virgo cluster show asymmetric polarized intensities in the outer edges of the galactic disks due to interaction with the cluster environment \citep{Vollmer2007}. Galaxy clusters at $z\sim0.6-0.9$ show similar B-field strengths to those in the Virgo Cluster \citep{DiG2021}. Galaxies up to $z\sim2$ have been observed to have coherent B-fields with strengths similar to those in the local universe \citep{Bernet2008,Mao2017}. As the largest structures of the universe, filamentary structures connect galaxies forming the cosmic web, which is thought to be permeated by gas and B-fields. At these scales, recent radio observations using the LOw-Frequency ARray (LOFAR) have obtained upper-limit constraints on the median intergalactic B-field of $\le 0.25~\mu$G at scales of $10$~Mpc \citep{Locatelli2021}.

%%%%%%%%%%%%%%%%%%%%%%
\subsection{Optical and near-infrared observational signatures of magnetic fields in galaxies}\label{subsec:OIRObs}
%%%%%%%%%%%%%%%%%%%%%%

Studies of interstellar polarization at optical and near-infrared ($1-2$ \um; NIR) wavelengths can reveal a) the B-field geometry as a result of magnetically aligned elongated dust grains by radiative alignment torques \citep[RATs; for a review][and Section \ref{subsec:BFIR} for a summary]{HL2016, Andersson2015}, and b) scattering properties of dust and/or electrons. Both polarization mechanisms have distinguishable polarization signatures. For absorptive polarization, the polarization fraction, $P$, is expected to be of a few $\%$ with a position angle ($PA$) of polarization parallel to the orientation of the local B-field. For scattering off the dust and electrons, the polarization fraction is expected to be up to tens $\%$ and the $PA$ of polarization is perpendicular to the last direction of flight of the radiation to our line-of-sight (LOS). Scattering processes produce an azimuthal (centrosymmetric) pattern in the $PA$ of polarization centered on the object radiating the ISM. 

Pioneering work by \citet{Elvius1962,Elvius1964} using optical polarimetric observations of nearby spiral and starburst galaxies showed a diversity of polarization mechanisms and physical components. Some edge-on spiral galaxies (e.g., Centaurus A and NGC~4631) show $PA$ of polarization parallel to the dust lane. \citet{Elvius1964} interpreted these results as absorptive polarization arising from magnetically aligned dust grains. These authors invoked a large-scale B-field parallel to the galaxy's disks following the velocity field of the galaxy. Further optical \citep[e.g.,][for a review]{Berry1985,Scarrott1996a,Scarrott1996} and NIR \citep{Jones2000} polarization observations of several edge-on spiral galaxies found a $PA$ of polarization parallel to the disk of galaxies. These results confirmed the interpretation of a large-scale B-field parallel to the dust lane. However, dust scattering was also found to be the dominant polarization mechanism at optical wavelengths in several other edge-on galaxies \citep{Fendt1996}. 

For the starburst galaxy M82, \citet{Elvius1962} measured strong polarization fractions, $P\ge16$\%, in regions up to $\sim2$ kpc above and below the disk. The $PA$ of polarization was measured to have an axisymmetric pattern centered at the core of the galaxy. This polarization pattern is also measured at the outskirts, $\ge2$ kpc, of the galaxy's disk. In addition, low polarization fractions, $P\le2$\%, and a $PA$ of polarization parallel to the disk was found within the central $\sim1.5$~kpc of the galaxy. The polarization in the outskirts of the galaxy was interpreted as scattering off the dust grains directly radiated by the central starburst. The polarization at the core of the galaxy may be interpreted as the galactic B-field in the disk arising from absorptive polarization of magnetically aligned dust grains. Further $1.65$ and $2$~\um~observations confirmed the contribution of a large-scale B-field across the galaxy's disk and dust scattering above and below the disk of the starburst galaxy M82 \citep{Jones2000}. This result was found after the scattering pattern was removed by subtracting an azimuthal pattern centered at the starburst. Another starburst galaxy NGC~1808 \citep{Scarrott1993} shows azimuthal patterns in the $PA$ of polarization, which indicates that dust scattering is the dominant mechanism at optical wavelengths.

For face-on spiral galaxies, optical polarimetric observations using 4-m class telescopes of M51 \citep{Scarrott1987} and NGC~6946 \citep{Fendt1998} show similar patterns in the $PA$ of polarization to those measured at radio wavelengths \citep{Klein1982,Beck1987,Beck2007,Fletcher2011}. Specifically, M51 shows $P\sim2$\% and up to $5$\% in NGC~6946 with a pattern of the $PA$ of polarization mostly azimuthal and centered at the galaxy's core. \citet{King1986} performed optical polarimetric observations of NGC~7331 and also found a dominant asymmetric pattern along the major axis of the galaxy. Absorptive polarization may be observed in the eastern outskirts of NGC~7331. These observations were performed without a filter covering the $0.45-1.0$ \um~wavelength range  with a peak at $0.85$ \um. Specifically, the observations were performed using the full wavelength response of the CCD, thus the final measured  polarization can be a mix of multiple polarization mechanisms (i.e. dust/electron scattering, dichroic absorption. Using $1.6$ \um~imaging polarization observations of M51, \citet{Pavel2012} measured an upper-limit of $\sim0.05$\% in the polarization fraction across the full galaxy's disk. Scattering cross section declines much faster, $\propto \lambda^{-4}$, than absorption, $\propto \lambda^{-1}$, between $0.55$ and $1.65$ \um~\citep{JW2015}. These observations support dust scattering as the most likely polarization mechanism at optical and NIR wavelengths.

Given the contamination of dust scattering at optical and NIR wavelengths, the search for B-fields in galaxies is very complex. Data analysis techniques, such as those developed by \citet{Jones2000}, that aim to remove the azimuthal $PA$ of polarization in M82 may minimize the dust scattering pattern at optical and NIR wavelengths. However, the resultant polarization pattern, expected to be arising from absorptive polarization, may be jeopardized by a residual contribution from the highly polarized scattering mechanism. The results of the B-field orientation and strength in galaxies using these residuals need to be interpreted with caution. Deeper polarimetric observations at optical and NIR wavelengths are still required to disentangle the contribution of dust scattering and absorptive polarization to measure the contribution, if any, of B-fields in galaxies.

%%%%%%%%%%%%%%%%%%%%%%
\subsection{Observational definitions of magnetic fields in galaxies}\label{subsec:BDef}
%%%%%%%%%%%%%%%%%%%%%%

Before we continue with the origin and role of B-fields in galaxy evolution, we must define the B-field components in galaxies from an observational framework. Here, we follow the common nomenclature used in galaxies described by \citet{Beck2019}. Following these definitions, the total synchrotron emission traces the total B-field component in the plane of the sky, which depends on the total B-field strength and the cosmic ray electron density. The total B-field can be separated into a regular (or coherent) component and a random component. The random B-field can be a turbulent one with a Kolmogorov power spectrum and a driving scale of $50-100$\,pc, or a tangled regular B-field that reveals a different power spectrum with larger spatial scales \citep{Gent2021}. Present-day observations with a spatial resolution of a few 100\,pc cannot distinguish between turbulent B-fields and tangled regular B-fields. A well-defined B-field direction within the beam size of the observations is described as a regular B-field. These regular B-fields can be traced using Faraday rotation measures, which are sensitive to the direction of the B-field along the LOS. 

The random B-fields may have spatial reversals within the beam of the observations, which can be isotropic or anisotropic. The directions of the isotropic random B-fields have the same dispersion in all spatial dimensions. If an averaged B-field orientation remains after several reversals within the beam of the observations then the B-fields are known as anisotropic random B-fields. Polarized synchrotron emission traces the ordered B-fields in the plane of the sky, which depends on the strength and geometry of the B-fields, and cosmic ray electron density. The anisotropic random B-fields and the regular B-fields both contribute to the polarization and are referred to as ordered B-field. Thus, the polarized synchrotron emission increases when the averaged anisotropic random B-field increases. Unpolarized synchrotron emission can be identified with isotropic random B-fields within the beam of the observations. The PA of polarization, corrected for Faraday rotation, traces the B-field orientation in the plane of the sky with a $180^{\circ}\times n$ ambiguity. 

We use these definitions to describe the B-fields traced by FIR polarimetric observations in Section \ref{subsec:BFIR}. Note the different nomenclature in the literature when describing B-fields in galaxies and the Milky Way, that is anisotropic random fields, tangled fields, striated, ordered random \citep{Jaffe2019}.

%%%%%%%%%%%%%%%%%%%%%%
\subsection{The galactic dynamo theory}\label{subsec:BOri}
%%%%%%%%%%%%%%%%%%%%%%

How B-fields have evolved from primordial seeds into their present-day configuration in galaxies remains an open question in astrophysics. The leading theory is based on dynamos that rely on the turbulent nature of the plasma flows, the so-called turbulent dynamos \citep{Subramanian1998,BS2005}. Turbulent dynamos are classified as either mean-field or fluctuation dynamos. The regular B-fields are thought to be generated by the mean-field dynamo that relies on differential rotation of the galactic disk and turbulent helical motions to amplify and order a `seed' B-field. The turbulent B-fields are thought to be generated by turbulent gas motions at scales smaller than the energy-carrying eddies \citep{BS2005}, driven by supernova explosions with the coherence turbulent scales of $l\sim50-100$ pc \citep{Ruzmaikin1988,BS2005,Haverkorn2008,ss21}. The correlation length of the turbulent or random B-fields is comparable to or smaller than the coherent turbulent scale, $l$.  The fluctuation dynamos are also named small-scale dynamos, while mean-field dynamos are also referred to as large-scale dynamos. This depends on whether the B-field scale is well below or above the turbulent scale, respectively. In general, the dynamo action can work as follows: the turbulent motions driven by supernova explosions tangle B-fields in the ISM, causing a transfer of turbulent magnetic and kinetic energy to smaller scales, at which they are converted into heat (i.e. turbulent diffusion). The same turbulent motions can also amplify the B-fields via electromagnetic induction \citep{Zeldovich1990}.
 
How these dynamo processes can explain the wide range of observational features (Section \ref{subsec:RadioObs} and \ref{subsec:OIRObs}) remains poorly understood.  For example, \citet{VanEck2015} \citep[and more recently][]{Beck2019} compared the B-field strengths and morphology with the properties of the ISM for 20 nearby galaxies. They found that the total B-field strength is correlated with the star formation rate, $B_{\rm{tot}}\propto \rm{SFR}^{0.34\pm 0.04}$. These results confirm the theoretical prediction of $B_{\rm{tot}} \propto \rm{SFR}^{1/3}$ \citep{Schleicher2013,Schober2016}. These models were computed for a galaxy dominated by turbulence driven by supernova explosions and where the energy of the turbulent B-field is a fixed fraction of the turbulent energy. In addition, the ratio between the regular and ordered B-field is found to increase radially in most galaxies. This result implies that the relative contribution of the fluctuation dynamo becomes less efficient than the mean-field dynamo with the galactocentric radius.

Furthermore, the magnetic pitch angle, $\Psi_{\rm B}$, provides information about the large-scale B-field that can be used to refine dynamo models. From theory, the $\Psi_{\rm B}$ involves the ratio of the radial, $B_{r}$, and azimuthal, $B_{\phi}$, components of the B-field, such as $\Psi_{\rm B}= \arctan(B_{r}/B_{\phi})$ \citep{Krasheninnikova1989}. From observations, the $\Psi_{\rm B}$ is defined as the angle between the B-field orientation and the tangent to the local circumference at a given distance from the galaxy's center. The $\Psi_{\rm B}$ can be estimated using the inferred B-field orientation from radio polarimetric observations and corrected by Faraday rotation. \citet{Beck2019} compiled the magnetic and optical arm pitch angles of 19 galaxies and found that, for the most well-studied spiral galaxies, the magnetic pitch angle is mostly constant within the central $\sim5-10$ kpc with values in the range of $20-35^{\circ}$. They also found that the pitch angles of regular B-fields vary from those of the ordered B-fields, which may indicate a higher relative presence of the fluctuation dynamo in the pitch angle of the ordered B-fields. Interestingly, the magnetic pitch angle seems to be systematically offset by $5-10^{\circ}$ when compared with the molecular (CO) spiral arms \citep{VanEck2015}. Note that this systematic offset is derived using a logarithmic spiral function fitted to the CO spiral arms and a multi-mode logarithmic spiral B-field fitted to the magnetic arms. Then, a median pitch angle value of the entire galaxy is estimated and compared between tracers. A wavelet-based analysis of the spiral pattern in the galaxy M83 revealed that the average pitch angle of the ordered magnetic field is about $20^{\circ}$ larger than that of the material spiral arms \citep{Frick2016}. A homogeneous analysis of the pitch angles  as a function of the galactrocentric radius and galaxy structures is required to refine these results. 

The polarized intensity can be used as a proxy of the large-scale ordered B-field. Using this relation, \citet{Tabatabaei2016} found that the large-scale ordered B-field strength is proportional to the rotational speed of 26 nearby galaxies. This result may be due to a potential coupling between the large-scale B-field and the dynamical mass of galaxies. This result leads to the argument that the large-scale B-field is enhanced and ordered due to gas compression and local shear in the disk of galaxies, and hence is dominated by anisotropic random fields.  Finally, \citet{Chyzy2017} applied a principal component analysis (PCA) to several properties of the B-field and ISM using a sample of 55 nearby galaxies (i.e., low-mass dwarf, Magellanic-types, spirals, and starbursts). They also found that the B-field strength is tightly related to the surface density of the star formation rate, but it is also related to the gas density. Weak relationships are found between the B-field strength and the dynamic mass of galaxies and their rotation velocity. These results suggest that the B-field generation is related to the local star-formation activity, which involves the action of fluctuation dynamos. 

Although theoretical predictions and observations have been compared, the turbulent dynamo is still poorly constrained. To understand turbulent dynamos in galaxies, we are required to measure the strength and morphology of B-fields in the multi-phase ISM of several galaxy types across their evolution. However,  the angular resolution of the observations is typically larger than the coherence length of the turbulence in nearby galaxies. Thus, direct comparisons between observations and realistic magnetohydrodynamical (MHD) simulations are required to characterize the physical mechanisms producing the aforementioned observational results.

%%%%%%%%%%%%%%%%%%%%%%
\subsection{Magnetic fields in galaxy evolution}\label{subsec:BEvo}
%%%%%%%%%%%%%%%%%%%%%%

Despite the origin of a seed B-field at the early stage of the universe during its accelerated expansion \citep[e.g.][]{Widrow2002,Subramanian2019},  B-fields had to be amplified across cosmic time to explain the present-day B-fields in galaxies. In the cosmic web, B-fields may be amplified by fluctuation dynamos driven by accretion shocks. This mechanism transfers gravitational collapse into turbulent magnetic energy \citep[e.g.,][]{Ryu2008}, and may explain the measured B-field in cosmic filaments and galaxy clusters \citep{Vazza2014,Vazza2018}. Following the hierarchical structure of galaxy formation by the standard $\Lambda$ cold dark matter ($\Lambda$CDM) cosmological model, dwarf galaxies may be one of the first galaxies to form in the universe. Dwarf galaxies have more turbulent gas and slower rotations than nearby spiral galaxies. However, the measured B-fields in local dwarf galaxies have strengths and kpc-scale structures similar to spiral galaxies \citep{Chyzy2000,Chyzy2003,Kepley2010,Chyzy2016}. The B-fields in dwarf galaxies are puzzling, but they are crucial to understanding the evolution and amplification of B-fields during the early stages of the universe.

Minor and major mergers are thought to be a main mechanism for galaxy growth during the early stages of the universe \citep[e.g.,][]{Kerev2009}. These mergers affect the ISM of the galaxies, perturbing their B-fields. Studies on 16 interacting systems have shown that a) the mean ordered B-field strength, $14\pm5~\mu$G, is stronger than in non-interacting galaxies ($\sim5\pm2~\mu$G), b) the ratio of the ordered to random B-field is lower than in non-interacting galaxies, and c) the ordered B-fields are strong along the tidal tails of the interacting galaxies \citep{Chyzy2004,Drzazga2011}. A recent study \citep{ELR2021} on Centaurus~A, a remnant galaxy of a major merger between an elliptical and a spiral galaxy, has shown a highly perturbed B-field along the warped disk of the galaxy. The observed B-field is thought to be dominated by the fluctuation dynamo at scales $\le120$~pc. This result indicates that the measured B-field is the perturbed large-scale B-field of the original spiral galaxy after the merger. These results show that B-fields may be amplified by the enhancement of turbulent motions in the mergers driven by galaxy interaction. These strong B-fields may quench star formation \citep{Tabatabaei2018} and drive the gas flows in tidal tails. However, it remains unclear how the B-fields are amplified and when the B-field dissipates in mergers.

Mergers can trigger starbursts activity \citep[e.g.,][]{Elbaz2003,Tacconi2010,Pearson2019}, where the already amplified B-field from mergers can be further enhanced via turbulent dynamos due to supernova explosions \citep{Schober2013}. Indeed, the strongest B-fields, $50-300~\mu$G, are found at the core of starburst galaxies with their ordered B-fields following the galactic outflows \citep[e.g.,][]{Thompson2006,Heesen2011,Adebahr2013,Adebahr2017,ELR2021a}. B-field strengths of $0.5-18$~mG are measured in starbursts using the Zeeman effect in the OH megamasers \citep{Robishaw2008}. Furthermore, the tight FIR-radio correlation applies to redshifts of at least $z\sim3$ \citep{Murphy2009}. This result requires that the turbulent B-field and star formation are linked during the peak of star formation at $z\sim1$ \citep{Heavens2004,Seymour2008}. The strong and ordered B-fields found in starbursts have effects on galaxy evolution. Galactic outflows can drag the B-fields from the galaxy's disk away to the intergalactic medium (IGM), which is then permeated by B-fields  \citep[e.g.,][]{Bertone2006,Samui2018,Pakmor2020,MA2021,ELR2021a}. Galaxy formation simulations also find temporary merger-induced amplification, and attribute this amplification to a combination of compressional and merger-driven turbulence \citep{Kotarba2011,MA2018,Whittingham2021}. These simulations also find some resolution-dependent B-field dissipation once the merging process concludes. A variety of mechanisms (e.g., winds, B-fields, cosmic-rays, diffusion losses) are interconnected and still under investigation \citep[e.g.,][]{Lacki2010,Birnboim2015}.

The subsequent story of galaxy evolution from post-starburst \citep{Caldwell1999} to quiescent galaxies starts with a strong and ordered B-field. These B-fields are in close equipartition with the turbulent kinetic energy in the galactic outflows \citep{Thompson2006,ELR2021a}, which makes the B-fields dynamically important during, and after, the starburst activity. Large reservoirs of turbulent molecular gas and quenched star formation have been found in post-starburst galaxies \citep[e.g.,][]{Falgarone2017,Wild2020,Suess2022,French2022}. Strong B-fields have been found to be responsible for reducing star formation rates in molecular clouds \citep{MTK2006,Pattle2022} and modifying the ISM by forming filamentary structures \citep{Hennebelle2019}. Furthermore, strong B-fields in the starburst ring of NGC~1097 have been found to quench the formation of massive stars \citep{Tabatabaei2018}. These results may indicate that strong B-fields in post-starburst galaxies may quench star formation by keeping the molecular gas from gravitationally collapsing to form massive stars in the galaxy after the cease of starburst activity. To characterize the effect of B-fields on the reservoirs of molecular gas and the subsequent star formation activity, it is required to perform studies of the ordered B-fields in post-starburst galaxies.

The evolutionary stage toward quiescent galaxies is the longest timescale to order B-fields in the present-day spiral galaxies. The observed ordered B-field in nearby galaxies may be explained by an interplay between mean-field dynamos driven by differential rotation (and helical turbulence) in galaxies, fluctuation dynamos driven by feedback-driven mechanisms (from stellar and/or active galaxies), and/or magneto-rotational instability-driven dynamo \citep{Kitchatinov2004,WA2009,Arshakian2009,Rieder2016,Bambic2018,MA2018,Ntormousi2018,Ntormousi2020}. Recent efforts using MHD simulations have shown that the structure of a galactic disk can be modified when B-fields are included. For example, smaller disks and higher accretion flows toward their cores are found if strong primordial B-field strengths are assumed in MHD simulations \citep{MA2020}. More disk-dominated and more massive black holes are found if B-fields are added in Milky Way-mass like galaxies \citep{VandenVoort2021}. However, the exact physical processes of B-field amplification to reach present-day B-field measurements remain an open question.

To sum up, B-fields are amplified from seed values as a consequence of galaxy formation and turbulence-driven dynamos. Galaxy evolution strongly depends on the physics of the ISM. In addition, the ISM is permeated by B-fields, in which magnetic energy is in close equipartition with the kinetic energy. The B-fields can be dynamically important at several stages of galaxy evolution. These B-fields generated via MHD dynamos can affect gas flows in the ISM and drive gas inwards toward the galaxy's center and outwards toward the intergalactic medium via galactic outflows. Thus, B-fields remain an important, but still largely ignored, element for understanding the evolution of galaxies across cosmic time.

%%%%%%%%%%%%%%%%%%%%%%%%%%%%%
%%%%% LEGACY'S OVERVIEW %%%%%
%%%%%%%%%%%%%%%%%%%%%%%%%%%%%

\section{Overview of the Legacy Program}\label{sec:LegOver}

We present the galaxy sample, observing strategy, and scientific goals of the program.

%%%%%%%%%%%%%%%%%%%%%%
\subsection{Measuring magnetic fields using far-infrared polarimetry}\label{subsec:BFIR}
%%%%%%%%%%%%%%%%%%%%%%

The results from this project rely on the fact that the dust grains are magnetically aligned with the local B-field. We summarize the fundamentals of dust-grain magnetic alignment theory. Then, we provide the observational signatures to study the B-fields in galaxies using FIR polarimetric observations.

Theories of dust grain alignment mechanisms have been in active development since the first discovery of the polarization signature of aligned elongated dust grains in the ISM by \citet{Hiltner1949} and \citet{Hall1949}. The Radiative Alignment Torques (RATs) is the leading theory describing dust grain alignment  \citep[e.g.,][]{Andersson2015,HL2016}. The RATs theory can be summarized as follows: Starlight radiation transfers angular momentum to the dust grains in the ISM. The dust grains absorb more radiation along one of their axes due to their asymmetric geometrical nature and end up spinning along the axis with the greatest moment of inertia (i.e., the minor axis of the dust grain). Since most dust grains are paramagnetic, they acquire a magnetic moment experiencing the fast Larmor precession. After a relaxation time the dust grains will be precessing along the orientation of the local B-field. The final configuration is one where the long axis of the dust grains is perpendicular to the orientation of the local B-field. 

The absorptive polarization has a $PA$ of polarization parallel to the local B-field, while the emissive polarization has a $PA$ of polarization perpendicular to the local B-field. This mechanism is known as B-RATs, and it is the most common property for interpreting polarimetric observations tracing the absorptive/emissive polarization in the IR and sub-mm wavelength range. Indeed, the induced polarization from dust grain alignment by RATs has been extensively used to interpret observations of the ISM in our Galaxy \citep[][for a review]{Andersson2015} and other galaxies \citep[e.g.][]{Elvius1962,Elvius1964,Jones2000,ELR2018,Jones2019,ELR2020,Jones2020,ELR2021a,ELR2021,Borlaff2021,Pattle2021b}. 

Polarized thermal FIR emission arising from magnetically aligned dust grains traces the ordered B-fields in the plane of the sky (POS). The inferred B-field orientation from the HAWC+ observations is computed after the measured polarization angle (i.e., E-vector) is rotated by $90^{\circ}$. Note that the angle of the measured B-field has a $180^{\circ}$ ambiguity due to the fact that FIR polarization is insensitive to the B-field direction. Thus, we only measure the B-field orientations on the POS using FIR polarimetric observations. In addition, the measured B-field orientation is the weighted-averaged column density along the LOS and on the POS within the several hundred pc size of the resolution elements of the observations. The angular resolutions of HAWC+ are in the range of $5-18$\arcsec, which correspond to spatial scales of $\sim500$~pc for a typical nearby galaxy at $\sim10$ Mpc and at an angular resolution of $10$\arcsec. The total FIR emission of galaxies has a typical vertical height of $0.1-0.6$~kpc \citep{Verstappen2013}. Furthermore, the FIR polarimetry is not directly sensitive to the B-field strength, and the spatial resolution of the observations, few $100$~pc, is larger than the turbulent scales, $l\sim50-100$ (Section \ref{subsec:BOri}), driven by supernova explosions in galaxies. Thus, the B-field strength cannot be estimated using the current set of observations. Except for the rare occasion of M82 at $53$~\um, where \citet{ELR2021a} modified the Davis-Chandrasekhar-Fermi method \citep{Davis1951,CF1953} to account for the large-scale galactic outflow. 

The total intensity is sensitive to the dust grain column density and dust temperature. The polarization fraction is sensitive to the turbulent and/or tangled (i.e., random) B-fields along the LOS, the B-field orientation with the plane of the sky, dust grain column density, and dust temperature. For a `perfect' condition where the B-field orientation is parallel to the POS and there is no variation in neither tangled B-fields, turbulence, or inclination effects, then the polarization fraction is constant for an ordered B-field (i.e., maximum dust grain alignment efficiency). The polarization fraction decreases when the averaged anisotropic random B-field decreases, and turbulence, tangled B-fields and inclination effects increase \citep[i.e.,][]{Chen2019,King2019}. The FIR polarization fraction also decreases when the angle of the B-field and the plane of the sky increases (Section  \ref{subsec:DIS_Inc}). Unpolarized thermal emission can be attributed to isotropic B-fields and/or a B-field oriented perpendicular to the plane of the sky.

%%%%%%%%%%%%%%%%%%%%%%
\subsection{Science goals}\label{subsec:SciCas}
%%%%%%%%%%%%%%%%%%%%%%

The science goals driving SALSA are motivated by the characterization of the B-fields in the multi-phase ISM of galaxies. As mentioned in Section \ref{sec:Back}, our knowledge about extragalactic magnetism has been purely based on radio polarimetric observations tracing the B-fields in the warm and diffuse ISM. SALSA provides a new ingredient by doing what only HAWC+/SOFIA can do: measuring the B-fields in the dense and cold ISM of galaxies. Since the gas dynamics and star formation activity are the main drivers of the amplification and evolution of B-fields, our analysis is supported with observations of neutral and molecular gas as well as star formation activity tracers.  The sample selection (Section \ref{subsec:DataSample}) was designed to address these core goals:

\begin{itemize}
    \item[1.] \textit{Characterize the dependence of the large-scale ordered B-fields on the probed ISM phase, and provide measurements for galactic dynamo theories.}
\end{itemize}

Spiral structures in galaxies can be described as density perturbations propagating in a multi-phase ISM and multi-component stellar disk \citep[][for a review]{Shu2016}. Several theories have been suggested to explain the formation of large-scale spiral arms: a) as dynamic transients, where the spiral arms can change during a galaxy's lifetime \citep[e.g.,][]{Sellwood1984,Elmegreen1993,Michikoshi2016}, b) as constant density waves, where the spiral arms are stationary patterns that last most of the galaxy's lifetime \citep[e.g.,][]{Bertin1989,Shu2016}, and/or c) as the result of tidal encounters \citep[e.g.,][]{Toomre1972,Elmegreen1991,Pettitt2017}. Density wave theory predicts that stars form in the arms as gas moves into a wave that is compressed by its gravitational potential. Under this scheme, the pitch angle should be different for different tracers of star formation (e.g., molecular clouds, HII regions, and newly formed stars) because they appear at different phases of a wave. The pitch angle at optical/NIR wavelengths is expected to trace already born stars. FIR wavelengths trace spiral density waves colocated with newly born stars where the radiation arises from thermal emission from dust grains. Radio wavelengths trace synchrotron emission from a volume-averaged electron density region along the halo and disk of the galaxy. Despite that a unique, or a combined, theory can describe the evolution of spiral galaxies, these theories have to predict the large-scale ordered B-fields found in all spiral galaxies using polarimetric observations \citep[e.g.,][]{BW2013,Beck2019}. 

B-fields should be compressed and sheared by the action of these density waves, adding another observable tracer to the previous photometric ones. Constant offsets within the full disk between photometric and polarimetric tracers are expected for a density wave and were indeed observed in M51 \citep{Patrikeev2006}. Galaxy dynamo theories also have to predict the differences between the morphological spiral arms traced by neutral and molecular gas and the large-scale ordered spiral B-fields traced by radio polarimetric observations \citep{VanEck2015,Beck2019}. HAWC+ polarimetric observations at $89$ \um~of NGC 1068 showed the first detection of a $\sim3$ kpc large-scale ordered spiral B-field structure \citep{ELR2020}. The B-fields were traced by means of thermal polarized emission arising from magnetically aligned dust grains. The measured ordered B-field morphology was characterized by a two-dimensional logarithmic spiral B-field model with a pitch angle of $16.9^{+2.7\circ}_{-2.8}$ and a disk inclination of $48\pm2^{\circ}$. The matter sampled by the estimated $89$~\um~magnetic pitch angle follows the starbursting regions with a pitch angle of $15^{\circ}$ using H$_{\alpha}$ along the spiral arms \citep{Emsellem2006}. The magnetic pitch angle shows an offset from that of the molecular gas with a pith angle of $7-10^{\circ}$ \citep{Planesas1991}. These results show a constant offset between the FIR and radio magnetic arms and the molecular gas arms using an average pitch angle across the galaxy's disk.

The wavelet-based analysis in M83 showed that one gaseous arm is displaced from the magnetic arm, while the other magnetic arm overlaps with the associated gaseous arm \citep{Frick2016}. This result was interpreted as the interaction between a galactic dynamo with a transient spiral arm. More recently, \citet{Borlaff2021} measured that the ordered large-scale B-fields in M51 do not necessarily have the same morphology using  FIR, radio, CO, and HI tracers. Specifically, the spiral arms are wrapped tighter at HI and CO than at FIR and radio wavelengths. These authors performed a study of the radial profiles of the spiral arms as a function of the galactrocentric radius. Results showed that the $154$~\um~ordered large-scale B-field is wrapped tighter at the outer parts, $r>5$ kpc, of the galaxy than those traced by radio polarimetric observations. This result has been interpreted as a consequence of galaxy interaction and/or star formation activity in the outerskirts of M51. Furthermore, several modes of an axysymmetric B-field toward the galactic center of  NGC~1097 were measured using FIR and radio polarimetric observations \citep{ELR2021c}. Specifically, a constant B-field traced at $89$ \um~was measured to be associated with the contact regions between the bar and the starburst ring within the central 1 kpc of NGC1097. A superposition of spiral B-fields at $3$ and $6$~cm was measured outside and within the starburst ring. These results deviate from the constant density wave theory in terms of how the spiral arms behave as a function of the galactrocentric radius and the specific tracer used to describe the spiral arms. Other physical mechanisms, such as transient arms and/or galaxy interaction may need to be taken into account to explain these measurements. However, only a handful of spiral galaxies have been studied thus far. 

In general, density wave theories do not include angular momentum transfer through the action of B-fields. However, a galaxy like NGC~1097 suggests that B-fields drive inflows on kpc scales via MHD dynamos, and galaxies like M51 and M83 suggest that magnetic pitch angles change with the galactrocentric distance due to galaxy interaction and/or star formation activity. It is mandatory to review the relative contribution of gravitational and magnetic forces in the global build up of spiral arms. We need a comprehensive spiral density wave theory that includes these two forces.

One of the goals of SALSA is to provide the B-field structure along the ongoing star-forming regions traced by FIR thermal polarized emission arising from magnetically aligned dust grains. These results are complementary to the B-fields structures traced by radio polarimetric observations where the polarized emission is mostly spatially located with the interarm regions and is strongest at the inner edges of arms (Section \ref{subsec:RadioObs}). In combination with neutral and molecular gas observations, SALSA aims to trace the B-field and morphological structures of the spiral arms and interarm regions as a function of the multi-phase ISM and gas dynamics in the disk of spiral galaxies. This study provides empirical inputs for dynamo models that can be used to obtain a conclusive theory of the evolution of spiral galaxies.

\begin{itemize}
    \item[2.] \textit{Quantify the relative contributions of the mean-field and fluctuation dynamos across a range of galaxy types and dynamical states.}
\end{itemize}

Present-day B-fields in galaxies are a consequence of turbulence-driven dynamos (Section \ref{subsec:BOri}). However, the relative contribution of the fluctuation and mean-field dynamos, and the physical conditions in the galaxy at which B-fields are amplified and dissipated are still under study. The current angular resolutions of the observations in radio and FIR, $5-20$\arcsec\ ($\ge 100$ pc), are too large to resolve the fluctuation dynamo (i.e., coherence length of the turbulence) in nearby galaxies. Our measurements rely on the relative contribution of the fluctuation and mean-field dynamo across the several structures within the disk of galaxies.

\citet{Krause2020} detected polarized synchrotron emission of $35$ edge-on galaxies with vertical scale heights in the range of $1-6$ kpc. The large-scale B-fields are dominated by an X-shape morphology above and below the galaxy's disk with a mostly unpolarized disk due to strong Faraday rotation. The thermal FIR total emission along the vertical scale height of edge-on galaxies has been measured to be in the range of $0.1-0.6$~kpc at $100-160$~\um~\citep{Verstappen2013}. The FIR thermal emission arises from dust grains heated by star-forming regions in the galaxy's disks. As an example, the thermal FIR polarized emission of the edge-on spiral galaxy NGC~891 has a vertical height of $\le0.5$~kpc \citep{Jones2020}.  Thus, radio polarimetric observations trace the B-fields as a weighted-average of the electron column density along the LOS across several kpcs above and below the galaxy's disk. The radio polarimetric observations suffer from Faraday rotation. FIR polarized emission traces the weighted-average column density B-field in the midplane of the galaxy and does not suffer Faraday rotation. Continuing with the example of NGC~891, the vertical height of the molecular gas, $^{12}$CO(1-0) and $^{12}$CO(3-2), is $\sim0.4$ kpc \citep{Hughes2014}, while the warm ionized gas extends up to $\sim2$ kpc beyond the disk \citep{Rand1990,Reach2020}. The bulk of star-forming regions and molecular gas is located close to the midplane of the galaxy, as well as the bulk of the thermal FIR emission. Thus, FIR polarimetric observations are a powerful tool for studying the effect of the gas kinetic energy in the ISM and star-forming regions on the relative contribution of the turbulence-driven dynamos across the galaxy's disk. 

Using HAWC+ polarimetric observations of the face-on spiral galaxy M51, \citet{Borlaff2021} showed that the arms have a relative increase of small-scale turbulent B-fields in regions with increasing column density and $^{12}$CO(1-0) velocity dispersion. The column density and velocity dispersion of the molecular gas were used as proxies of the turbulent kinetic energy of the gas in the ISM across the disk. In addition, these authors found that a) the polarized synchrotron emission is strongly correlated with star formation rate except in the FIR, and b) the polarized fraction decreases faster at FIR than at radio wavelengths with the star formation rate. These results imply an increase of the fluctuation dynamo in the star-forming regions, specifically, an increase of isotropic turbulent B-fields traced by FIR and an increase of anisotropic turbulent B-fields traced by radio. 

For the merger remnant Centaurus A, \citet{ELR2021} showed that the measured B-field in the warped disk is dominated by the fluctuation dynamo. This result was based on the decrease of the polarization fraction with a) increasing column density, used as a proxy for the turbulent and/or tangled B-fields along the LOS, b) increasing $^{12}$CO(1-0) velocity dispersion, used as a proxy of the turbulent kinetic energy of the molecular gas, and c) increasing dust temperature, used as a proxy of the radiation field in the ISM. In addition, the measured angular dispersion between the beam of the observations were larger than those associated with the individual uncertainties of the measured polarization angles. The measured position angles of polarization across the warped disk of Centaurus A were also not fully explained by a three-dimensional model of an ordered B-field. These early results have shown the potential of using FIR polarimetric observations to estimate the contribution of the fluctuation dynamo in the disk of galaxies. 

SALSA aims to provide a statistical sample of nearby galaxies with resolved FIR polarimetric observations. These observations can be used to analyse the effects of star formation and tidal interactions on the turbulent dynamos across the disk of galaxies and galaxy types. These results may need further realistic MHD simulations to disentangle the dominant physical mechanism associated with the small-scale turbulent B-field below the resolution elements of our observations \citep[e.g.,][]{Pakmor2014,Ntormousi2018,MA2018,Steinwandel2019,Ntormousi2020,MA2022}.

\begin{itemize}
    \item[3.] \textit{Measure the effect of galactic outflows on the galactic B-field in starburst galaxies, and quantify how the B-field is transported to the circumgalactic medium.}
\end{itemize}

Starburst galaxies may be important contributors to the magnetization of the IGM in the early universe \citep[e.g.,][]{Kronberg1999,Bertone2006}. Starbursts can amplify the galactic B-field and drag it away, magnetizing the circumgalactic medium (CGM) by means of fluctuation dynamos driven by supernovae (Section \ref{subsec:BEvo}). Nearby starburst galaxies are excellent astrophysical laboratories due to the presence of enhanced star formation and the associated galactic outflows that may be magnetizing and enriching the IGM with new elements formed in the star-forming regions. These structures can be resolved with current optical/NIR, FIR and radio polarimetric observations. 

Optical polarimetric observations are highly contaminated by dust and electron scattering, making it very complicated to characterize the signature of magnetically aligned dust grains in starbursts (Section~\ref{subsec:OIRObs}). Radio polarimetric observations of M82 show a dominant magnetized bar along the galaxy's disk \citep{Heesen2011,Adebahr2013,Adebahr2017}. However, these observations suffer from strong Faraday rotation and only hints of the helical B-fields along the outflow were detected. 

Preliminary HAWC+ results at $53$ \um~showed that the B-field structure traced by magnetically aligned dust grains is perpendicular to the disk and parallel to the galactic outflow in M82 \citep{Jones2019}. A further detailed analysis of these data found that the turbulent kinetic energy and the turbulent magnetic energy are in close equipartition within the central $\sim2$ kpc along the galactic outflow \citep{ELR2021c}. The B-fields in the galactic outflow of M82 remain `open' up to $10$ kpc scales above and below the galactic disk. These results indicate that the B-fields in the galactic outflow of M82 are frozen into the ionized outflowing medium and driven away kinetically. In addition, small-scale turbulent B-fields arising from a bow-shock are present within the outflow. These observations show that FIR polarimetric observations can trace and characterize the physical mechanisms of the B-field along the galactic outflow in starbursts.

SALSA aims to perform a multi-wavelength analysis of the nearest and brightest starburst galaxies. These new observations can be used to measure the effect of the galactic outflow properties (i.e., mass, ejected energy), dust properties (i.e., density, alignment efficiency, temperature), and CGM properties (i.e., tidal tails, gas velocity and composition) on the galaxy's B-field. These results are crucial to understanding the mechanisms of B-field amplification driven by supernovae and magnetization of the CGM, where nearby starburst can be used as analogues of galaxies at the peak of star formation at $z\sim 1-2$.

\begin{itemize}
    \item[4.] \textit{Measure the B-field structures of interacting galaxies.}
\end{itemize}

Some of the fundamental questions in galaxy evolution can be distilled to: what happens when galaxies collide and merge? What is the effect of this interaction on star-forming regions? Major and minor merging events may perturb the B-field and enhance it by fluctuation dynamos (Sections~\ref{subsec:RadioObs} and \ref{subsec:BEvo}). If the B-field is compressed, it will lead to strong radio polarized emission \citep{Arshakian2009}. Here, we are interested in characterizing how the ordered B-fields of galaxies are distorted due to the interaction and merger of galaxies and their influence in star formation activity. Given the analysis on spiral galaxies and starbursts in this project (Goals 1--3), SALSA provides the foundational framework to characterize more complex galaxies. 

As mentioned in Section \ref{subsec:BEvo}, the measured turbulent B-field in Centaurus A can be interpreted as the disturbed large-scale ordered B-field of the spiral galaxy after the major merger \citep{ELR2021}. Unfortunately, radio polarimetric observations of the entire galaxy are not available, as most of the analysis at radio wavelengths have been focused on the active galactic nucleus. Thus, FIR polarimetric observations provide a new window to characterize the B-fields in the dense ISM of highly turbulent galactic disks. In addition, the differences in the B-field between FIR and radio wavelengths in M51 \citep{Borlaff2021} can be a consequence of the interaction between M51a and M51b. Further observations of other interacting systems are required to understand their effect on the B-field morphology.

Merging galaxies are typically too faint for the HAWC+ sensitivities, or too close to each other for the angular resolutions ($5-18\arcsec$) of HAWC+. Based on the selection criteria explained in Section \ref{subsec:DataSample}, the Antennae galaxy is the only merging galaxy that can be measured with HAWC+. This galaxy is a well-studied, bright, and the nearest merger with extensive archival data. Radio polarimetric observations of the Antennae galaxy have been used to estimate the B-field strength in the diffuse ISM \citep{Chyzy2004}. However, the densest regions of the merger are depolarized. Thus, FIR polarimetric observations are needed. SALSA will measure the B-field structure in the areas of interaction between both galaxies in the Antennae system. These observations can be used to characterize the B-fields with the ongoing star formation along the gas streams between the merging galaxies.

%%%%%%%%%%%%%
\subsection{Galaxy sample}\label{subsec:DataSample}

Our galaxy sample represents the most accessible and technically feasible galaxies in the nearby ($<20$~Mpc) universe to be observed with HAWC+. These galaxies are observed within a maximum on-source time of $7$h per object per band to optimize, and be compatible with, the flight planning with other HAWC+ proposals within the multi-year period of this SOFIA Legacy Program. The full program was maximized to be within the 200h of execution time offered per SOFIA Legacy Program. Within this time, we selected a flux-limited galaxy sample of bright and nearby galaxies of several representative galaxy types to accomplish the scientific goals described in Section~\ref{subsec:SciCas}. Table~\ref{tab:GalaxySample} shows the properties of the galaxy sample. The selection criteria was based on the following: 

\begin{itemize}
\item The available data in the literature that support the physical interpretation of the HAWC+ observations. The available data include observations from VLA/Effelsberg, \textit{Herschel}, \textit{HST}, and ALMA,  
\item The HAWC+ sensitivity to observe extended polarized emission within $\sim2-10$~kpc diameter in $\le7$ hours of observations,
\item Having several representative galaxy types that can be used to characterize the B-fields from spiral galaxies to more complex galaxies (i.e., interaction, starbursts, active galactic nuclei), 
\item Having complementary FIR polarimetric observations of the brightest galaxies already observed from previous HAWC+ programs. These new observations can be used to perform multi-wavelength analysis.
\end{itemize}

The galaxy sample was originally selected from the Atlas of Galaxies\footnote{Atlas of Galaxies observed with radio polarimetric observations can be found at \url{https://atlas-of-galaxies.mpifr-bonn.mpg.de/introduction}} led by Maja Kierdorf and Rainer Beck at the Max Planck Institute for Radio astronomy (MPIfR) in Bonn, Germany. These galaxies are bright, nearby, and well-studied using radio polarimetric observations (i.e., VLA, and Effelsberg) with a broad wavelength coverage from $3$ to $23$~cm with angular resolutions of $8-20$\arcsec. Radio polarimetric observations have similar angular resolution as those from HAWC+ (Table \ref{tab:HAWC}).

%%%%%%%%%%%%%%%%%
%%%% TABLE 1 %%%%
%%%%%%%%%%%%%%%%%
\begin{deluxetable*}{lcccccl}
\centering
\tablecaption{Galaxy Sample. \emph{Columns, from left to right:} (a) Galaxy name, (b) galaxy distance in Mpc, (c) physical scale in pc per arcsec., (d) galaxy type, (e) inclination of the galaxy in degrees, (f) position angle of the long axis of the galaxy in the plane of the sky, (g) references associated to the distance, inclination, and tilt angles. 
\label{tab:GalaxySample} 
}
\tablecolumns{6}
\tablewidth{0pt}
\tablehead{\colhead{Galaxy} & 	\colhead{Distance$^{1}$}  & \colhead{Scale} & \colhead{Type$^{\star}$} & 
\colhead{Inclination ($i$)$^{2}$} &	\colhead{Tilt ($PA$)$^{2}$} &  \colhead{References} \\ 
 	&  \colhead{(Mpc)}	& \colhead{(pc/\arcsec)}	&
\colhead{($^{\circ}$)} & \colhead{($^{\circ}$)} & \colhead{($^{\circ}$)} \\
\colhead{(a)} & \colhead{(b)} & \colhead{(c)} & \colhead{(d)} & \colhead{(e)} & \colhead{(f)} & \colhead{(g)}} 
\startdata
Centaurus A 	&	$3.42$	&	$16.42$	&	S0pec/Sy2/RG	&	$83\pm6$		&	$114\pm4$	&	
$^{1}$\citet{Ferrarese2007}; $^{2}$\citet{Quillen2010}	\\
Circinus 		&	$4.20$	&	$20.17$	&	SA(s)b/Sy2	&	$40\pm10$	&	$205\pm10$	&	
$^{1}$\citet{Tully2009}; $^{2}$\citet{Jones1999}			\\
M51 		&	$8.58$	&	$41.21$	&	Sa			&	$22.5\pm5$	&	$-7\pm3$		&
$^{1}$\citet{McQuinn2017}; $^{2}$\citet{Colombo2014}	\\
M82 		&	$3.85$	&	$18.49$	&	I0/Sbrst		&	$76\pm1$		&	$64\pm1$		&
$^{1}$\citet{Vacca2015}; $^{2}$\citet{Mayya2005}		\\
M83 		&	$4.66$	&	$22.38$	&	SAB(s)c		&	$25\pm5$		&	$226\pm5$	&
$^{1}$\citet{Tully2013}; $^{2}$\citet{Crosthwaite2002}	\\
NGC~253 	&	$3.50$	&	$16.81$	&	SAB(s)c/Sbrst	&	$78.3\pm1.0$	&	$52\pm1$		&
$^{1}$\citet{RS2011}; $^{2}$\citet{Lucero2015}		\\
NGC~1068 	&	$14.40$	&	$69.16$	&	(R)SA(rs)b/Sy2	&	$40\pm3$		&	$286\pm5$	&
$^{1}$\citet{BH1997}; $^{b}$\citet{Brinks1997}			\\
NGC~1097 	&	$19.10$	&	$92.21$	&	SB(s)b/Sy1	&	$41.7\pm0.6$	&	$133.0\pm0.1$	&
$^{1}$\citet{Willick1997}; $^{2}$\citet{Hsieh2011} 		\\
NGC~2146 	&	$17.20$	&	$82.61$	&	SB(s)ab/Sbrst	&	$63\pm2$		&	$140\pm2$	&
$^{1}$\citet{Tully1988}; $^{2}$\citet{Tarchi2004} 		\\
NGC~3627 	&	$8.90$	&	$42.75$	&	SAB(s)b		&	$52\pm1$		&	$176\pm1$	&
$^{1}$\citet{Kennicutt2003}; $^{2}$\citet{Kuno2007} 		\\
NGC~3628 	&	$10.95$	&	$52.60$	&	Sb pec	&	$91.6\pm2.0$	&	$90\pm2$		&	
$^{1,2}$\citet{Shinn2015}\\
NGC~4038 	&	$22$		&	$105.67$	&	SB(s)m pec	&	$59\pm10$	&	$55\pm10$	&
$^{1}$\citet{Schweizer2008}; $^{2}$\citet{Amram1992} \\
NGC~4631 	&	$7.6$ 	&	$36.51$	&	SB(s)d	&	$89\pm1$	&	$85\pm1$	&
$^{1}$\citet{Seth2005}; $^{2}$\citet{MP2019}\\
NGC~4736 	&	$5.3$	&	$25.46$	&	SA(r)ab		&	$36\pm7$		&	$292\pm2$	&	
$^{1}$\citet{Kennicutt2003}; $^{2}$\citet{Dicaire2008} 		\\
NGC~4826 	&	$5.60$	&	$26.89$	&	(R)SA(rs)ab	&	$65\pm5$		&	$125\pm5$	&
$^{1}$\citet{Kennicutt2003}; $^{2}$\citet{Braun1994} 		\\
NGC~6946 	&	$6.80$	&	$32.66$	&	Sc			&	$38.4\pm3.0$	&	$239\pm1$	&
$^{1}$\citet{Karachentsev2000}; $^{2}$\citet{Daigle2006} \\
NGC~7331 	&	$15.7$	&	$75.40$	&	SA(s)b		&	$78.1\pm2.7$	&	$165\pm1.2$	&
$^{1}$\citet{Kennicutt2003}; $^{2}$\citet{Daigle2006}		\\	
\enddata
\tablenotetext{{\star}}{Galaxy type from NASA/IPAC Extragalactic Database (NED; \url{https://ned.ipac.caltech.edu/})}
\end{deluxetable*}

%%%%%%%%%%%%%
%%%% TABLE 2 %%%%
%%%%%%%%%%%%%
\begin{deluxetable}{cccccc}
\centering
\tablecaption{HAWC+ configuration. \emph{Columns, from left to right:} a) Band name, b) central wavelength of the band in \um, c) FWHM bandwidth in \um, d) pixel scale of the band in \arcsec, e) FWHM in \arcsec, and f) FOV for polarimetric observations in \arcmin.
\label{tab:HAWC} 
}
\tablecolumns{6}
\tablewidth{0pt}
\tablehead{\colhead{Band}	&	\colhead{$\lambda_{\rm c}$} &	\colhead{$\Delta \lambda$}	&	\colhead{Pixel scale}	&	\colhead{Beam size}	& \colhead{Polarimetry FOV}  \\ 
 	& \colhead{(\um)}	& \colhead{(\um)} &  \colhead{ (\arcsec)} & \colhead{(\arcsec)}  & \colhead{(\arcmin)} \\
\colhead{(a)} & \colhead{(b)} & \colhead{(c)} & \colhead{(d)} & \colhead{(e)} & \colhead{(f)} 
}
\startdata
A	& $53$	&	$8.7$&	$2.55$	&	$4.85$	&	$1.4\times1.7$ \\
C	& $89$	&	$17$	&	$4.02$	&	$7.8$	&	$2.1\times2.7$	\\
D	& $154$	&	$34$	&	$6.90$	&	$13.6$	&	$3.7\times4.6$	\\
E	& $214$	&	$44$	&	$9.37$ 	&	$18.2$	&	$4.2\times6.2$  \\	
\enddata
\end{deluxetable}
%%%%%%%%%%%%%%%%%%%%%%%%%%%%%

\textit{Herschel} data are used to a)  estimate the feasibility of the observations with HAWC+ in the $53-214$ \um\ wavelength range, and b) compute the column density and dust temperature across the galaxy disk by using $70-250$ \um\ PACS and SPIRE observations. Those galaxies without available data in the \textit{Herschel} Archive\footnote{\textit{Herschel} Archive can be found at \url{http://archives.esac.esa.int/hsa/whsa/}} were removed from the galaxy sample. 

Since one of the goals of this project is to characterize the B-field orientation with the underlying gas dynamics of the ISM, we selected galaxies with available neutral and molecular gas observations at similar, or better, angular resolution as those from HAWC+. ALMA observations provide the kinematics of the molecular gas, CO(1-0) and CO(2-1), of the galaxy disks. Objects with ALMA observations from the Physics at High Angular resolution in Nearby Galaxies (PHANGS) project\footnote{PHANGs project can be found at \url{https://sites.google.com/view/phangs/home}} \citep{Leroy2021}, and Nobeyama observations from the CO multi-line imaging of nearby galaxies (COMING) project \citep{Sorai2019} were selected. Spiral galaxies without HI data from The HI Nearby Galaxy Survey \citep[THINGS\footnote{THINGS project: \url{https://www2.mpia-hd.mpg.de/THINGS/Data.html}};][]{Walter2008} were removed from the sample. In addition, ultraviolet (UV) emission can be used as a proxy for star formation activity. Galaxies without available UV imaging observations (i.e., \textit{HST}/WFC3/F438W) from the \textit{HST} Archive\footnote{\textit{HST} Archive can be found at \url{https://archive.stsci.edu/missions-and-data/hst}} were removed from the sample. 

Given the total surface brightness of the galaxies using the \textit{Herschel} observations, we finally selected galaxies that can be observed within $\le7$h on-source time with the sensitivity criteria described below. The $7$h on-source time limit is based on how the observations are scheduled by SOFIA flight planning. The $7$h limit allows SOFIA to obtain the full observations of a galaxy within two flights, with an observing time leg of $3.5$h each. The maximum leg of a single flight is $\sim4$h, as SOFIA has to return to the same base, i.e., Palmdale in California, at the end of the flight. An optimal flight has two science legs of $3.5$h and a single leg of $0.5$h for calibrations, yielding a total time of $8$h of science observations per flight. Using this scheme, a minimum of two galaxies can be scheduled for observations within the same flight. Note that the aircraft has to return to the same base, so each of the 3.5h science legs point to opposite parts of the sky. In addition, given that the galaxies are uniformly distributed in the sky, this program is ideal for flight planning compatible with over-subscribed regions like the Galactic Center. This approach helps flight planners who schedule SOFIA flights.

The individual integration times for the imaging polarimetric mode were obtained using the SOFIA Instrument Time Estimator (SITE\footnote{SITE can be found at \url{https://dcs.arc.nasa.gov/proposalDevelopment/SITE/index.jsp}}) at $53$, $89$, $154$ and $214$ \um\ using the chop-nod (C2N) polarimetric technique. To obtain statistically significant polarimetric measurements of the diffuse polarized emission across the host galaxy, a minimum signal-to-noise radio (SNR) of 4 in the polarization fraction is assumed. We assumed an expected polarization fraction of $\sim3$\% at the lowest level total surface brightness of each galaxy, yielding an uncertainty polarization fraction $\le0.75$\% across the final polarization map. Based on HAWC+ observations of M82 \citep{Jones2019}, NGC~1068 \citep{ELR2020}, and Centaurus~A \citep{ELR2021}, the host galaxy typically shows levels of polarization fraction $\ge3$\%. We took a conservative approach for the estimation of the lowest level total surface brightness expected to be observed. The lowest total surface brightnesses that accomplish the above requirements are $0.053$~Jy~sqarcsec$^{-1}$ at 53~\um, $0.016$~Jy~sqarcsec$^{-1}$ at $89$~\um, and $0.005$ Jy sqarcsec$^{-1}$ at $154$~\um, and $0.0029$ Jy sqarcsec$^{-1}$ at $214$~\um.  Objects in the Atlas of Galaxies with \textit{Herschel} observations whose extended emission with a diameter of $2$~kpc is lower than the surface brightness quoted above were removed from the final sample of this project. 

The final sample (Table~\ref{tab:GalaxySample}) is comprised of $17$ bright and nearby ($\le 20$ Mpc) galaxies with a total execution time of $155.70$h. We requested observations of $15$ galaxies, as high quality observations of the other two galaxies (i.e., M51 and Circinus) are already available. Of these $15$ galaxies, $9$ galaxies are new, and we propose deeper and multi-wavelength polarimetric observations of $6$ galaxies to obtain high-quality data products to analyze the polarization spectrum of galaxies. In general, the large angular resolution, $18.2$\arcsec\ full-width at half-maximum (FWHM), achieved at $214$ \um\ is insufficient to resolve the galactic structure in most of our targets. Thus, $214$~\um\ was not requested, except for the starburst galaxies M82 and NGC~2146.

%%%%%%%%%%%%%
\subsection{Observing strategy}\label{subsec:OBS}

All the observations in this program were performed using the newly implemented on-the-fly-mapping (OTFMAP) polarization mode with HAWC+. In Paper~III \citep{ELR2022}, we described the data processing and presented the pipeline steps to obtain homogeneously reduced high-level data products of the galaxies in our sample. In summary, the  OTFMAP  observing strategy  moves  the  telescope following a parametric curve (i.e., Lissajous curve) at a constant mapping speed within a specific scan size. For all observations, the objects are always within the field-of-view (FOV) of the HAWC+ array in any given band. This configuration maximizes the on-source time of the object.  Table~\ref{tab:HAWC} shows the HAWC+ configuration per band used in this program. The most interesting result is that by using the OTFMAP polarization mode, we improved the observing overheads by a factor 2.34, and the sensitivity by a factor of 1.80 when compared to the commonly used C2N polarization mode. The OTFMAP is a significant optimization of the polarimetric mode of HAWC+ as it ultimately reduces the cost of operations of SOFIA by increasing the science collected per hour of
observation up to an overall factor of 2.49. All observations of galaxies, except for M51, were, and will be, performed using the OTFMAP polarimetric mode of HAWC+ (Paper~III). 

The final data products contain the Stokes $IQU$, polarization fraction ($P$), position angle of polarization ($PA$), polarized intensity ($PI$), and its associated uncertainties with a pixel scale equal to the detector pixel scale (i.e., Nyquist sample) in any given band. The polarization fraction has been debiased and corrected by instrumental polarization and polarization efficiency. The released files have the same format as shown in Table~2 of \citet{Gordon2018}. We point the reader to Paper~III for a full description of the observing strategy and data reduction of the data release. All datasets are available in the Legacy Program website \footnote{SALSA data products are available at: \url{http://galmagfields.com/\#data-section}}.

Note that at the time of writing the original SOFIA Legacy Program proposal, the OTFMAP polarization mode was under commissioning as part of an engineering time proposal for HAWC+. With the expectation that the OTFMAP polarization mode would improve the sensitivity of the polarimetric observations, the estimated on-source times were a conservative estimation for each galaxy. As part of the Legacy Program, we characterized the OTFMAP polarization mode for objects with an extension within the field-of-view (FOV) of the array in any given band of HAWC+ (Paper~III). As shown in Paper~III, the OTFMAP polarization mode overperforms the C2N polarimetric mode. The delivered data products have better image and polarimetric quality than those that would have been originally requested using the C2N polarization mode.

%%%%%%%%%%%%%%%%%%
%%%% FIRST RESULTS %%%%
%%%%%%%%%%%%%%%%%%

\section{First Results}\label{sec:DR1}

This first data release comprises 33\% (51.34h out of 155.70h) of the awarded total exposure time of the SOFIA Legacy Program taken from January 2020 to November 2021 \citep[][Paper III]{ELR2022}. Table \ref{tab:DR1} shows the current status of the released observations. We recently published results of M51 \citep[][Paper I]{Borlaff2021} and NGC~1097 \citep[][Paper II]{ELR2021c}. Here, we present the new data products associated with this data release with the goal of quantifying general trends in the polarization fraction of the galaxies in our sample. The characterization of the B-field orientation is presented in Paper V of the series.

%%%%%%%%%%%%%
%%%% TABLE 2 %%%%
%%%%%%%%%%%%%
\begin{deluxetable}{lcccc}
\tablecaption{Status of released galaxy sample. \emph{Columns, from left to right:} a) Galaxy name. b) Central wavelength of the band in \um. c) Requested on-source time in hours. d) Observed on-source time in hours. e) Completed fraction per object per band.
\label{tab:DR1} 
}
\tablecolumns{6}
\tablewidth{0pt}
\tablehead{\colhead{Galaxy} & 	\colhead{Band}  & \colhead{Requested} & \colhead{Observed} & 
\colhead{Completed$^{\star}$}  \\ 
 &		&	\colhead{Time$^{\dagger}$}	&	\colhead{Time$^{\dagger}$} 	 \\
  &	\colhead{(\um)}	&	\colhead{(h)}	&	\colhead{(h)} 	 \\
\colhead{(a)} & \colhead{(b)} & \colhead{(c)} & \colhead{(d)} & \colhead{(e)} } 
\startdata
Centaurus A &	89	&	-	    &	0.89	&	-	\\
Circinus 	&	53	&	-	    &	0.11	&	-	\\
			&	89	&	-	    &	0.89	&	-	\\
			&	214	&	-	    &	0.29	&	-	\\	
M51 		&	154	&	-	    &	2.78	&	-	\\
M82 		&	53	&	2.00	&	1.60	&	\checkmark	\\
			&	89	&	2.00	&	1.65	&	\checkmark	\\
			&	154	&	2.00	&	1.99	&	\checkmark	\\
			&	214	&	2.00	&	1.13	&	\checkmark	\\
M83 		&	154	&	6.80	&	6.60	&	\checkmark	\\
NGC~253 	&	89	&	3.00	&	1.00	&	33\% 	\\
			&	154	&	5.00	&	0.89	&	18\% 	\\
NGC~1068 	&	53	&	3.07	&	1.06	&	35\%		\\
			&	89	&	7.07	&	2.00	&	28\% 	\\
NGC~1097 	&	53	&	-	    &	1.87	&	-	\\
			&	154	&	-	    &	0.27	&	- 	\\
NGC~2146 	&	53	&	3.00	&	1.60	&	53\%		\\
			&	89	&	3.00	&	2.02	&	67\%		\\
			&	154	&	3.00	&	2.40	&	80\%		\\
			&	214	&	3.00	&	2.33	&	\checkmark	\\
NGC~3627 	&	154	&	6.80	&	4.53	&	67\%		\\
NGC~4736	&	154	&	6.80	&	2.22	&	33\%		\\
NGC~4826 	&	89	&	6.80	&	1.22	&	18\%		\\
NGC~6946 	&	154	&	6.80	&	7.20	&	\checkmark	 \\
NGC~7331 	&	154	&	6.80	&	6.10	&	\checkmark	\\	
\enddata
\tablenotetext{{\dagger}}{On-source times. The observing overheads are 1.08 for OTFMAP observations (Section \ref{subsec:OBS}), and 2.59 for the C2N observations of M51 \citep{Borlaff2021}. The observations from other SOFIA programs are labeled as `-'.}
\tablenotetext{{\star}}{The observations are considered as completed (i.e. \checkmark) when a) completion $>85$\% or b) a desired signal-to-noise ratio was reached. The observations from other SOFIA programs are labeled as `-'.}
\end{deluxetable}

%%%%%%%%%%%%%%%%%%%%%%
\subsection{Polarization fraction of galaxies}\label{subsec:Polmaps}
%%%%%%%%%%%%%%%%%%%%%%

\begin{figure*}[ht!]
\centering
\includegraphics[angle=0,width=\textwidth]{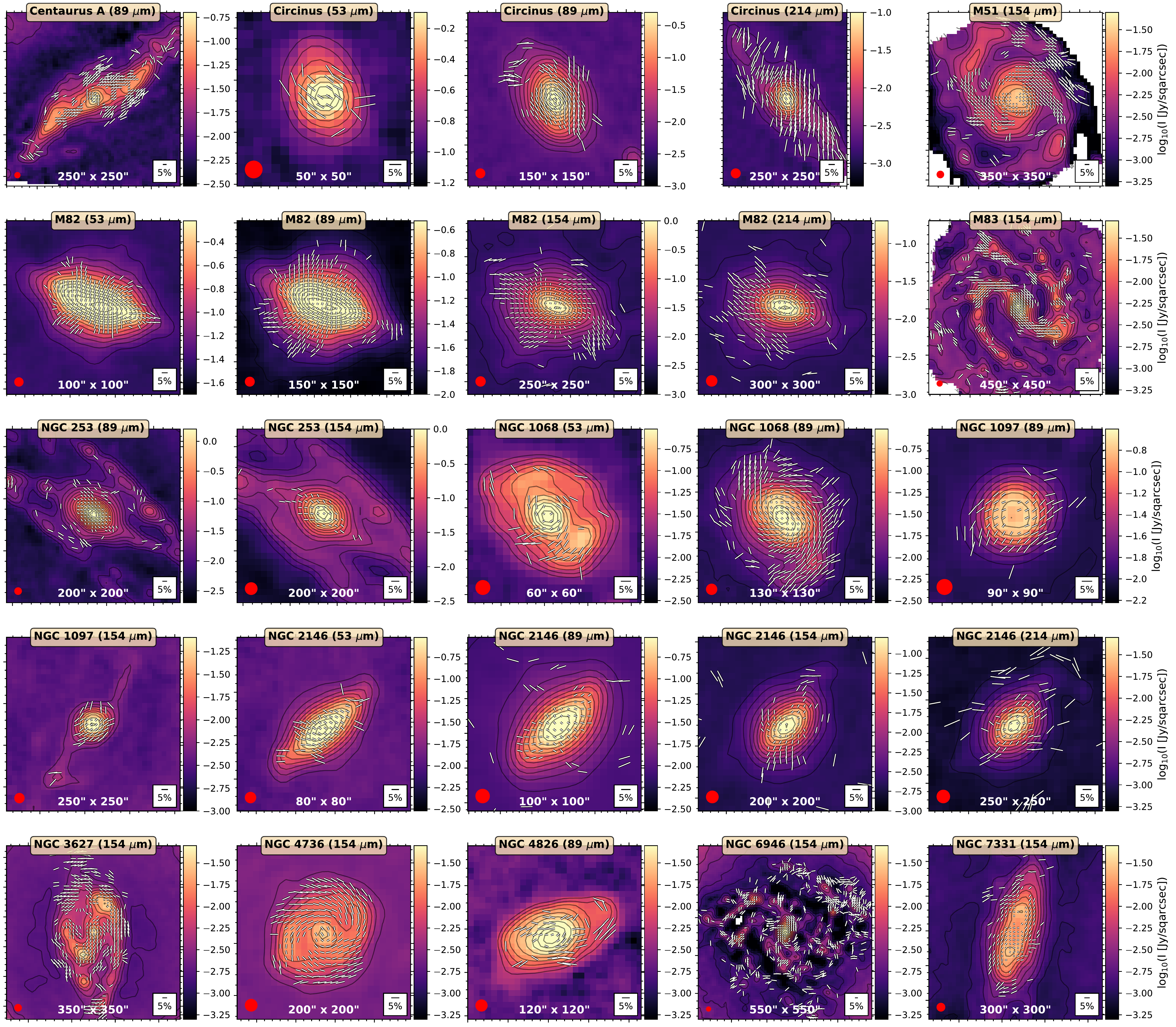}
\caption{Polarization maps of galaxies. Surface brightness (colorscale) in logarithmic scale with contours starting at $5\sigma$ increasing in steps of $2^{n}\sigma$, with $n= 5, 5.5, 6, \dots$ The polarization fraction measurements (white lines) increase as $\sqrt{P}$ (for visualization purposes), with a legend of $5$\% shown at the bottom right of each panel. The polarization measurements with $PI/\sigma_{PI} \ge 3.0$, $P\le20$\% were selected. We also performed quality cuts at several levels of $I/\sigma_{I}$ as shown in Appendix~\ref{A1:maps}. The FOV and beam size (red circle) are shown at the bottom center and left of each panel, respectively. Appendix~\ref{A1:maps} shows the polarization fraction measurements in linear scale with the coordinates of each galaxy over the total and polarized surface brightness.
\label{fig:fig1}}
\epsscale{2.}
\end{figure*}

\begin{figure*}[ht!]
\centering
\includegraphics[angle=0,width=\textwidth]{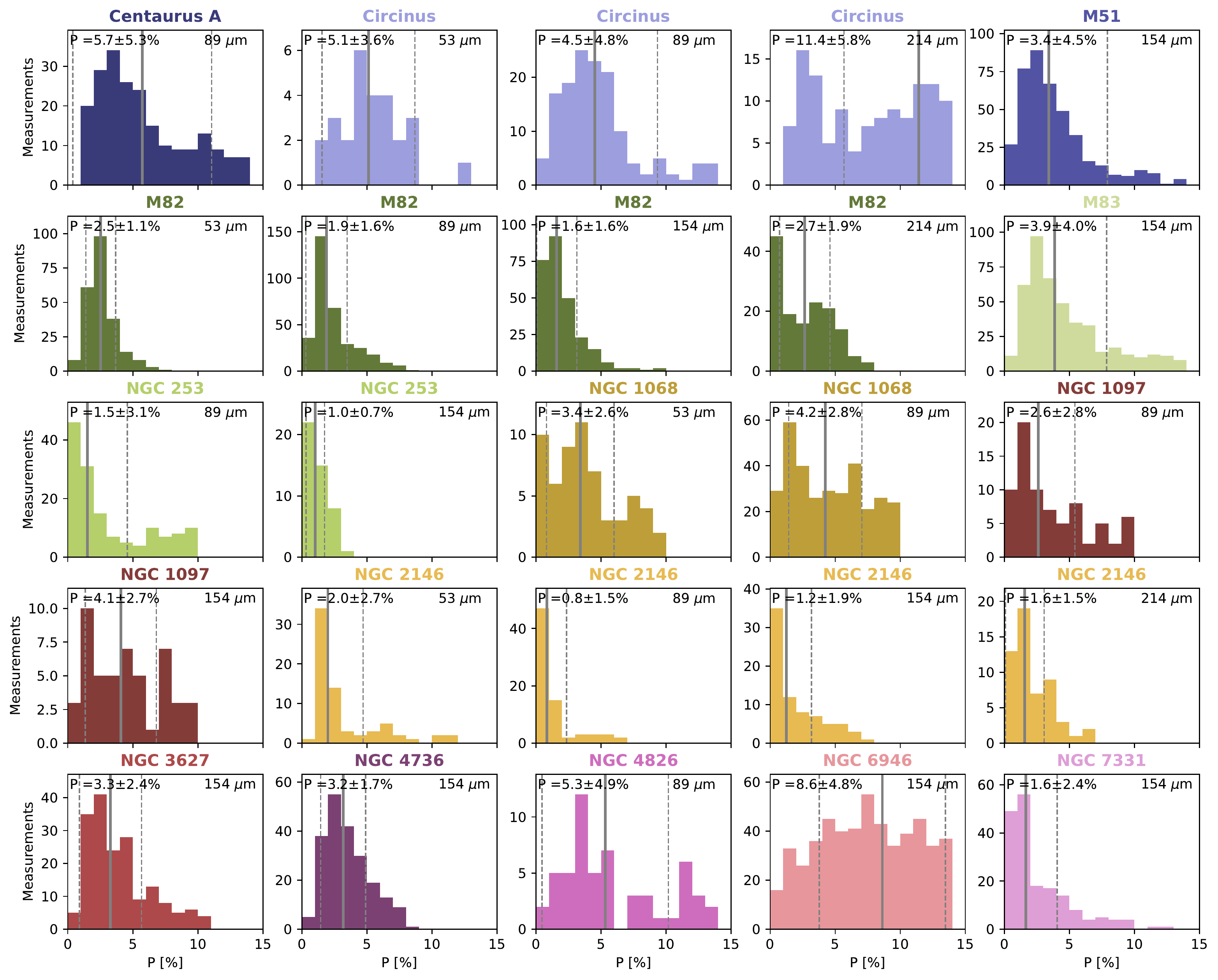}
\caption{Histograms of the polarization fractions per galaxy per band. The histograms employ $1$\% bins, each galaxy is shown in a unique color, the median $\langle P^{\rm{hist}} \rangle$ (grey solid line, Eq.~\ref{eq:Phist}), and $1\sigma$ uncertainty (grey dashed line) are shown. Median values displayed in upper right corners are also provided in Table~\ref{tab:PPA}.
\label{fig:fig2}}
\epsscale{2.}
\end{figure*}

Figure~\ref{fig:fig1} shows the polarization maps of the galaxies. For all galaxies, the color scale shows the surface brightness of the HAWC+ observations with overlaid polarization measurements. The length of the polarization fraction measurements is proportional to $\sqrt{P}$ (for visualization purposes), where a legend of 5\% is shown for reference. The polarization measurements with $PI/\sigma_{PI} \ge 3.0$ and $P\le20$\% were selected for all galaxies with varying  $I/\sigma_{I}$ criteria, as shown in Appendix~\ref{A1:maps}, where $\sigma_{PI}$ and $\sigma_{I}$ are the uncertainties per pixel of the polarized flux density and Stokes $I$, respectively. The maximum polarization fraction of $P=20$\% is given by the maximum polarization fraction, $19.8$\%, measured by \textit{Planck} observations \citep{Planck_Int_XIX_2015}. The polarization maps with a length proportional to the polarization fraction and the coordinate system of individual galaxies are shown in Appendix \ref{A1:maps} (Figures~\ref{fig:figA1} to \ref{fig:figA14}). Figure~\ref{fig:fig2} shows the histograms of the polarization fraction in bins of $1$\% for all the individual polarization measurements per galaxy per band.

%%%%%%%%%%%%%%
%%%% FIGURE 1 %%%%
%%%%%%%%%%%%%%
%%%%%%%%%%%%%%

%%%%%%%%%%%%%%
%%%% FIGURE 2 %%%%
%%%%%%%%%%%%%%
%%%%%%%%%%%%%%

We estimate the median of the polarization fraction measurements, $\langle P^{\rm{hist}} \rangle$, per galaxy and per band as

\begin{equation}\label{eq:Phist}
    \langle P^{\rm hist} \rangle = \frac{\sqrt{\langle Q^2 + U^2 - \sigma_{Q}\sigma_{U} \rangle}}{\langle I \rangle}
\end{equation}
\noindent
for all individual measurements shown in Figure \ref{fig:fig1}, where $\langle I \rangle$ is the median of Stokes $I$, and $\sigma_{Q}$ and $\sigma_{U}$ are the uncertainties per pixel of the Stokes $QU$. The $1\sigma$ uncertainty is estimated as the standard deviation of the histogram (Figure~\ref{fig:fig2}).

Since the expected polarization fraction from an unresolved galaxy may be scientifically useful, we estimate the integrated polarization fraction per galaxy and per band. To account for the vector quantity of the polarization measurements, we estimate the integrated polarization fraction, $\langle P^{\rm int}\rangle$, of each galaxy as 

\begin{equation}\label{eq:Pint}
    \langle P^{\rm int} \rangle = \frac{\sqrt{\langle Q \rangle^2 + \langle U \rangle^2 -\mbox{bias}}}{\langle I \rangle}
\end{equation}
\noindent
where $\langle I \rangle$, $\langle Q \rangle$ and $\langle U \rangle$ are the median of the Stokes $IQU$ for pixels with $I/\sigma_{I} \ge 20$, and $\mbox{bias}$ is the standard error of the median using the uncertainties of the Stokes $QU$, $\sigma_{Q}$ and $\sigma_{U}$, for these pixels. $\langle P^{\rm{hist}} \rangle$ and $\langle P^{\rm{int}} \rangle$ are shown in Figure~\ref{fig:fig3}, and the tabulated data can be found in Appendix~\ref{app:A2} (Table~\ref{tab:PPA}).

For all galaxies, we find that the individual measurements of the polarization fraction ranges from $0$ to $15$\% (Figure~\ref{fig:fig2}). The galaxies with the lowest median polarization fraction measurements are the starburst galaxies M82, NGC~253, and NGC~2146 (Figure~\ref{fig:fig2} and Table \ref{tab:PPA}). The highest polarization fraction measurements are found in the interarms and outskirts of spiral galaxies. The spiral galaxies have the lowest polarization fractions colocated with star-forming regions across the galaxy's disk. 

We use our $50-220$~\um~polarimetric observations to compute the polarized spectrum of galaxies. Figure~\ref{fig:fig4} shows the polarized spectrum of all galaxy types, as well as for spiral and starburst galaxies separately. Table~\ref{tab:PPA_all} shows the estimated polarization fractions per galaxy group and wavelength. We estimate the median of individual polarization fraction measurements for all galaxy types to be fairly constant $\langle P^{\rm hist}_{\rm all,50-220\mu m}\rangle=3.1\pm0.3$\% across the $50-220$~\um\ wavelength range (Figure~\ref{fig:fig4}-blue open circles with solid line and Table~\ref{tab:PPA_all}). $\langle P^{\rm hist}_{\rm all,50-220\mu m}\rangle$ was estimated using the median of the Stokes $IQU$ at each wavelength. The largest uncertainty is found at $214$~\um, which is driven by the small galaxy sample (3 galaxies: Circinus, M82, and NGC~2146). Large polarization fractions, $\ge8$\%, are measured along the bar of Circinus (Fig.~\ref{fig:fig1}), while M82 and NGC~2146 show $\langle P^{\rm hist} \rangle = 1.6$\% and $2.7$\% at $214$ \um, respectively (Figure \ref{fig:fig2}, Table~\ref{tab:PPA}). The median integrated polarization fraction for all galaxies types also remains constant  $\langle P^{\rm int}_{\rm all,50-220\mu m} \rangle=1.3\pm0.2$\% across the $50-220$ \um\ wavelength range (Figure~\ref{fig:fig4}-blue filled circles with solid line and Table~\ref{tab:PPA_all}). $\langle P^{\rm int}_{\rm all,50-220\mu m}\rangle$ was estimated using the median of the Stokes $IQU$ at each wavelength.

%%%%%%%%%%%%%
%%%% FIGURE 3 %%%%
%%%%%%%%%%%%%%
\begin{figure}[ht!]
\centering
\includegraphics[angle=0,width=\columnwidth]{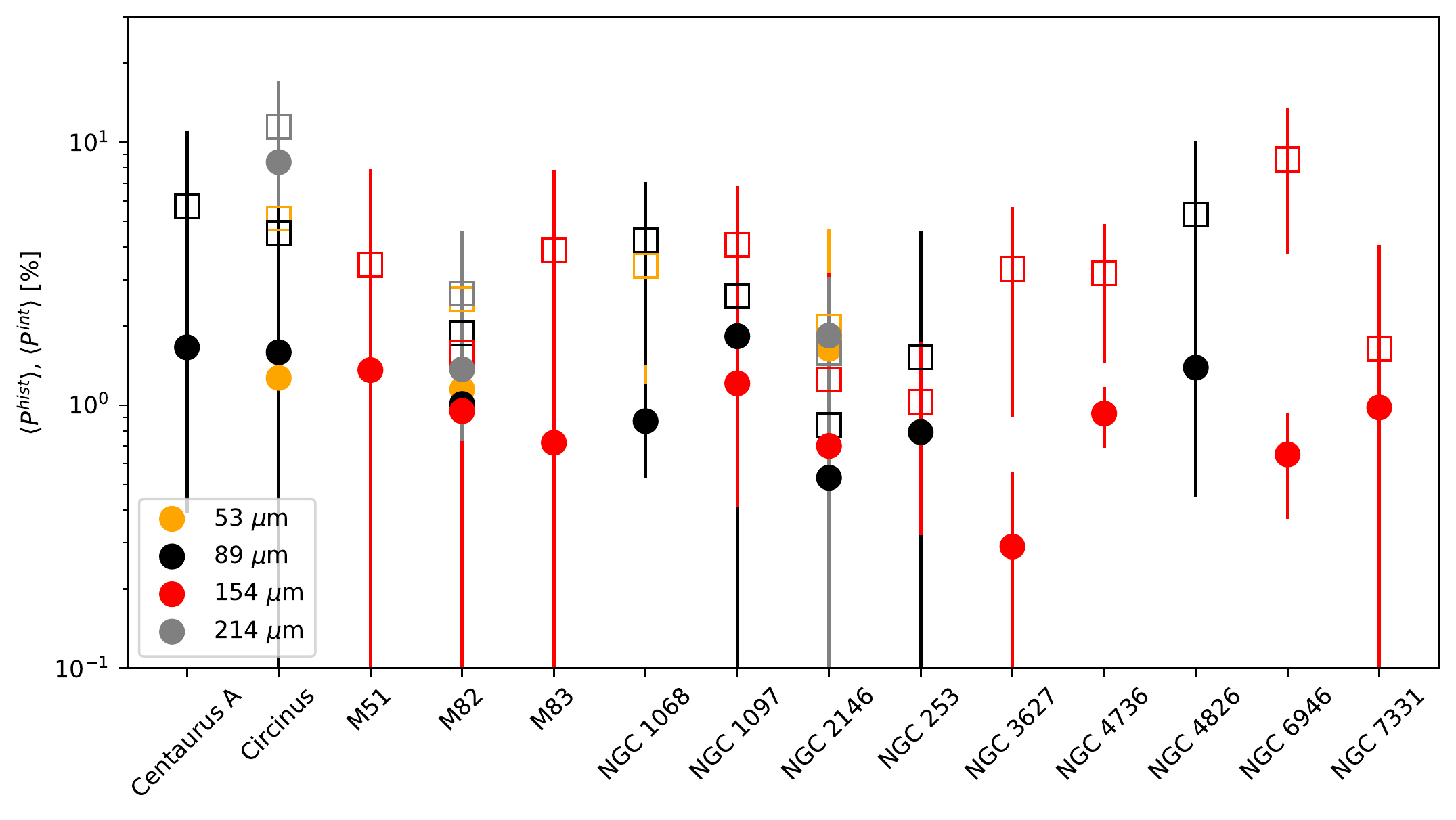}
\caption{Median and integrated polarization fractions of galaxies. Polarization fractions from individual measurements, $\langle P^{\rm{hist}} \rangle$ (open square, Eq.~\ref{eq:Phist}), and from the integrated Stokes $IQU$, $\langle P^{\rm{int}} \rangle$ (filled circle, Eq.~\ref{eq:Pint}), are shown. Tabulated values are shown in Table~\ref{tab:PPA}.
\label{fig:fig3}}
\epsscale{2.}
\end{figure}
%%%%%%%%%%%%%%
 
We separate the sample in starburst galaxies (M82, NGC~253, NGC~2146) at all wavelengths and spiral galaxies (M51, M83, NGC~3627, NGC~4736, NGC~6946, and NGC~7331) at $154$~\um. Then, we estimate the median of the individual polarization fractions, $\langle P^{\rm{hist}} \rangle$, and the integrated polarization fraction across the full galaxy, $\langle P^{\rm{int}} \rangle$ (Table \ref{tab:PPA_all}). 
 
For starburst galaxies, we find that the polarized spectrum varies as a function of wavelength with a minimum located in the range of $89-154$~\um\ (Figure~\ref{fig:fig4}-orange dashed line). The variation of the polarization spectrum of starbursts is measured for both $\langle P^{\rm hist}_{\rm starburst} \rangle$, within the range of $[1.3,2.3]$\% and $\langle P^{\rm int}_{\rm starburst} \rangle$, within the range of $[0.7,1.6]$\%. This result represents the first polarized spectrum of starburst galaxies in the FIR wavelength range. We discuss this result in Section~\ref{subsec:DIS_Pspec}.

 %%%%%%%%%%%%%%
%%%% FIGURE 4 %%%%
%%%%%%%%%%%%%%
\begin{figure}[ht!]
\centering
\includegraphics[angle=0,width=\columnwidth]{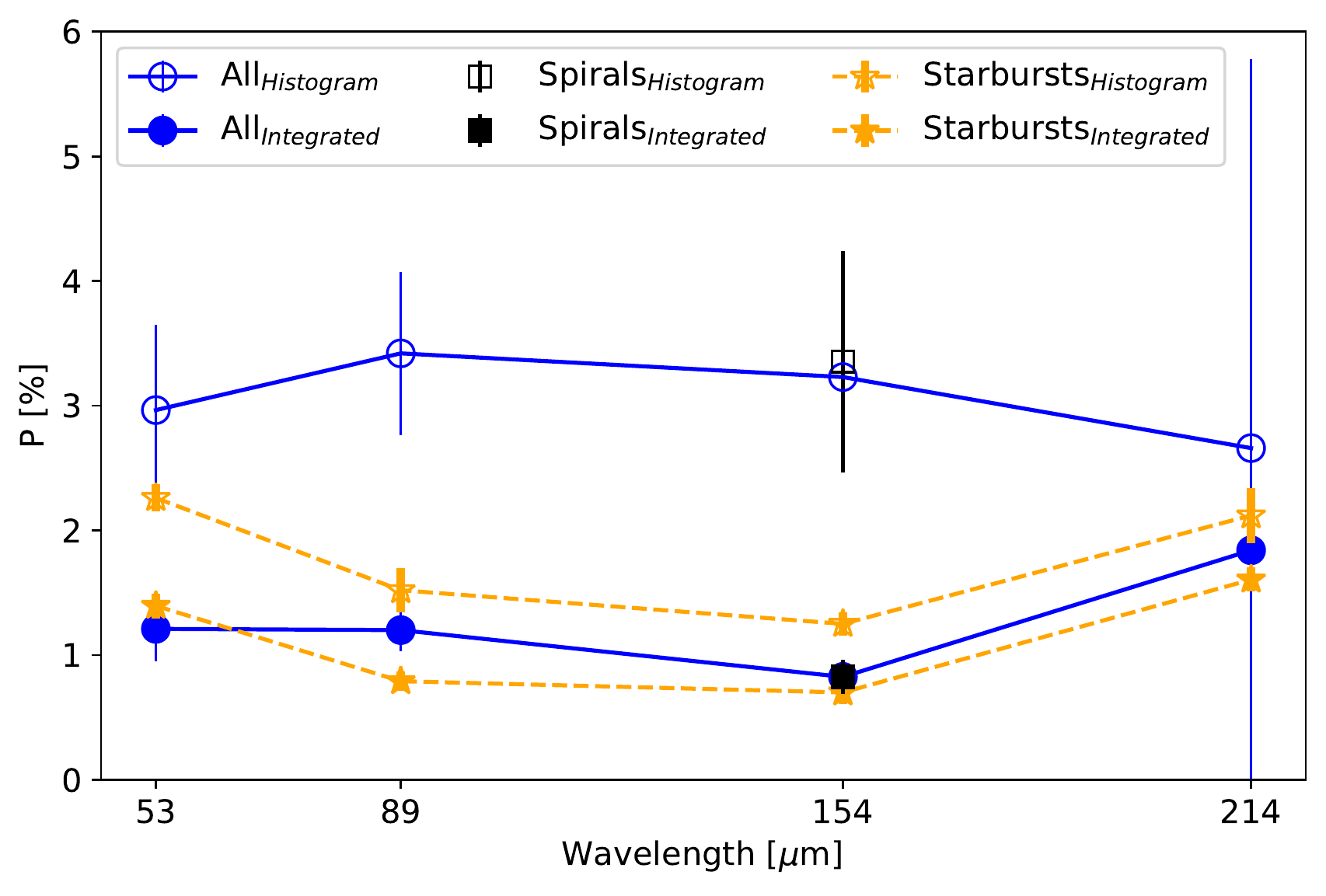}
\caption{Polarized spectrum of galaxies. Polarization fraction as a function of the wavelength in the range of $50-220$ \um\ for all galaxies (blue circle-solid line), spiral (black squares), and starburst (orange star-dashed line) galaxies in our sample. The median polarization fraction from the individual measurements (open symbols) across the galaxy and the integrated polarization fraction from the Stokes $IQU$ (filled symbols) are shown. See legend for description and Table~\ref{tab:PPA_all} for measurements.
\label{fig:fig4}}
\epsscale{2.}
\end{figure}
%%%%%%%%%%%%%%
 
The spiral galaxies have a median polarization fraction of $\langle P^{\rm hist}_{\rm spirals, 154\mu m}\rangle = 3.3\pm0.9$\% (Figure \ref{fig:fig4}-single black square dots at 154 \um). This value is more than a factor of two larger than the median polarization fraction of starbursts, $\langle P^{\rm hist}_{\rm starburst,154\mu m}\rangle= 1.3\pm0.1$\%, at $154$~\um. This result causes the $50-220$ \um~polarized spectrum to appear flat when accounting for all the galaxy types (Figure \ref{fig:fig4}-blue open circles with solid line). This is an effect of how $\langle P^{\rm hist}\rangle$ is measured. When the polarization fraction is measured using the integrated intensities of the Stokes $IQU$ within the disk, the integrated polarization fraction of spiral galaxies are measured to be $\langle P^{\rm int}_{\rm spirals,154\mu m}\rangle = 0.8\pm0.1$\% at $154$~\um. This value is comparable to the integrated polarization fraction, $\langle P^{\rm int}_{\rm starburst,154\mu m} \rangle = 0.7\pm0.1$\% at $154$ \um, of the starburst galaxies (Figure \ref{fig:fig4} and Table~\ref{tab:PPA_all}). The main reason of the decrease in polarization fraction from $\langle P^{\rm hist}_{\rm spirals, 154\mu m} \rangle$ to $\langle P^{\rm int}_{\rm spirals, 154\mu m} \rangle$ is due to the vector-like nature of the polarization measurements. The net polarization fraction tends to nullify the integrated polarization fraction due to a) the spiral pattern measured in spiral galaxies, and b) the several polarization components (i.e., outflow and disk) measured in starburst galaxies.
 
%%%%%%%%%%%%%
%%%% TABLE 5 %%%%
%%%%%%%%%%%%%
\begin{deluxetable}{lccccc}[ht!]
\tablecaption{Median and integrated polarization fraction per band of the full galaxy sample, and spiral and starburst galaxies. \textit{From left to right:} a) central wavelength of the band in \um, b) galaxy type, c) median polarization fraction based on individual measurements, and d) integrated polarization fraction of the full galaxy.
\label{tab:PPA_all} 
}
\tablecolumns{4}
\tablewidth{0pt}
\tablehead{\colhead{Band}  & \colhead{Galaxy type} & \colhead{$\langle P^{\rm hist} \rangle$} & \colhead{$\langle P^{\rm int} \rangle$} \\ 
\colhead{(\um)}  &	& \colhead{(\%)}	& \colhead{(\%)}	 	 \\
\colhead{(a)} & \colhead{(b)} & \colhead{(c)} & \colhead{(d)} } 
\startdata
53 	&	All 		            &	$2.9\pm0.7$ 	& $1.3\pm0.1$	\\
	&	Starbursts$^{\dagger}$	&	$2.3\pm0.1$		& $1.4\pm0.1$	\\
89 	& 	All 		            &	$3.4\pm0.7$ 	& $1.2\pm0.2$	\\
	&	Starbursts	            &	$1.5\pm0.2$		& $0.8\pm0.1$	\\
154 	& 	All		            &	$3.3\pm0.7$ 	& $0.9\pm0.1$	\\
	&	Spirals$^{\star}$	    &	$3.3\pm0.9$		& $0.8\pm0.1$	\\
	&	Starbursts	            &	$1.3\pm0.1$		& $0.7\pm0.1$	\\	
214 	&	All 		        &	$2.7\pm3.1$ 	& $1.8\pm2.3$	\\
	&	Starbursts	            &	$2.1\pm0.2$		& $1.6\pm0.1$	\\
\enddata
\tablenotetext{{\dagger}}{Polarization fraction for the starburst galaxies M82, NGC~253, NGC~2146}
\tablenotetext{{\star}}{Polarization fraction for the spiral galaxies M51, M83, NGC~3627, NGC~4736, NGC~6946, and NGC~7331}
\end{deluxetable}
%%%%%%%%%%%%%%%%%

%%%%%%%%%%%%%%%%%%%%%%
\subsection{Polarization fraction and column density relation}\label{subsec:PIPlots}
%%%%%%%%%%%%%%%%%%%%%%

%%%%%%%%%%%%%%
%%%% FIGURE 5 %%%%
%%%%%%%%%%%%%%
\begin{figure*}[ht!]
\centering
\includegraphics[angle=0,scale=0.17]{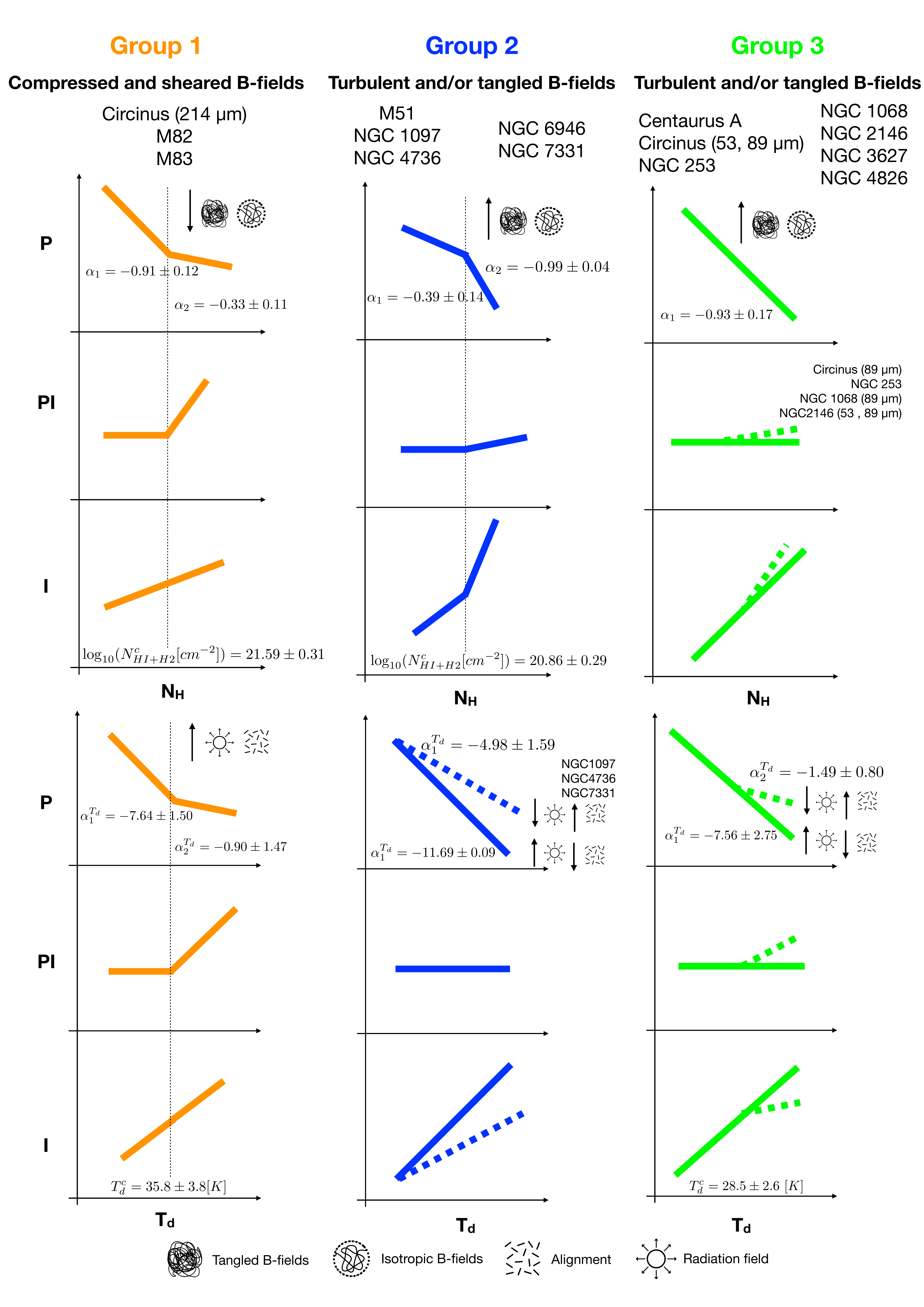}
\caption{Summary of the trends in $P$, $PI$, and $I$ as a function of column density, $N_{\rm H}$, and dust temperature, $T_{\rm d}$, and the physical mechanisms affecting their slopes.
Three groups are identified: Group 1 (orange), Group 2 (blue), and Group 3 (green) with their associated galaxies shown at the top of each group. For each group, the $P-N_{\rm H}$ (first row), $PI-N_{\rm H}$ (second row), $I-N_{\rm H}$ (third row), $P-T_{\rm d}$ (fourth row), $PI-T_{\rm d}$ (fifth row), and $I-T_{\rm d}$ (sixth right) relations are shown. The median values of the power-law indexes and cut-off column density and dust temperature are shown in the $P-N_{\rm H}$ and $P-T_{\rm d}$ relations. The physical mechanisms producing the trends are shown for each group (legend at the bottom). These trends are based on the data shown in Appendix \ref{App:A3} (Figures \ref{fig:A3_fig1}-\ref{fig:A3_fig3}) and described in Sections \ref{subsec:PIPlots}, \ref{subsec:PTdPlots}, and \ref{subsec:DIS_Bori}.
\label{fig:fig5}}
\epsscale{2.}
\end{figure*}
%%%%%%%%%%%%%%

%%%%%%%%%%%%%%
%%%% FIGURE 6 %%%%
%%%%%%%%%%%%%%
\begin{figure*}[ht!]
\centering
\includegraphics[angle=0,width=\textwidth]{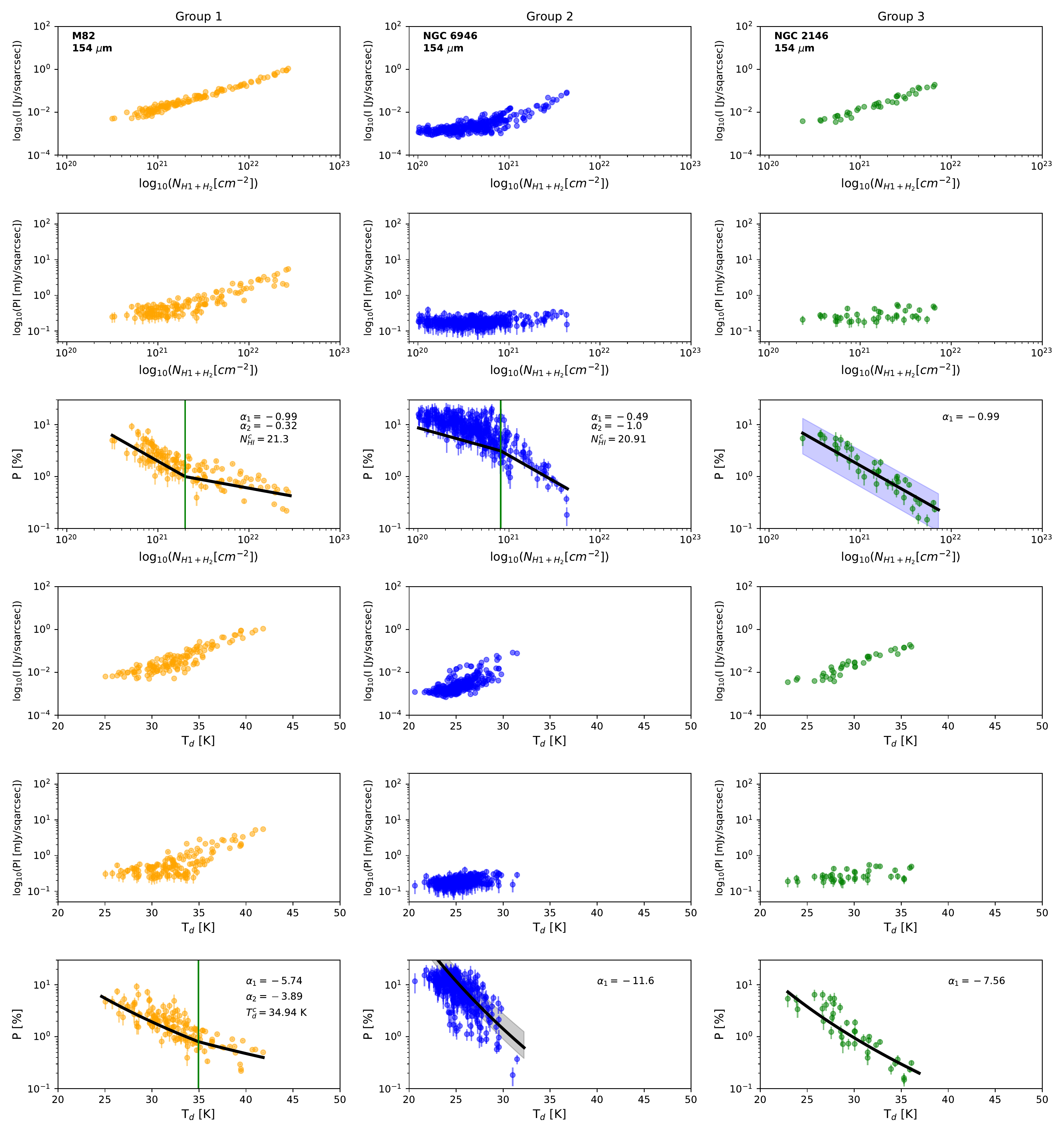}
\caption{Example of the variations of $P$, $PI$, and $I$ with  $N_{\rm HI+H_{2}}$ and $T_{\rm d}$ for the three groups shown in Figure \ref{fig:fig5}. The representative galaxies M82 for Group 1 (left), NGC~6949 for Group 2 (center), and NGC~2146 for Group 3 (right) are shown. The relations for all galaxies are shown in Appendix \ref{App:A3} (Figures \ref{fig:A3_fig1}-\ref{fig:A3_fig3}). These trends are discussed in Sections \ref{subsec:PIPlots}, \ref{subsec:PTdPlots}, and \ref{subsec:DIS_Bori}
\label{fig:fig6}}
\epsscale{2.}
\end{figure*}
%%%%%%%%%%%%%%

The polarization fraction in galaxies has been found to decrease with the total surface brightness at FIR wavelengths and column density, $N_{\rm H}$ \citep{Jones2019,ELR2020,Jones2020,ELR2021a,ELR2021c,ELR2021,Borlaff2021}. The $P\propto N_{\rm H}^{\alpha}$ relation arises from the separation of ordered and random (i.e. turbulent and/or tangled) B-fields in the ISM \citep[e.g.,][]{Jones1992,Planck_Int_XLIV_2016}. For the maximally aligned dust grains, $P$ is constant with $N_{\rm H}$, i.e., $P \propto N_{\rm H}^{0}$.  Any variation of the B-field orientation results in a decrease of the polarization fraction with column density. If  the  B-field  orientation  varies  completely  randomly  along  the  LOS,  then $P \propto N_{\rm H}^{-0.5}$. For  any  combination  of  random  and  ordered B-fields,  then  the  slope  is  between  these  two. However, as a) dust grain alignment efficiency decreases in denser regions,  b)  there  are  tangled  B-fields  along  the LOS, and c) there are regions of very high turbulence or very small scale lengths, $P$ can decrease faster than $-0.5$ \citep[e.g.][]{King2019}. Thus, the $P-N_{\rm H}$ relation can be used as a proxy to estimate the random B-fields in the ISM.

Figure~\ref{fig:fig5} serves as an illustration of the several physical mechanisms (turbulence, tangled B-fields, dust grain alignment efficiency, and isotropic random B-fields) responsible for the dependence between the polarization fraction and the column density. This figure shows the summary of the main results of this work, and we will refer the reader to Figure~\ref{fig:fig5} in this and the following sections.

We use the column density, $N_{\rm HI+H_{2}}$, estimated using the fitting of a modified blackbody function to the $70-250$~\um\ \textit{Herschel} images. Appendix~\ref{A1:maps} describes the details of the estimation of the column density, and we show the maps with overlaid polarization measurements. We use the dependence with $N_{\rm HI+H_{2}}$ as it provides a physical quantification of the ISM and it can be compared with the results of the Milky Way and other galaxies. Figure~\ref{fig:fig6} shows the total and polarized surface brightness and polarization fraction as a function of the column density, $N_{\rm HI+H_{2}}$, for a subsample of the galaxies at $154$~\um. The rest of the plots are shown in Appendix~\ref{App:A3} (Figures \ref{fig:A3_fig1}-\ref{fig:A3_fig3}). For all galaxies, we use the same polarization measurements shown in Section \ref{subsec:Polmaps}. 

The $P-N_{\rm HI+H_{2}}$ plots show a more complex relation than a single power-law. To characterize the $P-N_{\rm HI+H_{2}}$ relations, we use a broken power-law with two power-law indexes at a cut-off column density such as

\begin{align}
P(N_{\rm HI+H_{2}}) &= P_{0}
  \begin{cases}
    \left(\frac{N_{\rm HI+H_{2}}}{N_{\rm HI+H_{2}}^{\rm c}}\right)^{\alpha_{1}} & N_{\rm HI+H_{2}} \le  N_{\rm HI+H_{2}}^{\rm c}\\
    \left(\frac{N_{\rm HI+H_{2}}}{N_{\rm HI+H_{2}}^{\rm c}}\right)^{\alpha_{2}} & N_{\rm HI+H_{2}} > N_{\rm HI+H_{2}}^{\rm c}
  \end{cases}
\end{align}
\noindent
where $P_{0}$ is the scale factor, N$_{\rm HI+H_{2}}^{\rm c}$ is the cut-off of the broken power-law, and $\alpha_{1}$ and $\alpha_{2}$ are the power-law indexes before and after the break, respectively. Note that several galaxies (e.g., M82 at $53$~\um, M83 at $154$~\um, NGC~1068 at $89$~\um) show more complex trends than a broken power-law, however a detailed analysis of each galaxy will be the subject of future projects. Here, we focus on the general trends of the full galaxy sample.

%%%%%%%%%%%%%%
%%%% FIGURE 7 %%%%
%%%%%%%%%%%%%%
\begin{figure}[ht!]
\centering
\includegraphics[angle=0,width=\columnwidth]{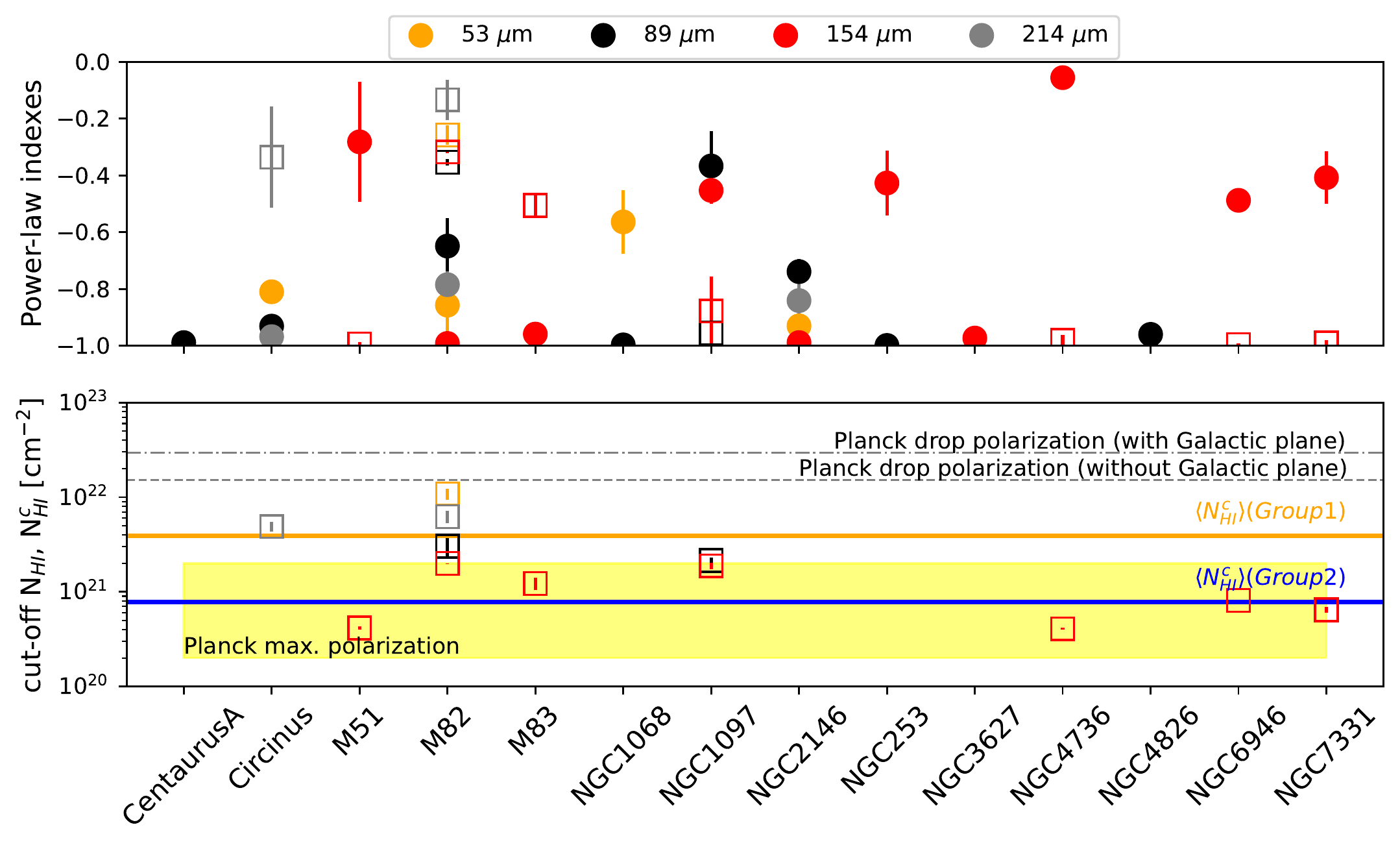}
\caption{Power-law indexes (top) and cut-off column densities (bottom) for all galaxies based on the fitting of the $P-N_{\rm HI+H_{2}}$ relations. Top panel: the first power-law, $\alpha_{1}$ (filled circles), and second power-law, $\alpha_{2}$ (open squares), indexes at $53$ (orange), $89$ (black), $154$ (red), and $214$ (grey) \um\ are shown. Bottom panel: the cut-off column densities, $N_{\rm HI+H_{2}}^{\rm c}$ (open squares), for those galaxies with a broken power-law are shown. The median cut-off column densities of  $\log_{10}(N_{\rm HI+H_{2}}^{\rm c} [{\rm cm}^{-2}]) = 21.59\pm0.31$ (orange solid line) and $20.86\pm0.29$ (blue solid line) for Groups 1 and 2, respectively are shown. The range of column densities (yellow shadowed region) for the maximum polarization fraction, and the column densities of $\log_{10}(N_{\rm H} [{\rm cm}^{-2}]) = 22.18$ (grey dashed line) and $22.47$ (grey dashed-dotted line) associated with a sharp drop in the polarization fraction with and without the Galactic plane of the Milky Way observed by \textit{Planck} \citep{Planck_Int_XIX_2015}  are shown.
\label{fig:fig7}}
\epsscale{2.}
\end{figure}
%%%%%%%%%%%%%%

We have four free model parameters: $P_{0}$, N$_{\rm HI+H_{2}}^{\rm c}$, $\alpha_{1}$, and $\alpha_{2}$. We perform a Markov Chain Monte Carlo (MCMC) approach  using the differential evolution metropolis sampling step in the \textsc{python} code \textsc{pymc3} \citep{pymc}. The prior distributions are set to flat within the range of $P_{0} = [0,15]$\%, $N_{\rm HI+H_{2}}^{\rm c} = [N_{\rm HI+H_{2}}^{\rm min}, N_{\rm HI+H_{2}}^{\rm max}]$ cm$^{-2}$, where $N_{\rm HI+H_{2}}^{\rm min}$ and $N_{\rm HI+H_{2}}^{\rm max}$ are the minimum and maximum column density values of each galaxy in any given band, $\alpha_{1} = [-1,0]$, and $\alpha_{2} = [-1,0]$. We run the code using 5 chains with 10000 steps and an extra 5000 burn-in per chain, which provides 50000 steps for the full MCMC code useful for data analysis. The best inferred model with the associated $1\sigma$ uncertainty is plotted for a subset of our sample in Figure~\ref{fig:fig6} and for each source in Appendix~\ref{App:A3} (Figures~\ref{fig:A3_fig1}-\ref{fig:A3_fig3}). Figure~\ref{fig:fig7} shows $\alpha_{1}$ for all galaxies, and $\alpha_{2}$ and cut-off column densities, N$_{\rm HI+H_{2}}^{\rm c}$, for those galaxies with a broken power-law. The tabulated values can be found in Appendix~\ref{app:A2} (Table~\ref{tab:PI_fits}). 

For all galaxies, the polarization fraction varies across the column density range of $\log_{10}(N_{\rm HI+H_{2}} [{\rm cm}^{-2}]) = [19.96,22.91]$. The highest polarization fractions, $\ge10$\%, are found at the lowest column density ranges, $\log_{10}(N_{\rm HI+H_{2}} [{\rm cm}^{-2}]) \le 21$. We find three main groups based on the fitting results of the $P-N_{\rm HI+H_{2}}$ relations (Figure~\ref{fig:fig5}). We describe the characteristics of each group below.

Group~1 is characterized by a flatter power-law, $\alpha_{2} > \alpha_{1}$, after the cut-off column density. Galaxies in this group are Circinus at $214$ \um, M83 at $154$ \um, and M82 at all bands. We estimate $\alpha_{1}$ in the range of $[-0.99,-0.65]$ with a median of $-0.91\pm0.12$, $\alpha_{2}$ in the range of $[-0.51,-0.13]$ with a median of $-0.33\pm0.11$, and a cut-off column density, $\log_{10}{N_{\rm HI+H_{2}}^{\rm c}[{\rm cm}^{-2}]}$, in the range of $[21.09,22.04]$ with a median of $21.59\pm0.31$ (Figures~\ref{fig:fig5} and \ref{fig:fig7}). Figures~\ref{fig:fig6} and \ref{fig:A3_fig1} show that the flatter $\alpha_{2}$ is due to a faster increase of the polarized intensity compared to the total intensity after the cut-off column density. For a visualization of the spatial region associated with both power-laws, the cut-off column density, $N_{\rm HI+H2}^{\rm c}$, is displayed over the column density maps in Appendix \ref{A1:maps}.  We find that the second power-law is required for the central starburst M82, the inner-bar and the inflection regions between the inner-bar and the spiral arms of M83, and the starburst ring of Circinus.

Group~2 is characterized by a steeper power-law, $\alpha_{2} < \alpha_{1}$, after the cut-off column density. Galaxies in this group are M51, NGC~4736, NGC~6946, and NGC~7331 at $154$~\um, and NGC~1097 at $89$ and $154$~\um. We estimate $\alpha_{1}$ in the range of $[-0.49,-0.06]$ with a median of $-0.39\pm0.14$, $\alpha_{2}$ in the range of $[-1.0,-0.88]$ with a median of $-0.99\pm0.04$, and a cut-off column density, $\log_{10}{N_{\rm HI+H_{2}}^{\rm c}[{\rm cm}^{-2}]}$, in the range of $[20.61,21.33]$ with a median of $20.86\pm0.29$ (Figures~\ref{fig:fig5} and \ref{fig:fig7}). Figures~\ref{fig:fig6} and \ref{fig:A3_fig2} show that the stepper $\alpha_{2}$ is due to the faster increase of the total intensity compared to the polarized intensity after the cut-off column density. We find that a second power-law is required for the central regions of galaxies associated with the galaxy's core and star-forming regions across the galaxy's disk (figures in Appendix~\ref{A1:maps}).

Group~3 is characterized by a single power-law. Galaxies in this group are shown in Figure~\ref{fig:fig5} and Figure~\ref{fig:A3_fig3}. We estimate $\alpha_{1}$ in the range of $[-1.0,-0.43]$ with a median of $-0.93\pm0.17$. The single power-law is characterized by the increase of the total intensity while the polarized flux remains constant compared to the full range of column densities. However, we find several galaxies where the polarized flux varies at a similar rate as the total intensity, which keeps a single power-law in $P-N_{\rm HI+H_{2}}$. These galaxies are Circinus, NGC~1068 at $89$~\um, NGC~253 at $89$ and $154$~\um, and NGC~2146 at $53$ and $89$~\um. This change in $PI$ occurs at the central star-forming regions of these galaxies. Figure~\ref{fig:fig5} shows this behaviour as a dashed green line in the $I-N_{\rm HI+H_{2}}$ and $PI-N_{\rm HI+H_{2}}$.

The physical mechanisms responsible for the trends described in each group are discussed in Section~\ref{subsec:DIS_Bori}.

%%%%%%%%%%%%%%%%%%%%%%
\subsection{Polarization fraction and dust temperature relation}\label{subsec:PTdPlots}
%%%%%%%%%%%%%%%%%%%%%%

\citet{DW1996} computed that the dust temperature, $T_{\rm d}$, can be estimated as a function of the energy density of the radiation field, $u_{\rm rad}$, such as $T_{\rm d} \propto (u_{\rm rad}/u_{\rm ISM})^{1/(4+\beta)}$ K, where $u_{\rm ISM}$ is the radiation field in the ISM of the Milky Way, and $\beta$ is the emissivity index. We took $\beta=1.60$ (Appendix \ref{A1:maps}), yields $4+\beta = 5.60$. Thus, we can use $T_{\rm d}$ as a proxy of the strength of the energy density of the radiation field producing the thermal emission from dust in the disk of galaxies. Specifically, we are interested in measuring the effect of the radiation field on the polarization efficiency within the disk of galaxies. We analyze the relation between the polarization fraction and the dust temperature, $P-T_{\rm d}$.  Appendix \ref{A1:maps} describes the details of the estimation of the dust temperature and shows the polarization maps over the dust temperature for the galaxies in our sample. Figure \ref{fig:fig5} shows the general trends found in the galaxy sample. Figure \ref{fig:fig6} shows the total and polarized surface brightness and polarization fraction as a function of the dust temperature of a representative galaxy at $154$ \um~in each group (full sample is shown in Appendix \ref{App:A3}). 

\begin{figure}[ht!]
\centering
\includegraphics[angle=0,width=\columnwidth]{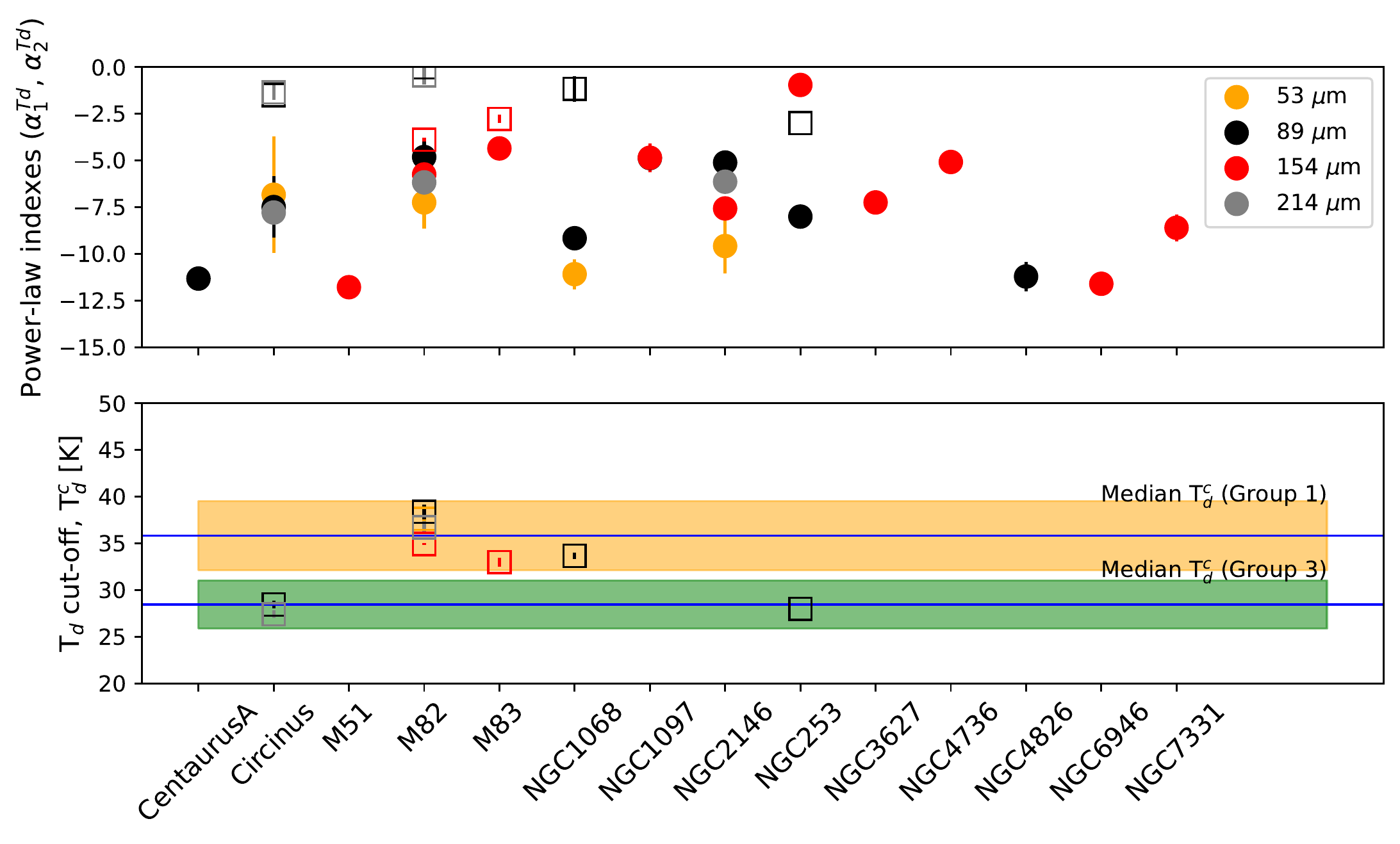}
\caption{Power-law indexes (top) and cut-off dust temperatures (bottom) for all galaxies based on the fitting of the $P-T_{\rm d}$ relations. Top panel: the first power-law, $\alpha_{1}^{T_{\rm d}}$ (filled circles), and second power-law, $\alpha_{2}^{T_{\rm d}}$ (squares), indexes at $53$ (orange), $89$ (black), $154$ (red), and $214$ (grey) \um\ are shown. Bottom panel: the cut-off dust temperatures, $T_{\rm d}^{\rm c}$ (squares), for those galaxies and wavelengths with a broken power law are shown. The median cut-off dust temperatures of  T$_{\rm d}^{\rm c} = 35.8\pm3.8$ K (orange),  $28.5\pm2.6$ K (green) and $1\sigma$ uncertainty (shadowed region) for Groups 1 and 3, respectively are shown. 
\label{fig:fig8}}
\epsscale{2.}
\end{figure}

To characterize the $P-T_{\rm d}$ relations, we use a broken power-law with two power-law indexes: 

\begin{align}
P(T_{\rm d}) &= P_{0}^{T_{\rm d}}
  \begin{cases}
    \left(\frac{T_{\rm d}}{T_{\rm d}^{\rm c}}\right)^{\alpha_{1}^{T_{\rm d}}} & T_{\rm d} \le  T_{\rm d}^{\rm c}\\
    \left(\frac{T_{\rm d}}{T_{\rm d}^{\rm c}}\right)^{\alpha_{2}^{T_{\rm d}}} & T_{\rm d} > T_{\rm d}^{\rm c}
  \end{cases}
\end{align}
\noindent
where $P_{0}^{T_{\rm d}}$ is the scale factor, $T_{\rm d}^{\rm c}$ is the cut-off dust temperature of the broken power-law, and $\alpha_{1}^{T_{\rm d}}$ and $\alpha_{2}^{T_{\rm d}}$ are the power-law indexes before and after the cut-off, respectively. We have four free model parameters: $P_{0}^{T_{\rm d}}$, $T_{\rm d}^{\rm c}$, $\alpha_{1}^{T_{\rm d}}$, and $\alpha_{2}^{T_{\rm d}}$ that we fit using the same procedure described in Section \ref{subsec:PIPlots} for the column density. Figure \ref{fig:fig8} shows $\alpha_{1}^{T_{\rm d}}$ for all galaxies, and $\alpha_{2}^{T_{\rm d}}$ and cut-off dust temperature,$T_{\rm d}^{\rm c}$, for those galaxies with a broken power-law. The tabulated values can be found in Appendix \ref{app:A2} (Table \ref{tab:PI_fits}).

%%%%%%%%%%%%%%
%%%% FIGURE 8 %%%%
%%%%%%%%%%%%%%
%%%%%%%%%%%%%%

For all galaxies, we find that the polarization fraction varies across the dust temperature range of $T_{\rm d} = [19,48]$ K. In general, we find that the polarization fraction decreases with increasing dust temperature. We find that the highest polarization fractions, $\ge10$\%, are found at the lowest dust temperature ranges, $T_{\rm d} \le 30$ K. The highest dust temperature are colocated with the star-forming regions and have the lowest polarization fractions, $<1$\%. We find the following trends for the $P-T_{\rm d}$ plots and discuss them together with the $P-N_{\rm HI+H2}$ relations in Section \ref{subsec:DIS_Bori}.

Group~1 is characterized by a flatter power-law, $\alpha_{2}^{T_{\rm d}} > \alpha_{1}^{T_{\rm d}}$, after the cut-off dust temperature. We estimate $\alpha_{1}^{T_{\rm d}}$ in the range of $[-7.78,-4.35]$ with a median of $-7.64\pm1.50$, $\alpha_{2}^{T_{\rm d}}$ in the range of $[-3.89,-0.01]$ with a median of $-0.90\pm1.47$, and a median $T_{\rm d}^{\rm c} = 35.8\pm3.8$~K in the range of $[27.4,38.4]$~K (Figure \ref{fig:fig8}). Figures \ref{fig:fig5} and \ref{fig:A3_fig1} show that a flatter $\alpha_{2}^{T_{\rm d}}$ is due to the faster increase of the polarized intensity compared to the total intensity after the cut-off dust temperature. We find that the second power-law is required for the regions associated with starburst regions and low polarization fractions, $P_{0}^{T_{\rm d}}\le1$\% (Appendix \ref{A1:maps}). 

Group~2 is characterized by a single power-law. We estimate $\alpha_{1}^{T_{\rm d}}$ in the range of $[-11.78,-4.85]$ with two different subgroups based on their slopes (Figures~\ref{fig:fig5} and \ref{fig:A3_fig2}). M51 and NGC~6946 have a steeper slope, $\alpha_{1}^{T_{\rm d}} = -11.69\pm0.09$, than NGC~1097, NGC~4736 and NGC~4826, $\alpha_{1}^{T_{\rm d}} = -4.98\pm1.59$. The single power-law is characterized by the increase of total intensity while the polarized flux remains constant across the full range of dust temperatures.

Group~3 is also characterized by single power-law with a median of $\alpha_{1}^{T_{\rm d}} = -7.56\pm2.75$ in the range of $[-11.32,-0.95]$. The exceptions are Circinus, NGC~253, and NGC~1068 at $89$~\um. These galaxies have a broken power-law with a median $\alpha_{2}^{T_{\rm d}} = -1.49\pm0.80$ in the range of $[-3.0,-1.16]$, and a median $T_{\rm d}^{\rm c} = 28.5\pm2.6$~K in the range of $[28.0,33.7]$~K. We find that the second power-law is associated with the central $\sim1$ kpc of the galaxies cospatial with star-forming regions in Circinus and NGC~253, and the hot environment surrounding the active galactic nucleus of NGC~1068.

The physical mechanisms responsible for the trends described for each group are discussed in Section \ref{subsec:DIS_Bori}.

%%%%%%%%%%%%%%%%%
%%%% DISCUSSION %%%%
%%%%%%%%%%%%%%%%%

\section{Discussion}\label{sec:DIS}

%%%%%%%%%%%%%%%%%
\subsection{Relative contribution of the ordered and random B-fields.}\label{subsec:DIS_Bori}
%%%%%%%%%%%%%%%%%

We have identified three groups based on the $P-N_{\rm HI+H_{2}}$ relation (Section \ref{subsec:PIPlots}). In general, we found a decrease in polarization fraction with increasing column density. As described in Section \ref{subsec:BFIR},  a maximum polarization is measured in the absence of tangled B-fields along the LOS, turbulence, and/or inclination effects. Under these `perfect' conditions, the polarization fraction should be expected to be constant with column density due to a dominant ordered B-field within the beam of the observations. Assuming that the galaxy disk is at a constant inclination and no other inclination effects are important, a decrease in polarization fraction with increasing column density is attributed to an increase in turbulent and/or tangled B-fields along the LOS (i.e., random B-fields) within the beam of the observations. The inclination effects on the polarization fraction are discussed in Section \ref{subsec:DIS_Inc}. We use the $P-N_{\rm HI+H_{2}}$ plots as a proxy for the relative contribution of the random-to-ordered B-fields within the beam of the observations.

We found a similar slope $\sim0.9$ for low column densities, $N_{\rm HI+H_{2}} < N_{\rm HI+H_{2}}^{\rm c}$, in galaxies with a broken power-law in Group 1, at high column densities, $N_{\rm HI+H_{2}} > N_{\rm HI+H_{2}}^{\rm c}$,  in galaxies with a broken power-law in Group 2, and at all column densities, $\log_{10}(N_{\rm HI+H_{2}} [{\rm cm}^{-2}]) = [19.96,22.91]$, in galaxies with a single power-law (Group 3). 

Using the $P-T_{\rm d}$ relation, we find that $\alpha_{1}^{T_{\rm d}}\sim -7$, within the uncertainties, for the three groups identified in this galaxy sample. For the galaxies with a broken power-law, we find that $\alpha_{2}^{T_{\rm d}} \sim -1$, within the uncertainties, independently of the group. These results indicate that the $P-T_{\rm d}$ relation is not sensitive to the physical conditions of the B-fields in the galaxy's disk, but rather to the ISM conditions (i.e. radiation field). The $P-T_{\rm d}$ relation provides information about the polarization efficiency toward star-forming regions. 

We now discuss the trends of each group based on the relative contribution of the ordered and random (i.e. turbulent and/or tangled) B-fields and dust grain alignment efficiency. We refer the reader to the summary of these trends shown in Figure \ref{fig:fig5}.

%%%%%%%%%%%%%%%%%
\subsubsection{Group 1: Compressed and/or sheared B-fields.}\label{subsubsec:DIS_G1}

Group~1 has a flatter power-law after the cut-off column density and dust temperature. This is a result of a faster increase in polarized intensity than in total intensity (Figure Figure \ref{fig:fig5} and \ref{fig:A3_fig1}). This result is interesting because, a priori, it may indicate that the relative contribution of turbulence and/or tangled B-fields along the LOS (i.e., random B-fields) is smaller in the regions associated with $\alpha_{2}$ than in the regions with $\alpha_{1}$, or that the relative contribution of an ordered B-field  is larger. The regions associated with $\alpha_{2}$ are spatially correlated with the central starburst of M82, the inner-bar and the inflection regions between the inner-bar and the spiral arms of M83, and the starburst ring of Circinus.  These areas are typically associated with regions of higher turbulence and/or tangled B-fields along the LOS than those in the diffuse ISM of galaxies. This result indicates that another physical mechanism may be flattening the $P-N_{\rm HI+H2}$ and $P-T_{\rm d}$ relations. 

For M82, \citet{ELR2021c} showed that the turbulent kinetic and magnetic energies are in close equipartition within the central $\sim2$~kpc radius of the starburst. Using 53~\um~polarimetric observations, these authors estimated a median turbulent B-field strength of $\sim300~\mu$G. The B-field orientation is parallel to the galactic outflow with an angular dispersion of $17.1^{\circ}$. Although M82 may have small-scale random B-fields in the outflow within the beam of the observations, the dominant component is the ordered B-field driven by the turbulent kinetic energy of the galactic outflow. The ordered B-field has a coherence length ($\delta = 73.6\pm5.6$~pc, $3.99\pm0.30$\arcsec) larger than the beam of the observations. The $53$ \um~polarimetric observations are sensitive to the compressed B-fields along the galactic outflow. For M83, \citet{Frick2016} showed that the measured B-fields may be dominated by the mean-field dynamo at a resolution of $520$~pc ($12$\arcsec) using $6-13$~cm polarimetric observations. These authors concluded that the measured B-fields are compressed and partially aligned with the material arms in the galaxy. Circinus galaxy has a starburst ring-like structure with a diameter of $\sim350$~pc ($\sim21$\arcsec) in $^{12}$CO(1-0) \citep{Elmouttie1998} and $\sim1$~kpc ($\sim61$\arcsec) in HI \citep{Jones1999}. A compressed B-field may be present within the central $1$~kpc of Circinus driven by galactic-shock dynamics in the inner-bar and starburst ring. This environment may be similar to that found in the central $1$~kpc of NGC~1097 using HAWC+ observations \citep{ELR2021c}.

We interpret the flatter $\alpha_{2}$ found in Group~1 as follows.  After the cut-off column density, $\log_{10}(N_{\rm H+H_{2}}^{\rm c} [{\rm cm}^{-2}]) \ge 21.59\pm0.31$ and dust temperature $T_{\rm d}^{\rm c} \ge 35.8\pm3.8$~K, the relative contribution of the ordered B-field of these galaxies increases faster than the random B-field component. This result is based on the fact that the B-field structure associated with $\alpha_{2}$ has a dominant large-scale B-field with low angular dispersion from beam to beam. The enhancement of the ordered B-fields may be produced by compression or shear in high density and high dust temperature regions of the galaxy associated with galactic outflows and starburst rings. This scenario drives a faster increase in $PI$ than in $I$. The physical reason is that compressed B-fields produce a relative decrease of the isotropic B-fields (i.e., an increase in anisotropic B-fields), yielding an increment of $PI$.

The $P-T_{\rm d}$ relation may be explained as a change in the dust alignment efficiency based on RATs \citep{HL2016}. The polarization fraction depends on the minimum size of aligned dust grains, $a_{\rm align}$. \citet{Hoang2021} provided an analytical function of $a_{\rm align}$ as a function of the volume density, $n_{\rm H}$, and dust temperature as $a_{\rm align} \propto n_{\rm H}^{2/7}T_{\rm d}^{-12/7}$. Higher polarization fractions are expected for smaller values of $a_{\rm align}$ due to the fact that the size distribution of aligned dust grains is broader. For the galaxies within this group, our results indicate that $\alpha_{2}^{T_{\rm d}}$ is flatter than $\alpha_{1}^{T_{\rm d}}$ after the cut-off dust temperature of $T_{\rm d}^{\rm c} = 37.8\pm3.8$ K. This behaviour may indicate a faster increase of $T_{\rm d}$ than the column density, which increases the alignment size, $a_{\rm align}$. The increase in dust grain alignment efficiency produces a flatter $P-T_{\rm d}$ relation. An enhancement of the dust grain alignment efficiency may be spatially located along the compression and/or shear in high density and high temperature regions.

%%%%%%%%%%%%%%%%%
\subsubsection{Group 2: Random B-fields and dust grain alignment efficiency}\label{subsubsec:DIS_G2}

Group~2 has a steeper power-law after the cut-off column density in $P-N_{\rm HI+H_{2}}$, and a single power-law in the $P-T_{\rm d}$ relation. For the $P-N_{\rm HI+H_{2}}$ relation, this is a result of a faster increase in total intensity than in polarized intensity (Figures~\ref{fig:fig5} and \ref{fig:A3_fig2}). We find that $\alpha_{2}$ is spatially associated with the spiral arms of M51, the star-forming regions in NGC~6946, and the central $\sim1$ kpc starburst rings of NGC~1097, NGC~4736, and NGC~7331. For the $P-T_{\rm d}$ relation, a single-power-law is the result of a constant $PI$ with increasing $I$. 

A study of the FIR-radio correlation in NGC~6946 showed that the star formation rate surface density is correlated with the turbulent B-field, $B_{\rm tur}\propto\Sigma_{SFR}^{0.16\pm0.01}$ \citep{Tabatabaei2013}. This result indicates that the increase of the gas turbulence associated with the star-forming regions is correlated with an increase of the isotropic random B-field. Supernova explosions increase turbulence in the ISM, which amplify the B-field strength via fluctuation dynamo (Section \ref{subsec:BOri}). In addition, the synchrotron polarized intensity is spatially located in the interarm regions of M51 and NGC~6949 \citep{Beck2007,Fletcher2011}. This result has been explained due to the fact that B-fields in the arms are more turbulent when compared to those in the interarms. \citet{Borlaff2021} measured a relative increase of turbulent B-fields in the spiral arms of M51. The increase of turbulent B-fields is associated with regions of high column density and high turbulent kinetic energy traced by the $^{12}$CO(1-0) molecular gas. Furthermore, the star formation rate was found to be uncorrelated with the FIR polarized flux but correlated with the radio polarized flux. These authors suggested a fluctuation dynamo-driven B-field amplification scenario associated with the star-forming regions in the disk of M51.

For NGC~1097, the FIR polarized intensity is spatially located with the warmest areas of shock-driven galactic-bar dynamics in the contact regions between the bar and the starburst ring within the central $1$~kpc \citep{ELR2021c}. The radio polarization observations show a superposition of ordered spiral B-fields across the starburst ring toward the core \citep{Beck2007,ELR2021c}. \citet{Tabatabaei2018} showed that the B-field strength decreases with the star formation efficiency, which is interpreted as a region dominated by the non-thermal pressure from the B-field, cosmic rays, and turbulence.

For NGC~4736, the radio polarimetric observations showed an ordered spiral B-field across the starburst ring toward the core \citep{Chyzy2008}. This B-field is not related to the star-forming activity but rather may be related to an evolutionary stage of the galaxy. Our FIR polarimetric observations also seem to show an ordered spiral B-field across the starburst ring (Figure \ref{fig:figA11}).

NGC~7331 suffers from strong Faraday depolarization due to its high inclination, $78.1\pm2.7^{\circ}$ (Table \ref{tab:GalaxySample}), and also suffers from high rotation measure, $-177\pm7$~rad~m$^{-2}$, arising from Galactic foreground \citep{Heald2009}. This situation makes the study of this galaxy at radio wavelengths very challenging. NGC~7331 has an axisymmetric gas inflow at the inner boundaries of the starburst ring with its ring rotating clockwise \citep{Battaner2003}. After a comparison of the UV, FIR, CII, and PAH emission, \citet{Sutter2022}  showed a complex environment due to extinction and directly radiated gas within the central $1$~kpc of this galaxy (similar FOV and angular resolution as those from HAWC+ were used).  We did not find a detailed analysis of the B-field strength with the star formation rate in this galaxy anywhere in the literature. Further quantitative analysis across the galactic disk of NGC~7331 is required using the information summarized above.

For M51 and NGC~6946, we interpret the steeper $\alpha_{2}$ as the result of the fact that the turbulent and/or tangled B-fields increase faster than the ordered B-fields in the star-forming regions and spiral arms. This scenario drives a faster increase in $I$ than in $PI$. The physical reason is that the B-fields become more isotropic in the star-forming regions at the resolution of our FIR observations. The isotropic B-field is driven by an increase of turbulent and/or tangled B-fields, which increase the unpolarized total intensity associated with the star-forming regions in the galaxy's disk. The relative contribution of the fluctuation dynamo increases toward the regions of star formation in M51 and NGC~6946. Furthermore, Group 1 has flatter slopes with higher column densities, higher dust temperatures, and higher star formation rates than Group 2. These results invite the question of how the stages of star-forming regions as a function of galaxy types affect the fluctuation dynamo in galaxies.

The lack of a relation between the FIR polarized flux and the star-forming regions in NGC~1097, NGC~4736, and NGC~7331 indicates that another physical mechanism to that invoked in M51 and NGC~6946 may be playing a role in the $P-N_{\rm HI+H_{2}}$ relation. Although we find a median of $\alpha_{2}=0.99\pm0.04$ in the $P-N_{\rm HI+H_{2}}$ relation for all galaxies in Group~2, this subset of galaxies (NGC~1097, NGC~4736, and NGC~7331) have flatter $\alpha_{1}^{T_{\rm d}}=4.98\pm1.59$ to those measured in M51 and NGC~6946, $\alpha_{1}^{T_{\rm d}}=11.69\pm0.09$, in the $P-T_{\rm d}$ relation (Figure~\ref{fig:fig5}, Table~\ref{tab:PI_fits}). As $PI$ remains constant with $T_{\rm d}$ for all galaxies in Group~2, the flatter $\alpha_{1}^{T_{\rm d}}$ is driven by a flatter $I$ across the full range of $T_{\rm d}$ when compared with M51 and NGC~6946 (Figure~\ref{fig:fig5}).  

We provide an alternative scenario in terms of the dust grain alignment efficiency. For this subset of galaxies, our results indicate that $T_{\rm d}$ increases faster than $N_{\rm HI+{H_{2}}}$, which increases the alignment size, $a_{\rm align}$. The increase in dust grain alignment efficiency  produces a flatter $P-T_{\rm d}$ relation for NGC~1097, NGC~4736, and NGC~7331. An enhancement of the dust grain alignment efficiency may be spatially located with the shock-driven regions of NGC~1097 and a large-scale ordered B-field associated with gas dynamics in an evolutionary stage of the galaxy NGC~4736. This interpretation may explain the ordered B-fields found in both objects. In addition, note that NGC~4736 has the flattest power-law, $\alpha_{1} = -0.06^{+0.04}_{-0.04}$, in the $P-N_{\rm HI+H_{2}}$ relation of our galaxy sample. This result implies that NGC~4736 has the largest relative contribution of an ordered B-field and/or highest dust grain alignment efficiency for galactrocentric radius $\ge1$~kpc of any of the galaxies observed so far. These results require a detailed study of the B-RAT alignment mechanism (Section~\ref{subsec:BFIR}) under shock-driven conditions and galaxy evolution stages.

%%%%%%%%%%%%%%%%%
\subsubsection{Group 3: Random B-fields}\label{subsubsec:DIS_G3}

Group~3 has a single power-law in both $P-N_{\rm HI+H_{2}}$ and $P-T_{\rm d}$ relations. This is a result of a constant polarized intensity with increasing  total intensity (Figures~\ref{fig:fig5} and \ref{fig:A3_fig3}). We attribute this result to an increase of isotropic random B-fields with increasing column density and dust temperature. The enhancement of the random B-fields may be produced by star-forming regions in the galaxy's disk. The physical reason is due to the fact that the isotropic B-fields are driven by an increase of unpolarized total intensity associated with denser and hotter regions across the galaxy's disk. 

We find several galaxies with a change in $PI$ similar to that in $I$ (Figure~\ref{fig:fig5}). These galaxies are Circinus and NGC~1068 at $89$~\um, NGC~253 at $89$ and $154$~\um, and NGC~2146 at $53$ and $89$~\um. For these galaxies, the PI remains flat up to $\log_{10}(N_{\rm HI+H_{2}} [{\rm cm}^{-2}]) \sim 21.30$ for all galaxies except for NGC~253 where the change in $PI$ occurs at $\log_{10}(N_{\rm HI+H_{2}} [{\rm cm}^{-2}]) \sim 22.00$. These galaxies also show a broken power-law in the $P-T_{\rm d}$ relation with a flatter $\alpha_{2}^{T_{\rm d}}=-1.49\pm0.80$ than those with a single power-law, $\alpha_{1}^{T_{\rm d}}=-7.56\pm2.75$ (Figure~\ref{fig:fig5} and \ref{fig:A3_fig3}). The flatter $\alpha_{2}^{T_{\rm d}}$ is driven by an increase in $PI$ relative to $I$ toward regions of higher dust temperatures. The regions with $\alpha_{2}^{T_{\rm d}}$ are associated with the starburst ring of Circinus at $89$ \um, the central starburst of NGC~253 at $89$~\um, and the hot environment affected by the radio jet of NGC~1068 at $89$~\um. We speculate that these regions may have an enhancement of dust grain alignment efficiency and/or a relative increase of the ordered B-field driven by the central starburst and the galaxy-jet interaction.

%%%%%%%%%%%%%%%%%
\subsection{Polarized spectrum of starburst galaxies}\label{subsec:DIS_Pspec}
%%%%%%%%%%%%%%%%%

\begin{figure}[ht!]
\centering
\includegraphics[angle=0,width=\columnwidth]{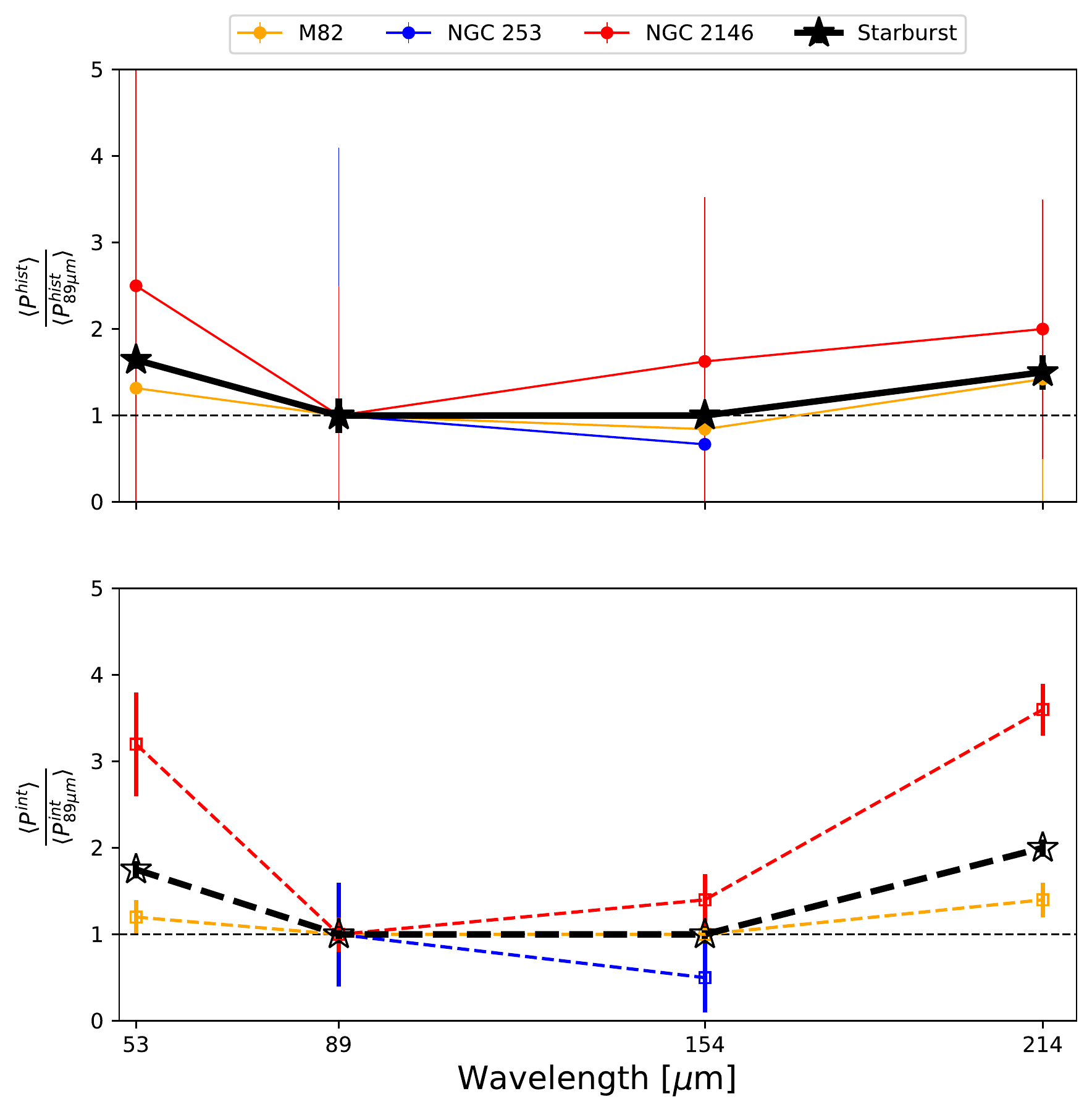}
\caption{Polarized spectrum of starburst galaxies. The normalized spectrum, $P_{\lambda}/P_{\lambda_0}$, to the reference wavelength of $\lambda_{0} = 89$~\um~for M82 (orange), NGC~253 (blue), NGC~2146 (red), and the median of the starburst galaxies (black stars) are shown. The  individual $\langle P^{\rm hist} \rangle$ (top panel; circles with solid lines) and integrated  $\langle P^{\rm int} \rangle$ (bottom panel; open squares with dashed lines) polarization fractions are shown. The horizontal black dashed line shows a flat normalized spectrum for comparison.
\label{fig:fig9}}
\epsscale{2.}
\end{figure}

We find that the polarized spectrum of starburst galaxies varies within the $50-220$~\um\ wavelength range (Figure~\ref{fig:fig4}). Figure~\ref{fig:fig9} shows the normalized spectrum, $P_{\lambda}/P_{\lambda_0}$, to the reference wavelength of $\lambda_{0} = 89$ \um~for individual starburst galaxies and the median spectrum of all starburst galaxies. The normalized spectrum minimizes depolarization effects (i.e., inclination, grain elongation, turbulence, tangled B-fields), which work at all wavelengths equally well \citep{Hildebrand1999}. Variations in the polarized spectrum are due to physical variations (i.e., dust temperature, dust grain alignment efficiency, dust grain composition, random B-fields) that affect each band differently. For individual starburst galaxies, the uncertainties in $\langle P^{\rm hist} \rangle$ are too large to provide any statistical trend, so we study $\langle P^{\rm hist}_{\rm starburst} \rangle$. We measure the median individual, $\langle P^{\rm hist}_{\rm starburst} \rangle$, and integrated, $\langle P^{\rm int}_{\rm starburst} \rangle$, polarization fractions in the range of $[1.3,2.3]$\% and $[0.7,1.6]$\%, respectively (Table \ref{tab:PPA_all}). In both cases the polarization fraction falls from $53$ to a minimum in the $89-154$~\um\ wavelength range and then increases to $214$~\um.

%%%%%%%%%%%%%%
%%%% FIGURE 9 %%%%
%%%%%%%%%%%%%%
%%%%%%%%%%%%%%

There are no polarization spectrum models available that consider the physical conditions of a starburst galaxy. However, several models have been produced for the diffuse ISM and molecular clouds in the Milky Way, and nearby galaxies. \citet{DF2009} generated the $2-3000$~\um~polarized spectrum using a mixture of spheroidal silicate and graphite grains. These authors produced four models: a) two of them (models 1 and 3) assumed spherical graphite grains incapable of being magnetically aligned and producing polarized emission, and oblate spheroids silicate grains magnetically aligned as a function of their grain size distribution, and b) the other two models (models 2 and 4) assumed both silicate and graphite grains to be oblate. For all models, the $50-220$ \um~polarized spectrum is flat and then increases with wavelength. For models 1 and 2, the polarized spectrum is $\sim0$\% up to $\sim60$~\um~and then increases to $\sim7$\% at $220$~\um. For models 2 and 4, the polarized spectrum is $\sim1$\% at $50$~\um~and then increases to $\sim7-9$\% at $\sim220$~\um. The increase in their polarized spectrum is expected as the optical depth decreases with wavelength, producing an increase in thermal polarized emission. Interestingly, \citet{Vandenbroucke2021} showed a polarized spectrum with a dip in the $60-90$~\um~wavelength range. They assumed  a linear mixture of perfectly aligned silicate dust grains and non-aligned graphite dust grains. Only dust grains with sizes $>0.1$~\um~were assumed to be aligned. However, the authors did not use this mixture for the rest of their simulations and the $60-90$~\um~polarization fraction dip is missing in their synthetic $50-220$~\um~polarimetric observations of a nearby galaxy. The synthetic observations were performed to simulate HAWC+ observations at all bands for a Milky Way-like galaxy at a distance of $10$~Mpc.

The FIR polarized spectrum of several molecular clouds and star-forming regions in the Milky Way have been measured using FIR polarimetric observations. \citet{Hildebrand1999} measured variations in the $60-350$~\um~polarized spectrum normalized at $100$~\um~of Galactic clouds. Dense cloud cores (Orion BN/KL) have a rising polarization spectrum, while cloud envelopes (M17) have a falling spectrum. These authors argued that the falling spectrum of cloud envelopes may be caused by a mix of dust grain composition, with their efficiencies changing as a function of dust temperature. The regions with warmer temperatures contain the aligned dust grains. For dense clouds, the rising polarized spectrum is in agreement with the expectation of an increasing polarization fraction as the optical depth decreases with increasing wavelength. \citet{Michail2021} measured a falling $50-220$~\um\ polarized spectrum normalized at $214$~\um~in the OMC-1 (including BN/KL) star-forming region. These authors concluded that the falling polarized spectrum is produced by variations in dust grain alignment efficiency due to a mixture of dust grain populations and variations in dust temperatures along the LOS. 

Our FIR polarized spectrum comprises all the LOS within the central several kpcs of the starburst galaxies. However, the starburst galaxies show several B-field structures: a) parallel to the galactic outflow, b)  parallel to the galactic plane, and c) potentially aligned with tidal tails (Figures \ref{fig:figA4}, \ref{fig:figA6}, and \ref{fig:figA9}). These structures have different dust temperature variations as a function of wavelength (Section \ref{subsec:DIS_Bori}). Thus, our polarized spectrum is the result of all these structures combined. Despite this effect, the polarized emission at $53$ \um~is dominated by the hot dust in the galactic outflow; at $214$~\um~is dominated by the cold dust in the plane of the galaxy; and at $89-154$ \um~is the result of a combination of both the galactic outflow and the galactic plane. With these caveats, it is plausible that the falling $53-154$~\um~polarized spectrum may be due to the hot dust temperature variations along the LOS in the galactic outflow. The rising $154-214$ \um~polarized spectrum may be due to the expected increase of the polarization fraction from cold dust and low optical depth in the galaxy plane. Any further analysis must be taken with caution as it is required to physically separate the contributions of the galactic outflow and the galactic disk as a function of wavelength.

%%%%%%%%%%%%%%
\subsection{Inclination effects}\label{subsec:DIS_Inc}
%%%%%%%%%%%%%%

\begin{figure}[ht!]
\centering
\includegraphics[angle=0,width=\columnwidth]{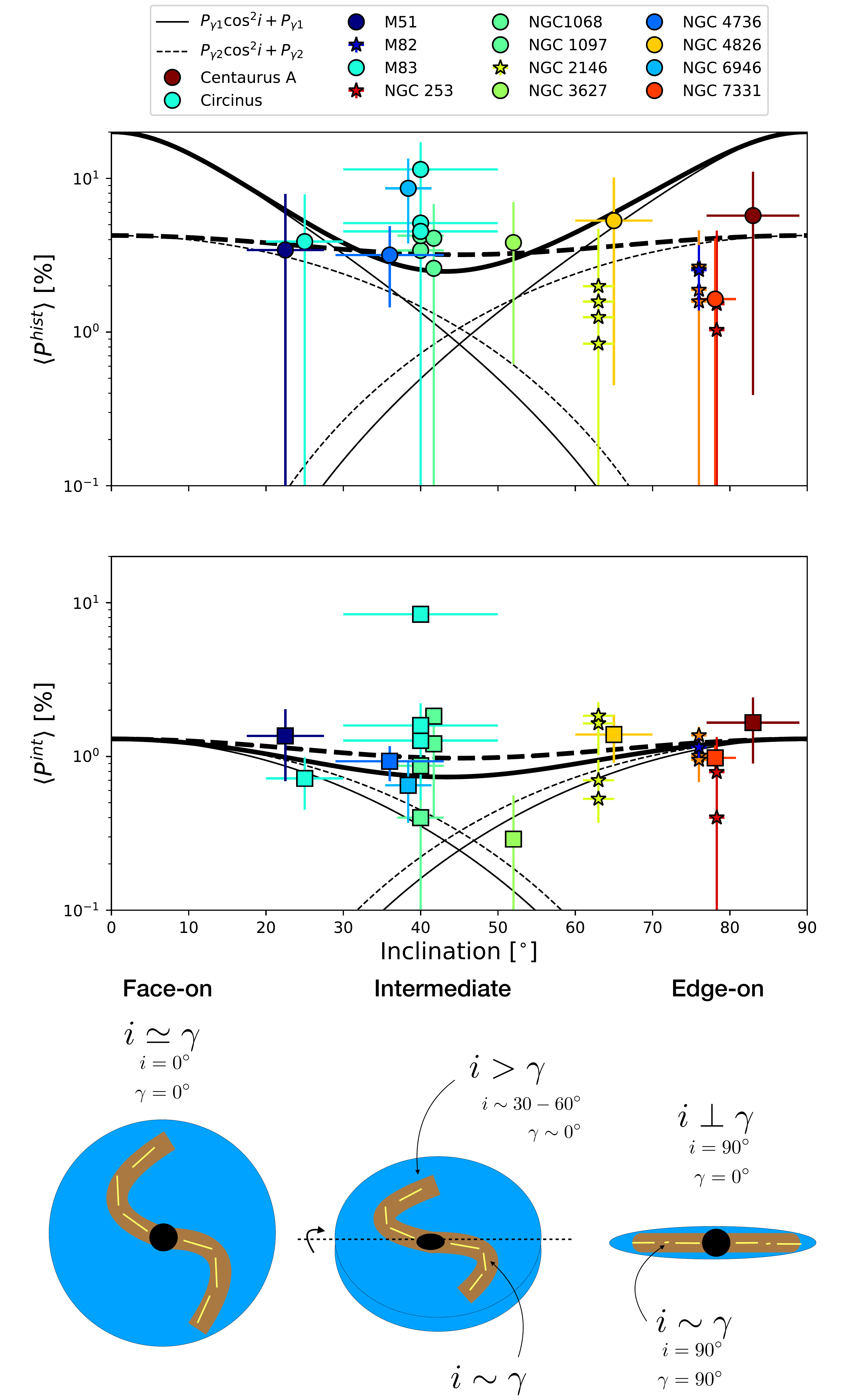}
\caption{Median and integrated polarization fraction as a function of the galaxy inclination and illustration of results. The median individual, $\langle P^{\rm hist} \rangle$ (top panel) and integrated, $\langle P^{\rm int} \rangle$, (middle panel) polarization fractions for each galaxy are shown. Starburst galaxies are displayed as stars while the other galaxies are displayed as circles for $\langle P^{\rm hist} \rangle$ and squares for $\langle P^{\rm int} \rangle$. The colors are sorted by the galaxy inclination.
The expected polarization using the physical conditions of the environment in which the dust grains are embedded (solid thick line) and a pure geometrical model (dashed thick line) are shown. See Section~\ref{subsec:DIS_Inc} for details about the thin lines of each model.
The illustration (bottom panel) show the change of polarization fraction (yellow lines) as a function of the galaxy's inclination $i$, and B$_{\rm POS}$ angle, $\gamma$, for face-on (left), intermediate (middle), and edge-on (right) views. For the intermediate view, the axis and direction of rotation in the plane of the sky (dashed thin line) are shown. The lower region of the galaxy is the near side of the galaxy, while the upper region is the far side.
\label{fig:fig10}}
\epsscale{2.}
\end{figure}

We analyze the dependence of the measured polarization fraction with the galaxy inclination, $i$. Figure~\ref{fig:fig10} shows the median individual, $\langle P^{\rm hist} \rangle$, and integrated, $\langle P^{\rm int} \rangle$, polarization fractions (Table \ref{tab:PPA}) as a function of the galaxy inclination (Table~\ref{tab:GalaxySample}). We estimate that the median individual polarization fraction is mostly flat, $\langle P^{\rm hist}_{\rm all}\rangle = 3.1\pm0.3$\%, across the $22.5-83.0^{\circ}$ range of galaxy inclinations. The uncertainties of the median polarization fractions of individual galaxies are too large to measure any potential variation across the galaxy inclination range (Figures~\ref{fig:fig2} and \ref{fig:fig10}, Table~\ref{tab:PPA}). The large uncertainties in $\langle P^{\rm hist} \rangle$ are due to the effect of the several physical components (i.e. galactic outflow, spiral arms, interarms, and star-forming regions) on the polarization fraction (Section~\ref{subsec:DIS_Bori}). The median integrated polarization fraction is also mostly flat, $\langle P^{\rm int}_{\rm all} \rangle = 1.3\pm0.2$\%, across the $22.5-83.0^{\circ}$ range of galaxy inclinations.

%%%%%%%%%%%%%%
%%%% FIGURE 10 %%%%
%%%%%%%%%%%%%%
%%%%%%%%%%%%%%

We estimate the expected variation of the polarization fraction as a function of the inclination and the B-field orientation in the POS using two approaches. The first approach uses analytical solutions assuming polarized emission in an optically thin medium, isothermal gas, and equilibrium between gas and dust grains. Under the perfect conditions of no dispersion along the LOS and a constant inclination, the tilt-corrected polarization fraction, $P_{\gamma 1}$, can be estimated as

\begin{equation}\label{eq:P1}
    P_{\gamma 1} = \frac{p_{0}\cos^{2} \gamma}{1-p_{0}(\frac{\cos^{2} \gamma}{2} -\frac{1}{3})}
\end{equation}
\noindent
\citep{Fiege2000,King2019,Chen2019}, where $p_{0}$ is the polarization efficiency, which is related to the dust grain properties, and $\gamma$ is the tilt angle between the ordered B-field and the plane of the sky. For a face-on view and assuming that the B-field orientation is in the plane of the sky, $\gamma=0^{\circ}$, the maximum polarization is then obtained to be $P_{\gamma 1,\rm max} = p_{0}/(1-p_{0}/6)$ \citep{Fiege2000}. $p_{0}$ is assumed to be the intrinsic polarization efficiency associated with the entire object with homogeneous alignment.

The second approach parametrizes the observed polarization as

\begin{equation}\label{eq:P2}
    P_{\gamma 2} = p_{0}' \mathcal{R} \mathcal{F} \cos^{2}\gamma 
\end{equation}
\noindent
\citep{LD1985}, where $p_{0}'$ is the intrinsic polarization arising from aligned dust grains given a set of dust grain properties, $\mathcal{R}$ is the Rayleigh reduction factor, which characterizes the efficiency of grain alignment with the local B-field orientation, $\mathcal{F}$ quantifies the effect of the tangled B-field along the LOS. Assuming no dispersion along the LOS and perfect alignment efficiency, then $P_{\gamma 2} = p_{0}' \cos^{2} \gamma$. For a face-on view and assuming that the B-field orientation is in the plane of the sky, $\gamma=0^{\circ}$, the maximum polarization is then obtained to be $P_{\gamma 2,\rm max} = p_{0}'$. $P_{\gamma 1}$ is a purely geometric definition of the observed polarization fraction, while $P_{\gamma 2}$ takes into account the physical conditions of the environment in which the dust grains are embedded. 

Figure~\ref{fig:fig10} shows the expected dependence of the polarization fraction as a function of $\gamma$ and $i$ using $P_{\gamma 1}\cos^{2} i$ and $P_{\gamma 2}\cos^{2} i$. For $\langle P^{\rm hist} \rangle$, we scaled $P_{\gamma 1, \rm max}$ to be equal to $20$\% given by the maximum polarization fraction \citep[][ our Section~\ref{subsec:DIS_Planck}]{Planck_Int_XIX_2015} allowed for an individual polarization  measurement across the galaxy disk. We scaled $P_{\gamma 2}\cos^{2} i$ to be equal to the measured median polarization fraction $\langle P^{\rm hist}_{\rm all}\rangle = 3.1\pm0.3$\% at the median galaxy inclination of $\langle i \rangle = 46.9^{\circ}$ from our sample. For $\langle P^{\rm int} \rangle$, we scale both $P_{\gamma 1,\rm max}$ and $P_{\gamma 2,\rm max}$ to be equal to the median integrated polarization fraction $\langle P^{\rm int}_{\rm all} \rangle = 1.3\pm0.2$\%. 

Both $P_{\gamma 1}\cos^{2} i$ and $P_{\gamma 2}\cos^{2} i$ decrease with increasing $\gamma$ and $i$, however the measured polarization fraction remains mostly flat at all inclinations. This implies that, in average, the B-field orientation remains fairly constant with respect to the plane of the sky within the galaxies across the inclination range of $i = 22.5-83.0^{\circ}$. This result is due to the fact that the B-field is parallel to the plane of galaxies at FIR wavelengths. As an example, for highly inclined ($i\sim90^{\circ}$) galaxies, the B-fields are expected to be parallel to the plane of the sky ($\gamma\sim0^{\circ}$), yielding a maximum polarization fraction. To account for this effect, we add the same $P_{\gamma 1}$ and $P_{\gamma 2}$ dependence but with decreasing inclination. We compute the total expected polarization fraction with inclination as the linear combination of $P_{\gamma}\cos^{2} i$ and $P_{\gamma}$ across the range of inclinations. We show the total expected polarization as a function of $i$ and $\gamma$ in Figure~\ref{fig:fig10} (thick black lines). The total expected polarization fraction is the combination of a) a decrease of the polarization fraction with inclination and tilt from low to intermediate inclinations, and then b) an increase of the polarization fraction with tilt from intermediate to high inclinations. This variation may be more or less severe depending on the B-field morphology and the galaxy type. We estimate the expected polarization fraction as a function of the $i$ and $\gamma$ below. 

For face-on views ($i=0^{\circ}$), taking into account that a) the disk of the galaxy is at a constant inclination in our LOS and  parallel to the plane of the sky, and b) the measured weighted-average column density B-field is parallel to the plane of the galaxy ($\gamma=0^{\circ}$), then we find that $\gamma \simeq i$ (Fig.~\ref{fig:fig10}). The effect caused by the tilt and inclination is the same across the galaxy disk. The expected polarization is close to its maximum given by $P_{\gamma 1,\rm max}$ and $P_{\gamma 2,\rm max}$. For low galaxy inclinations, tilt and inclination effects only displace the entire trend up or down in $P$ without affecting the slope of the $P-N_{\rm HI+H_{2}}$ and $P-T_{\rm d}$ relations. 

For edge-on views ($i=90^{\circ}$), taking that the B-field is parallel to the disk of spiral galaxies, then $i \perp \gamma$, where $\gamma = 0^{\circ}$ (Fig.~\ref{fig:fig10}). The expected polarization fraction is close to its maximum at both face-on and edge-on views. Indeed, we measure this effect at the two edges of our inclination ranges: M51 ($i=22.5\pm5^{\circ}$) and Centaurus~A ($i=83\pm6^{\circ}$). The measured polarization fractions along the several LOS of the resolved disks may vary due to the dependence with $P-N_{\rm HI+H_{2}}$ and $P-T_{\rm d}$ driven by turbulence and/or tangled B-fields along the LOS. An exception may be when the spiral arms are along our LOS, yielding the B-field orientation to be perpendicular to the plane of the sky, $\gamma = 90^{\circ}$. In this case, the polarization fraction may tend to zero. This example is displayed in Figure~\ref{fig:fig10}-bottom right. \citet{Jones2020} observed tentative evidence of this behaviour in NGC~891 using $154$~\um~polarimetric observations with HAWC+.

For intermediate inclinations ($i\sim 30-60^{\circ}$), the measured polarization is affected by both $\gamma$ and $i$ (Figure~\ref{fig:fig10}). This case creates a depolarization effect within the same galactrocentric distance across the galaxy disk. Along the axis of rotation in the plane of the sky the ordered B-field orientation and the inclination can be up to $\gamma \leq i$, yielding a decrease in polarization fraction driven by both $i$ and $\gamma$. Perpendicular to this axis, the ordered B-field is parallel to the plane of the sky but different to the inclination of the galaxy disk, i.e. $i>\gamma$, yielding the maximum polarization fraction in the galaxy disk. As the inclination is the same across the galaxy disk, the depolarization is driven by the change of $\gamma$ (Section \ref{App:A4}). This effect causes the dip in polarization at intermediate inclinations $i\sim30-60^{\circ}$ shown in Fig.~\ref{fig:fig10}. 

We provide an estimation of the depolarization effect due to the change of $\gamma$ across the disk for a constant inclination in Appendix~\ref{App:A4} and Figure \ref{fig:A4_fig1}. As an example, we use the galaxies in our sample within an inclination of $30-60^{\circ}$. These galaxies have a median  $i_{30-60^{\circ}}=40.0\pm2.8^{\circ}$ and  $\langle P^{\rm hist}_{30-60^{\circ}} \rangle = 4.3\pm2.8$\%. We assume that the angular difference between the B-field orientation and the plane of the sky can be as large as the galaxy inclination, i.e., $\Delta \gamma \le i$. Finally, we estimate that the regions of the galaxy along the axis of rotation in the plane of the sky with $\Delta \gamma \sim i$ can suffer a depolarization of up to a factor of $2.5$  assuming a maximum polarization fraction of $p_{\gamma 1, \rm max} = \langle P^{\rm hist}_{30-60^{\circ}} \rangle = 4.3\pm2.8$\% for our first approach ($P_{\gamma 1}$), and up to a factor of $1.7$ for the pure geometrical approach ($P_{\gamma 2}$). Under these conditions, the depolarized regions in the galaxy disk would have an expected polarization of $\langle P^{\rm hist}_{30-60^{\circ}} \rangle / 2.5 \ge 1.7$\% produced by $\gamma$ and $i$, which can still be measured using our observations. Note that this depolarization effect is different from the trends shown in $P-N_{\rm HI+H_{2}}$ and $P-T_{\rm d}$. These relations are driven by turbulence and/or tangled B-fields along the LOS and dust grain alignment efficiency.

\subsection{Comparison with \textit{Planck} observations of the Milky Way}\label{subsec:DIS_Planck}

We compare our results in spiral galaxies with those from  $Planck$ in the Milky Way. The $P-N_{\rm HI}$ relation in the Milky Way shows a maximum polarization fraction of $P\sim20$\% at 353 GHz within the column density range of $\log_{10}(N_{\rm HI} [{\rm cm}^{-2}]) = [20.30,21.30]$ and a sharp drop of polarization at $\log_{10}(N_{\rm HI} [{\rm cm}^{-2}]) = 22.18$ \citep[][see fig.~19]{Planck_Int_XIX_2015}. Note that the maximum polarization by \textit{Planck} is measured for diffuse sightlines with very ordered B-fields \citep[e.g.][]{Panopoulou2019}. When polarization measurements within the Galactic plane are included, a flatter $P-{N_{\rm HI}}$ relation is found at $\log_{10}(N_{\rm HI} [{\rm cm}^{-2}]) \ge 22.47$. Thus, an increase of $\sim0.3$~dex. in $\log_{10}(N_{\rm HI} [{\rm cm}^{-2}])$ and a flatter slope in the $P-N_{\rm HI}$ relation are found when high column density, high dust temperature, and high star formation rate are included. These values are shown in Figure~\ref{fig:fig7}.

Using our observations, we measure a maximum polarization fraction of $15$\% (Fig.~\ref{fig:fig2}) within the range of column densities of $\log_{10}(N_{\rm HI} [{\rm cm}^{-2}]) = [20.00,21.83]$. After correcting by inclination, the maximum polarization fraction reached $\sim18$\% for column densities $\log_{10}(N_{\rm HI+{H_{2}}} [{\rm cm}^{-2}]) \le 21.19$ in the interarms of the outskirts of NGC~6946. We estimate a difference of $0.73$ dex. between the measured cut-off column densities of $\log_{10}(N_{\rm HI+H_{2}}^{\rm c} [{\rm cm}^{-2}]) = 21.59\pm0.31$, which includes starburst galaxies, and $\log_{10}(N_{\rm HI+H_{2}}^{\rm c} [{\rm cm}^{-2}]) = 20.86\pm0.26$, which includes spiral galaxies. These cut-off densities are associated with both flatter and steeper slopes, respectively (Figure~\ref{fig:fig5}). These results show that an increase in the cut-off column density and star formation activity flatten the $P-N_{\rm HI}$ relation. This result is similar to that found in the Milky Way by \textit{Planck}, considering the large uncertainties in the \textit{Planck}'s measurements (which are dominated by zero-point uncertainty in Stokes I).

Using \textit{Planck} data \citep[][see fig.~19]{Planck_Int_XIX_2015}, we estimate an average polarization fraction of $\sim4$\% with a range of $[0,20]$\% for our estimated median cut-off column density of $\log_{10}(N_{\rm HI+H_{2}}^{\rm c} [{\rm cm}^{-2}]) = 20.86\pm0.29$ in spiral galaxies (i.e., M51, M83, NGC~4736, NGC~6946, and NGC~7331). Using HAWC+ data, the median polarization fraction is estimated to be $\langle P^{\rm hist}_{\rm spirals,154\mu m} \rangle = 3.3\pm0.9$\% (Section~\ref{subsec:Polmaps}) with individual polarization fractions in the range of $[0,15]$\% across the disks of these spiral galaxies. Figure~\ref{fig:fig7} shows that the FIR polarization measurements within the disk of spiral galaxies fall within the range of column densities associated with the maximum polarization fraction by means of magnetically aligned dust grains in the Milky Way. Our measured maximum polarization fraction and the trends in $P-N_{\rm HI+H_{_{2}}}$ are similar to those found in the Milky Way by \textit{Planck}. These results also show that the emissive polarization from these spiral galaxies may arise from a similar ISM composition to those in the Milky Way.

%%%%%%%%%%%%%%%%%
%%%% CONCLUSIONS %%%%
%%%%%%%%%%%%%%%%%

\section{Conclusions}\label{sec:CON}

We have presented SALSA, the first FIR polarimetric survey of nearby galaxies using HAWC+/SOFIA. We have provided a brief background on extragalactic magnetism (Section~\ref{sec:Back}) with the goal of introducing to the reader the definitions (Sections~\ref{subsec:BDef} and \ref{subsec:BFIR}) and the most current results. The main scientific goals of this program are to (Section \ref{subsec:SciCas}):

\begin{itemize}
    \item[1.] \textit{Characterize the dependence of the large-scale ordered B-fields on the probed ISM phase, and provide measurements for galactic dynamo theories.}
    \item[2.] \textit{Quantify the relative contributions of the mean-field and fluctuation dynamos across a range of galaxy types and dynamical states.}
    \item[3.] \textit{Measure the effect of galactic outflows on the galactic B-field in starburst galaxies, and quantify how the B-field is transported to the circumgalactic medium.}
    \item[4.] \textit{Measure the B-field structures of interacting galaxies.}
\end{itemize}

We summarize the first data release of the program (Sections~\ref{subsec:DataSample} and \ref{subsec:OBS}):

\begin{itemize}
\item The first data release comprises $33$\% ($51.34$h out of $155.70$h) of the awarded total exposure time of the SOFIA Legacy Program taken from January 2020 to November 2021. 
\item All observations in this program were, and will be, performed using the newly implemented OTFMAP polarization mode with HAWC+ (Section~\ref{subsec:OBS}).  A full characterization of the OTFMAP polarimetric mode is described in Paper~III \citep{ELR2022}.
\item The final data products of $14$ galaxies (Table~\ref{tab:DR1}) in the wavelength range of $53-214$~\um\ ready for scientific analysis as well as tabulated figures and codes are available on the Legacy Program website (\url{http://galmagfields.com/}).
\end{itemize}

The main results of this data release focused on the polarization fraction of galaxies are (Section~\ref{subsec:Polmaps}):

\begin{itemize}
    \item Polarization fractions across the galaxy disks range from $0$ to $15$\% (Fig.~\ref{fig:fig2}). 
    \item The lowest median polarization fraction measurements are found in the starburst galaxies M82, NGC~253, and NGC~2146 (Figure~\ref{fig:fig2} and Table~\ref{tab:PPA}). The lowest polarization fractions are colocated with star-forming regions across the disk of spiral galaxies. The highest polarization fraction measurements are found in the interarms and outskirts of spiral galaxies.
    \item We report the first polarized spectrum of starburst galaxies (Fig.~\ref{fig:fig4}). The polarized spectrum varies as a function of wavelength with a minimum located in the range of $89-154$~\um. The polarized emission at $53$~\um~is dominated by the hot dust in the galactic outflow, at $214$~\um~is dominated by the cold dust in the plane of the galaxy, and at $89-154$~\um~is the result of a combination of both the galactic outflow and the galactic plane (Fig.~\ref{fig:figA4}). It is plausible that the falling $53-154$~\um~polarized spectrum may be due to the hot dust temperature variations along the LOS in the galactic outflow. The rising $154-214$~\um~polarized spectrum may be due to the expected increase of the polarization fraction from cold dust and low optical depth in the galaxy plane (Section \ref{subsec:DIS_Pspec}).
\end{itemize}

The main results focused on the polarization fraction of galaxies with the column density and dust temperature are (Sections \ref{subsec:PIPlots}, \ref{subsec:PTdPlots}):

\begin{itemize}
\item For all galaxies, the polarization fraction decreases across the column density range of $\log_{10}(\rm N_{HI+H_{2}} [{\rm cm}^{-2}]) = [19.96,22.91]$, and dust temperatures of $T_{\rm d} = [19,48]$~K.
\item The highest polarization fractions, $\ge10$\%, are found at the lowest column density ranges, $\log_{10}(N_{\rm HI+H_{2}} [{\rm cm}^{-2}]) \le 21$, and dust temperature ranges, $T_{\rm d} \le 30$~K, typically located  in the outskirts of galaxies.
\item The spiral galaxies show a median polarization fraction of $3.3\pm0.9$\% at $154$~\um~using the individual measurements across the galaxy disk(Figure~\ref{fig:fig4} and Table~\ref{tab:PPA_all}), and an integrated polarization fraction of $\langle P^{\rm int}_{\rm spiral,154\mu m} \rangle = 0.8\pm0.1$\%.
\end{itemize}

Based on the trends of the $P-N_{\rm HI+H_{2}}$ and $P-T_{\rm d}$ relations, we found that 50\% (7 out of 14) of the galaxies require a broken power law in the $P-N_{\rm HI+H2}$ relation. As described in Section \ref{subsec:BFIR},  a maximum polarization is measured in the absence of tangled B-fields along the LOS, turbulence, and/or inclination effects. Under these `perfect' conditions, the polarization fraction should be expected to be constant with column density due to a dominant ordered B-field within the beam of the observations. Assuming that the galaxy disk is at a constant inclination, a decrease in polarization fraction with increasing column density is attributed to an increase in turbulent and/or tangled B-fields along the LOS (i.e., random B-fields) within the beam of the observations. The inclination effects on the polarization fraction are discussed in Section \ref{subsec:DIS_Inc}. We use the $P-N_{\rm HI+H_{2}}$ plots as a proxy for the relative contribution of the random-to-ordered B-fields within the beam of the observations. The $P-T_{\rm d}$ relation is not sensitive to the physical conditions of the B-fields in the galaxy's disk, but rather to the ISM conditions (i.e. radiation field). The $P-T_{\rm d}$ relation provides information about the polarization efficiency toward star-forming regions

We identified three main groups (Figure~\ref{fig:fig5} and Section~\ref{subsec:DIS_Bori}) and interpret them in terms of the B-field structure:

\begin{itemize}
\item[Group 1:] The $P-N_{\rm HI+H_{2}}$ and $P-T_{\rm d}$ relations have a flatter power-law, $\alpha_{2} > \alpha_{1}$, after the cut-off column density and dust temperature, respectively. Galaxies in this group are Circinus at $214$~\um, M83 at $154$~\um, and M82 at all bands (Figure~\ref{fig:A3_fig1}). We found that the second power-law is associated with the central starburst of M82, the inner-bar and the inflection regions between the inner-bar and the spiral arms of M83, and the starburst ring of Circinus. After the cut-off column density, $\log_{10}(N_{\rm HI+H_{2}}^{\rm c} [{\rm cm}^{-2}]) \ge 21.59\pm0.31$ and dust temperature $T_{\rm d}^{\rm c} \ge 35.8\pm3.8$ K, the ordered B-fields in these galaxies increase at a higher rate than the random B-fields. The physical reason is that compressed B-fields produce a relative decrease of the isotropic B-fields (i.e., an increase in anisotropic B-fields), yielding flatter slopes.

\item[Group 2:] The $P-N_{\rm HI+H_{2}}$ relation has a steeper power-law, $\alpha_{2} < \alpha_{1}$, after the cut-off column density, while a single power-law is found in the $P-N_{\rm HI+H_{2}}$ relation. Galaxies in this group are M51, NGC~4736, NGC~6946, and NGC~7331 at $154$~\um, and NGC~1097 at $89$ and $154$~\um~(Figure~\ref{fig:A3_fig2}). We found that a second power-law is required for the central regions of galaxies associated with the galaxy's core and star-forming regions across the galaxy's disk. We interpret the steeper $\alpha_{2}$ being due to a relative faster increase of the turbulent and/or tangled B-fields than the ordered B-fields in the star-forming regions and spiral arms.  The physical reason is that the B-fields become more isotropic in the star-forming regions at the resolution of our FIR observations. The isotropic B-field is driven by an increase of turbulent and/or tangled B-fields, which increase the unpolarized total intensity associated with the star-forming regions in the galaxy's disk.

\item[Group 3:] The $P-N_{\rm HI+H_{2}}$ and $P-T_{\rm d}$ relations have a single power-law. Galaxies associated with this group are shown in Figure~\ref{fig:A3_fig3}. This result is due to an increase of isotropic random B-fields with increasing column density and dust temperature. The enhancement of the random B-fields may be produced by star-forming regions in the galaxy's disk. The physical reason is that the isotropic B-fields are driven by an increase of unpolarized total intensity associated with denser and hotter regions across the galaxy's disk.
\end{itemize}

We have also compared our FIR polarimetric observations of spiral galaxies to those in the Milky Way by \textit{Planck} \citep{Planck_Int_XIX_2015}. The main results are (Figure~\ref{fig:fig7} and Section~\ref{subsec:DIS_Planck}):

\begin{itemize}
    \item Using our observations, we found that an increase in the cut-off column density and star formation activity flatten the $P-N_{\rm HI}$ relation. This result is similar to the trends found in the $P-N_{\rm HI}$ relation with and without the Galactic center by \textit{Planck}.
    \item The FIR polarization measurements within the disk of spiral galaxies fall within the range of column densities associated with the maximum polarization fraction by means of magnetically aligned dust grains in the Milky Way by \textit{Planck}.
\end{itemize}

As mentioned above, SALSA III released 33\% of the awarded total time of the SOFIA Legacy program  from January 2020 to November 2021. From November 2021 to September 2022 (the end of SOFIA operations), SOFIA is prioritizing the observations associated with SALSA, while maximizing the scientific output from multiple Legacy and high priority programs. Regardless of  the SOFIA output, this data release already conveys the power and potential of FIR polarimetric observations on extragalactic magnetism. Without SOFIA, the next opportunity for FIR polarimetry will arise somewhere between decades from now and infinity--no such capability is even being contemplated, as far as we know. To continue to push the boundaries of our understanding of extragalactic magnetism, FIR polarimetric observations must be included in the next NASA FIR probe mission class.


\begin{acknowledgments}

We are grateful to the anonymous referee, whom comments greatly helped to clarify and improve the manuscript. Based on observations made with the NASA/DLR Stratospheric Observatory for Infrared Astronomy (SOFIA) under the 07\_0034, 08\_0012 Program. SOFIA is jointly operated by the Universities Space Research Association, Inc. (USRA), under NASA contract NNA17BF53C, and the Deutsches SOFIA Institut (DSI) under DLR contract 50 OK 0901 to the University of Stuttgart. KT has received funding from the European Research Council (ERC) under the European Unions Horizon 2020 research and innovation programme under grant agreement No. 771282. EN acknowledges funding the Hellenic Foundation for Research and Innovation (Project number 224) and the ERC Grant "Interstellar" (Grant agreement 740120).

\end{acknowledgments}


\vspace{5mm}
\facilities{SOFIA (HAWC+), \textit{Herschel} (PACS, SPIRE)}


\software{\textsc{astropy} \citep{astropy}, 
\textsc{APLpy} \citep{aplpy},
\textsc{matplotlib} \citep{hunter2007}, 
\textsc{pymc3} \citep{pymc}
          }

\appendix

\section{Polarization, column density, and dust temperature maps}\label{A1:maps}

\begin{figure*}[ht!]
\includegraphics[angle=0,scale=0.31]{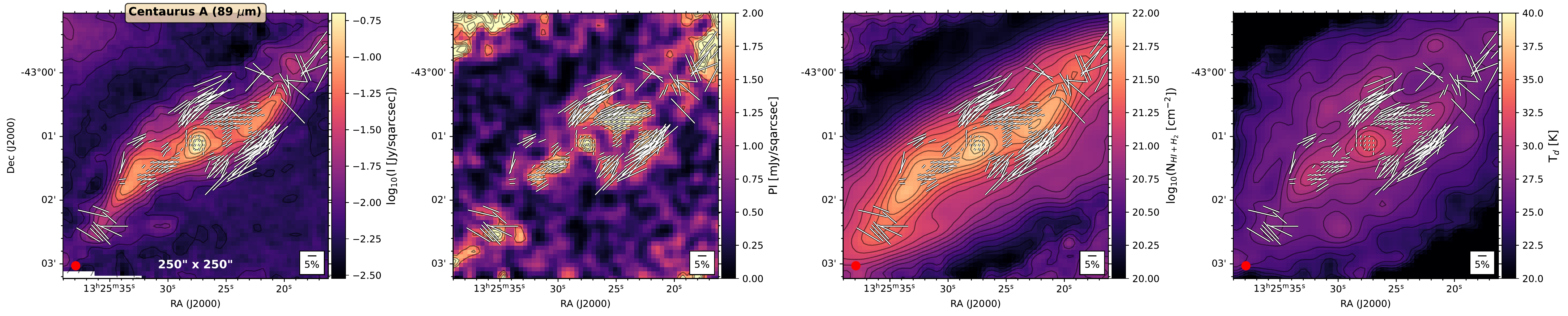}
\caption{Centaurus A at $89$~\um\ (Group 3). 
\textit{First column:} Total intensity (colorscale) in logarithmic scale with contours starting at $8\sigma$ increasing in steps of $2^{n}\sigma$, with $n= 3, 3.5, 4, \dots$ the polarization fraction measurements (white lines) with a legend of $5$\% (bottom right) are shown. The polarization fraction measurements with $PI/\sigma_{PI}\ge3$, $P\le20$\%, and $I/\sigma_{I} \ge 10$ were selected. The beam size (red circle) is shown.
\textit{Second column:} Polarization map over the polarized intensity (colorscale) in linear scale with contours starting at $3 \sigma$ increasing in steps of $1\sigma$.
\textit{Third column:} Polarization map over the column density (colorscale) in logarithmic scale with contours starting at $\log_{10}{(N_{\rm HI+H_{2}} [{\rm cm}^{-2}]) = 20}$ increasing in steps of $\log_{10}{N_{\rm HI+H_{2}} [{\rm cm}^{-2}] = 0.1}$.
\textit{Fourth column:} Polarization map over the dust temperature (colorscale) in linear scale with contours starting at $20$~K increasing in steps of $1$~K.
\label{fig:figA1}}
\epsscale{2.}
\end{figure*}

\begin{figure*}[ht!]
\includegraphics[angle=0,scale=0.37]{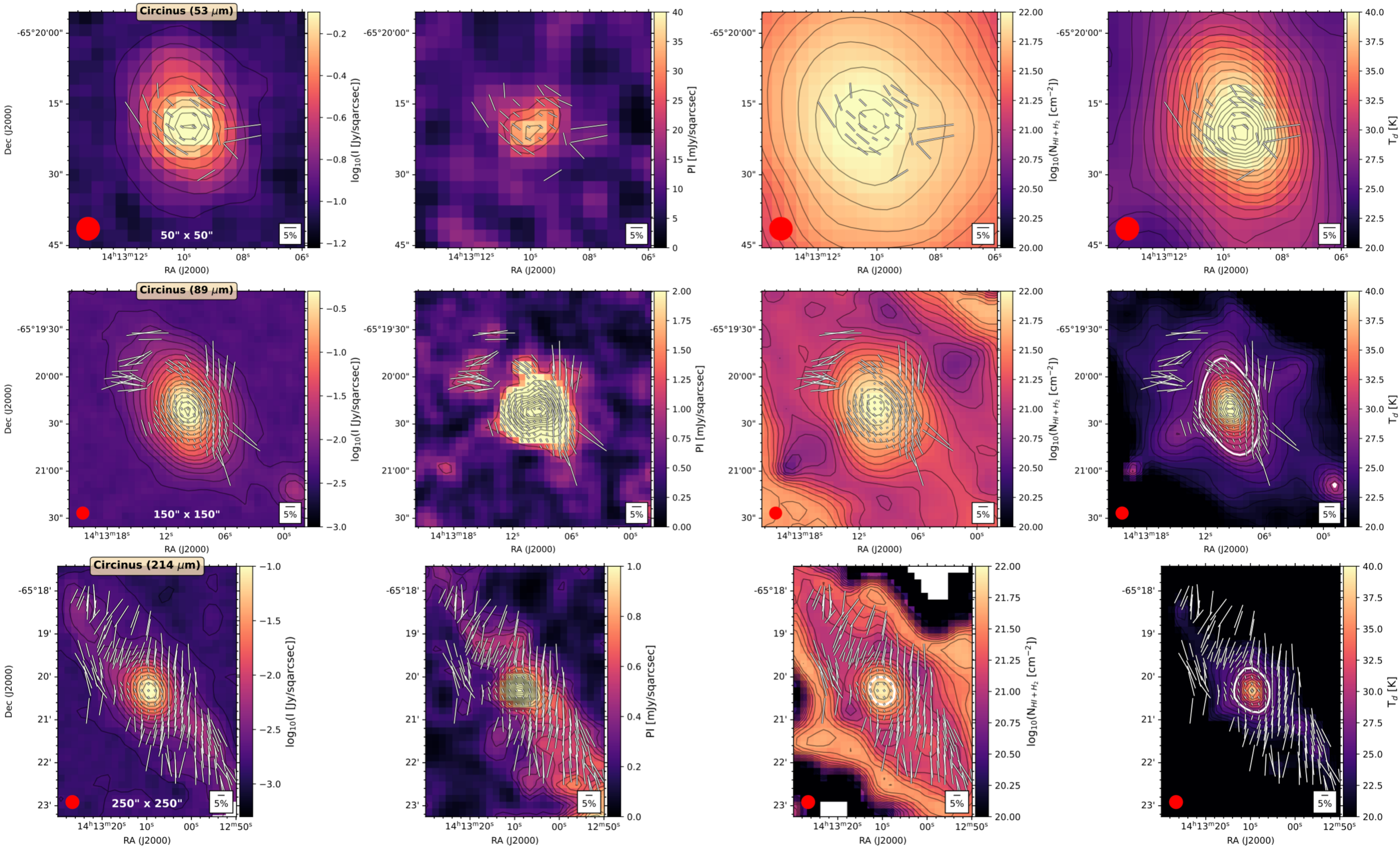}
\caption{Circinus. at $53$ (top), $89$ (middle) in Group 3, and $214$ (bottom) \um\ in Group 1. 
\textit{First column:} Total intensity (colorscale) in logarithmic scale with contours starting at $8\sigma$ increasing in steps of $2^{n}\sigma$, with $n= 3, 3.5, 4, \dots$ the polarization fraction measurements (white lines) with a legend of $5$\% (bottom right) are shown. The polarization fraction measurements with $PI/\sigma_{PI}\ge3$, $P\le20$\%, and $I/\sigma_{I} \ge 10$ were selected. The beam size (red circle) is shown.
\textit{Second column:} Polarization map over the polarized intensity (colorscale) in linear scale with contours starting at $3 \sigma$ increasing in steps of $1\sigma$ at $53$ and $214$ \um~and $5\sigma$ at $89$ \um.
\textit{Third column:} Polarization map over the column density (colorscale) in logarithmic scale with contours starting at $\log_{10}{(N_{\rm HI+H_{2}} [{\rm cm}^{-2}]) = 20}$ increasing in steps of $\log_{10}{N_{\rm HI+H_{2}} [{\rm cm}^{-2}] = 0.1}$.
\textit{Fourth column:} Polarization map over the dust temperature (colorscale) in linear scale with contours starting at $20$ K increasing in steps of $1$ K.
The white contours in the $N_{\rm HI+H_{2}}$ (at  $214$ \um) and $T_{\rm d}$ (at  $89$ and $214$ \um) maps show the cut-off column density and dust temperature values shown in Table \ref{tab:PI_fits} and described in Section \ref{subsec:PIPlots} and \ref{subsec:PTdPlots}, respectively.
\label{fig:figA2}}
\epsscale{2.}
\end{figure*}

\begin{figure*}[ht!]
\includegraphics[angle=0,scale=0.31]{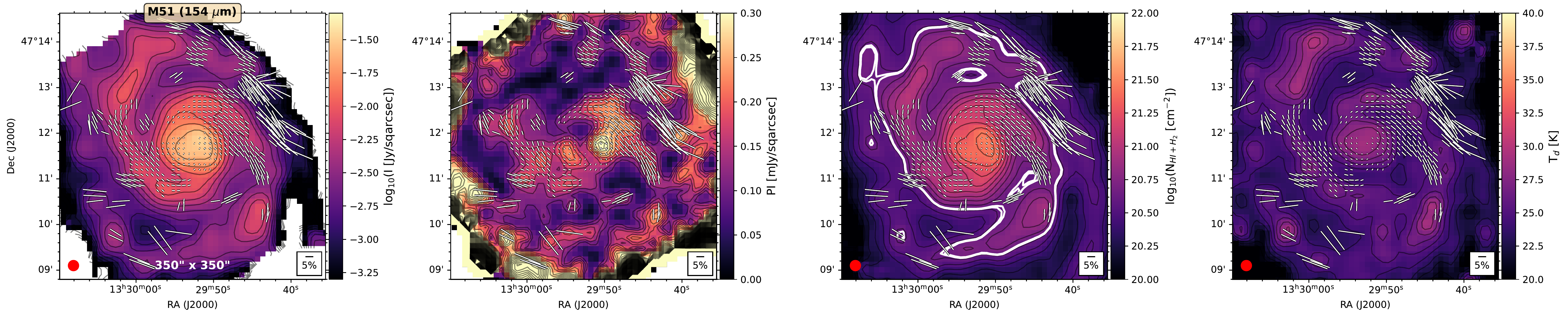}
\caption{M51 at $154$ \um\ (Group 2). 
\textit{First column:} Total intensity (colorscale) in logarithmic scale with contours starting at $8\sigma$ increasing in steps of $2^{n}\sigma$, with $n= 3, 3.5, 4, \dots$ the polarization fraction measurements (white lines) with a legend of $5$\% (bottom right) are shown. The polarization fraction measurements with $PI/\sigma_{PI}\ge3$, $P\le20$\%, and $I/\sigma_{I} \ge 10$ were selected. The beam size (red circle) is shown.
\textit{Second column:} Polarization map over the polarized intensity (colorscale) in linear scale with contours starting at $3 \sigma$ increasing in steps of $1\sigma$.
\textit{Third column:} Polarization map over the column density (colorscale) in logarithmic scale with contours starting at $\log_{10}{(N_{\rm HI+H_{2}} [{\rm cm}^{-2}]) = 20}$ increasing in steps of $\log_{10}{N_{\rm HI+H_{2}} [{\rm cm}^{-2}] = 0.1}$.
\textit{Fourth column:} Polarization map over the dust temperature (colorscale) in linear scale with contours starting at $20$ K increasing in steps of $1$ K.
The white contours in the $N_{\rm HI+H_{2}}$ maps show the cut-off column density and dust temperature values shown in Table \ref{tab:PI_fits} and described in Section \ref{subsec:PIPlots} and \ref{subsec:PTdPlots}, respectively.
\label{fig:figA3}}
\epsscale{2.}
\end{figure*}

\begin{figure*}[ht!]
\includegraphics[angle=0,scale=0.48]{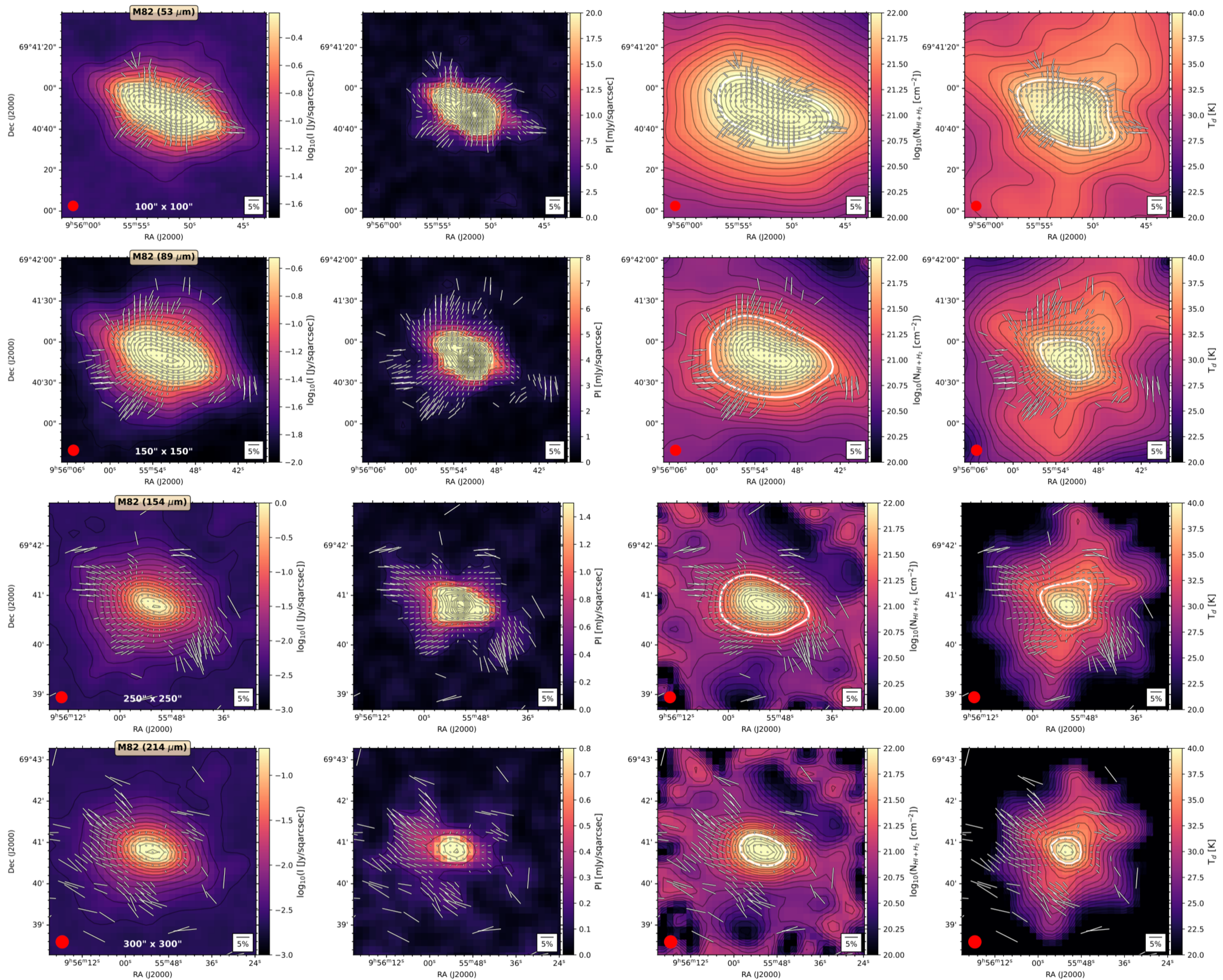}
\caption{M82 at $53$ (first row), $89$ (second row), $154$ (third row), and $214$ (Fourth row) \um\ (Group 1). 
\textit{First column:} Total intensity (colorscale) in logarithmic scale with contours starting at $8\sigma$ increasing in steps of $2^{n}\sigma$, with $n= 3, 3.5, 4, \dots$ the polarization fraction measurements (white lines) with a legend of $5$\% (bottom right) are shown. The polarization fraction measurements with $PI/\sigma_{PI}\ge3$, $P\le20$\%, and $I/\sigma_{I} \ge 80$ were selected. The beam size (red circle) is shown.
\textit{Second column:} Polarization map over the polarized intensity (colorscale) in linear scale with contours starting at $3 \sigma$ increasing in steps of $5\sigma$.
\textit{Third column:} Polarization map over the column density (colorscale) in logarithmic scale with contours starting at $\log_{10}{(N_{\rm HI+H_{2}} [{\rm cm}^{-2}]) = 20}$ increasing in steps of $\log_{10}{N_{\rm HI+H_{2}} [{\rm cm}^{-2}] = 0.1}$.
\textit{Fourth column:} Polarization map over the dust temperature (colorscale) in linear scale with contours starting at $20$ K increasing in steps of $1$ K.
The white contours in the $N_{\rm HI+H_{2}}$ (at  all wavelengths) and $T_{\rm d}$ (at  all wavelengths) maps show the cut-off column density and dust temperature values shown in Table \ref{tab:PI_fits} and described in Section \ref{subsec:PIPlots} and \ref{subsec:PTdPlots}, respectively.
\label{fig:figA4}}
\epsscale{2.}
\end{figure*}

\begin{figure*}[ht!]
\includegraphics[angle=0,scale=0.31]{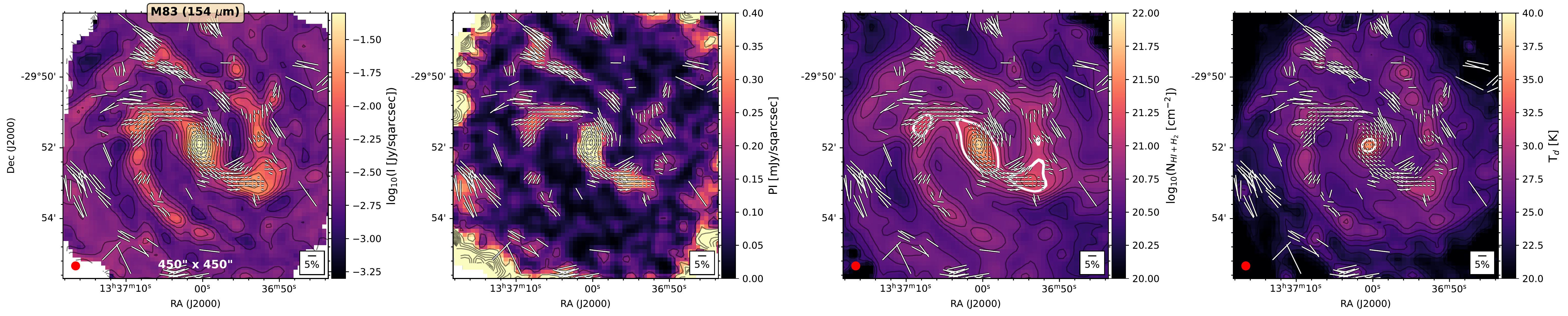}
\caption{M83 at $154$ \um\ (Group 1).
\textit{First column:} Total intensity (colorscale) in logarithmic scale with contours starting at $8\sigma$ increasing in steps of $2^{n}\sigma$, with $n= 3, 3.5, 4, \dots$ the polarization fraction measurements (white lines) with a legend of $5$\% (bottom right) are shown. The polarization fraction measurements with $PI/\sigma_{PI}\ge3$, $P\le20$\%, and $I/\sigma_{I} \ge 10$ were selected. The beam size (red circle) is shown.
\textit{Second column:} Polarization map over the polarized intensity (colorscale) in linear scale with contours starting at $3 \sigma$ increasing in steps of $5\sigma$.
\textit{Third column:} Polarization map over the column density (colorscale) in logarithmic scale with contours starting at $\log_{10}{(N_{\rm HI+H_{2}} [{\rm cm}^{-2}]) = 20}$ increasing in steps of $\log_{10}{N_{\rm HI+H_{2}} [{\rm cm}^{-2}] = 0.1}$.
\textit{Fourth column:} Polarization map over the dust temperature (colorscale) in linear scale with contours starting at $20$ K increasing in steps of $1$ K.
The white contours in the $N_{\rm HI+H_{2}}$ and $T_{\rm d}$ maps show the cut-off column density and dust temperature values shown in Table \ref{tab:PI_fits} and described in Section \ref{subsec:PIPlots} and \ref{subsec:PTdPlots}, respectively.
\label{fig:figA5}}
\epsscale{2.}
\end{figure*}

\begin{figure*}[ht!]
\includegraphics[angle=0,scale=0.34]{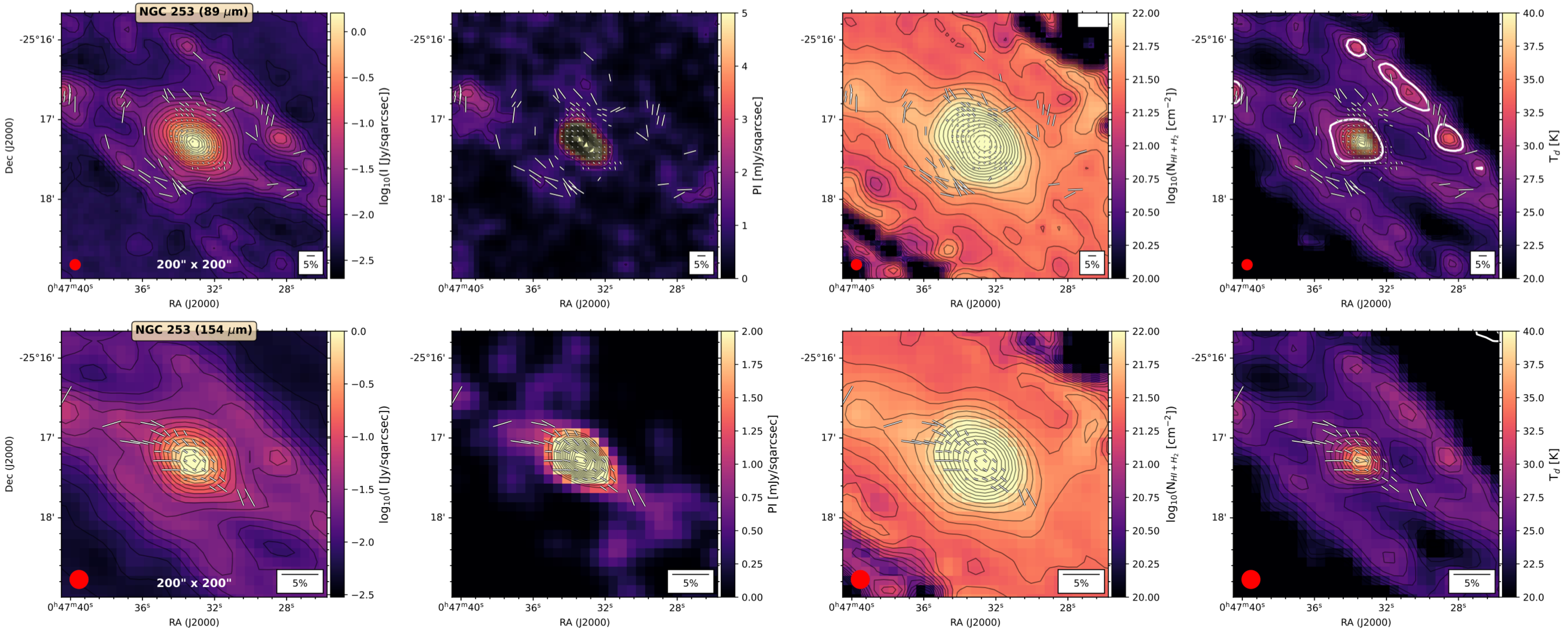}
\caption{NGC~253 at $89$ (top) and $154$ (bottom) \um\ (Group 3). 
\textit{First column:} Total intensity (colorscale) in logarithmic scale with contours starting at $8\sigma$ increasing in steps of $2^{n}\sigma$, with $n= 3, 3.5, 4, \dots$ the polarization fraction measurements (white lines) with a legend of $5$\% (bottom right) are shown. The polarization fraction measurements with $PI/\sigma_{PI}\ge3$, $P\le20$\%, and $I/\sigma_{I} \ge 10$ were selected. The beam size (red circle) is shown.
\textit{Second column:} Polarization map over the polarized intensity (colorscale) in linear scale with contours starting at $3 \sigma$ increasing in steps of $5\sigma$ at $89$ \um, and $2\sigma$ at 154 \um.
\textit{Third column:} Polarization map over the column density (colorscale) in logarithmic scale with contours starting at $\log_{10}{(N_{\rm HI+H_{2}} [{\rm cm}^{-2}]) = 20}$ increasing in steps of $\log_{10}{N_{\rm HI+H_{2}} [{\rm cm}^{-2}] = 0.1}$.
\textit{Fourth column:} Polarization map over the dust temperature (colorscale) in linear scale with contours starting at $20$ K increasing in steps of $1$ K.
The white contours in the $T_{\rm d}$ (at  $89$ \um) maps show the cut-off column density and dust temperature values shown in Table \ref{tab:PI_fits} and described in Section \ref{subsec:PIPlots} and \ref{subsec:PTdPlots}, respectively.
\label{fig:figA6}}
\epsscale{2.}
\end{figure*}

\begin{figure*}[ht!]
\includegraphics[angle=0,scale=0.34]{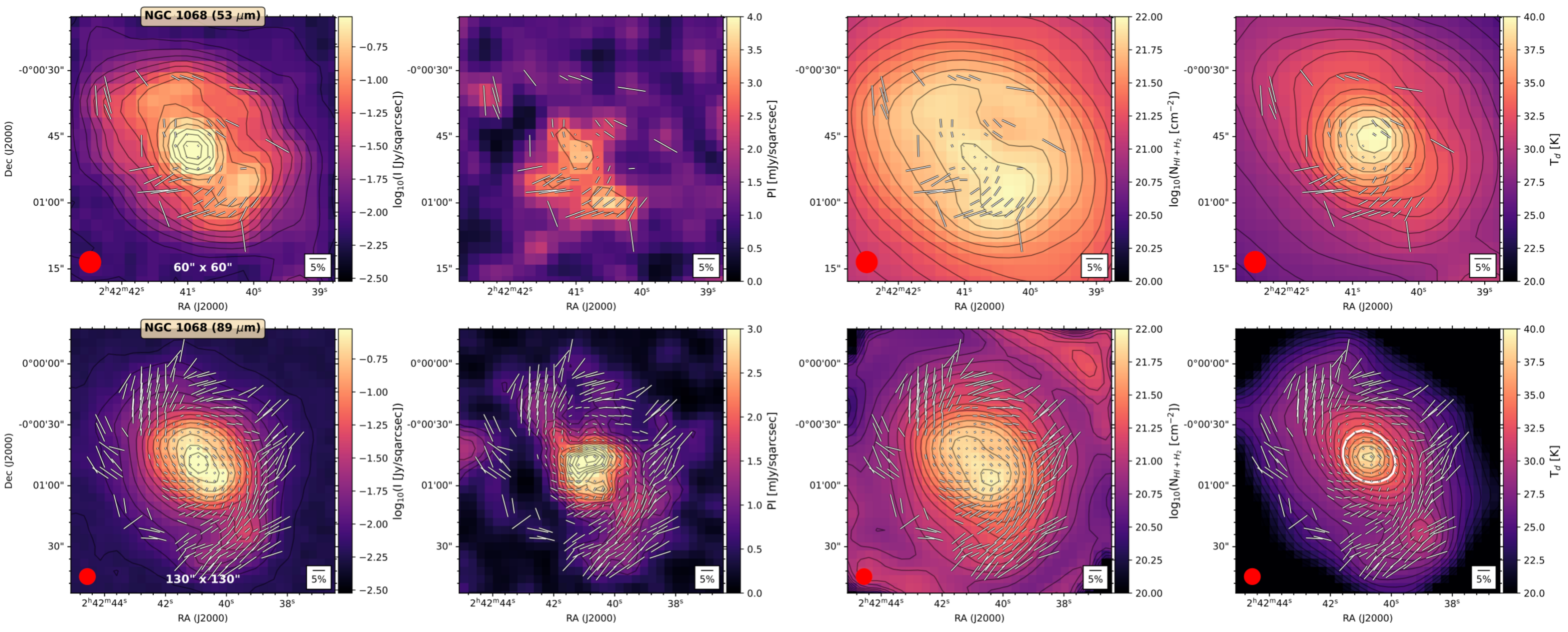}
\caption{NGC~1068 at $53$ (top) and $89$ (bottom) \um\ (Group 3). 
\textit{First column:} Total intensity (colorscale) in logarithmic scale with contours starting at $8\sigma$ increasing in steps of $2^{n}\sigma$, with $n= 3, 3.5, 4, \dots$ the polarization fraction measurements (white lines) with a legend of $5$\% (bottom right) are shown. The polarization fraction measurements with $PI/\sigma_{PI}\ge3$, $P\le20$\%, and $I/\sigma_{I} \ge 10$ were selected. The beam size (red circle) is shown.
\textit{Second column:} Polarization map over the polarized intensity (colorscale) in linear scale with contours starting at $3 \sigma$ increasing in steps of $1\sigma$.
\textit{Third column:} Polarization map over the column density (colorscale) in logarithmic scale with contours starting at $\log_{10}{(N_{\rm HI+H_{2}} [{\rm cm}^{-2}]) = 20}$ increasing in steps of $\log_{10}{N_{\rm HI+H_{2}} [{\rm cm}^{-2}] = 0.1}$.
\textit{Fourth column:} Polarization map over the dust temperature (colorscale) in linear scale with contours starting at $20$ K increasing in steps of $1$ K.
The white contours in the $T_{\rm d}$ (at  $89$ \um) maps show the cut-off column density and dust temperature values shown in Table \ref{tab:PI_fits} and described in Section \ref{subsec:PIPlots} and \ref{subsec:PTdPlots}, respectively.
\label{fig:figA7}}
\epsscale{2.}
\end{figure*}

\begin{figure*}[ht!]
\includegraphics[angle=0,scale=0.34]{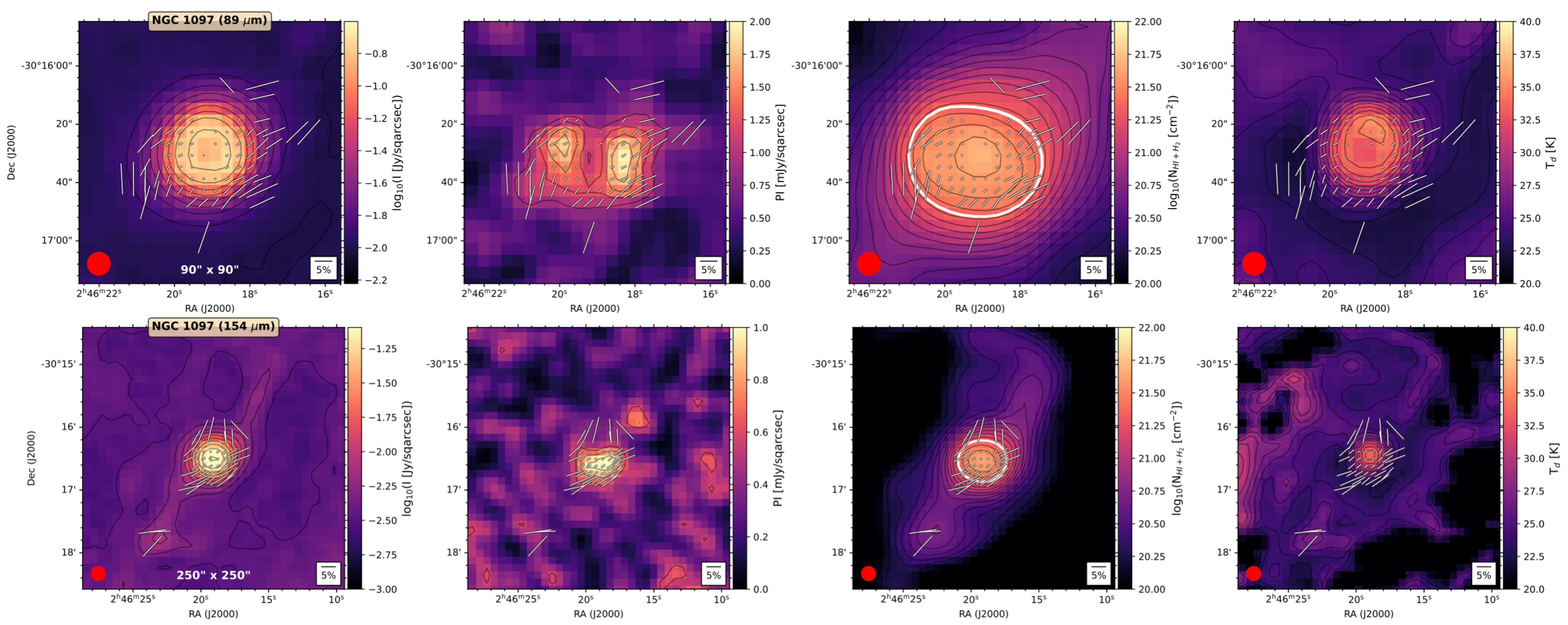}
\caption{NGC~1097 at $89$ (top) and $154$ (bottom) \um\ (Group 2).
\textit{First column:} Total intensity (colorscale) in logarithmic scale with contours starting at $8\sigma$ increasing in steps of $2^{n}\sigma$, with $n= 3, 3.5, 4, \dots$ the polarization fraction measurements (white lines) with a legend of $5$\% (bottom right) are shown. The polarization fraction measurements with $PI/\sigma_{PI}\ge3$, $P\le20$\%, and $I/\sigma_{I} \ge 10$ were selected. The beam size (red circle) is shown.
\textit{Second column:} Polarization map over the polarized intensity (colorscale) in linear scale with contours starting at $3 \sigma$ increasing in steps of $1\sigma$.
\textit{Third column:} Polarization map over the column density (colorscale) in logarithmic scale with contours starting at $\log_{10}{(N_{\rm HI+H_{2}} [{\rm cm}^{-2}]) = 20}$ increasing in steps of $\log_{10}{N_{\rm HI+H_{2}} [{\rm cm}^{-2}] = 0.1}$.
\textit{Fourth column:} Polarization map over the dust temperature (colorscale) in linear scale with contours starting at $20$ K increasing in steps of $1$ K.
The white contours in the $N_{\rm HI+H_{2}}$ (at  $89$ and $154$ \um) maps show the cut-off column density and dust temperature values shown in Table \ref{tab:PI_fits} and described in Section \ref{subsec:PIPlots} and \ref{subsec:PTdPlots}, respectively.
\label{fig:figA8}}
\epsscale{2.}
\end{figure*}

\begin{figure*}[ht!]
\includegraphics[angle=0,scale=0.48]{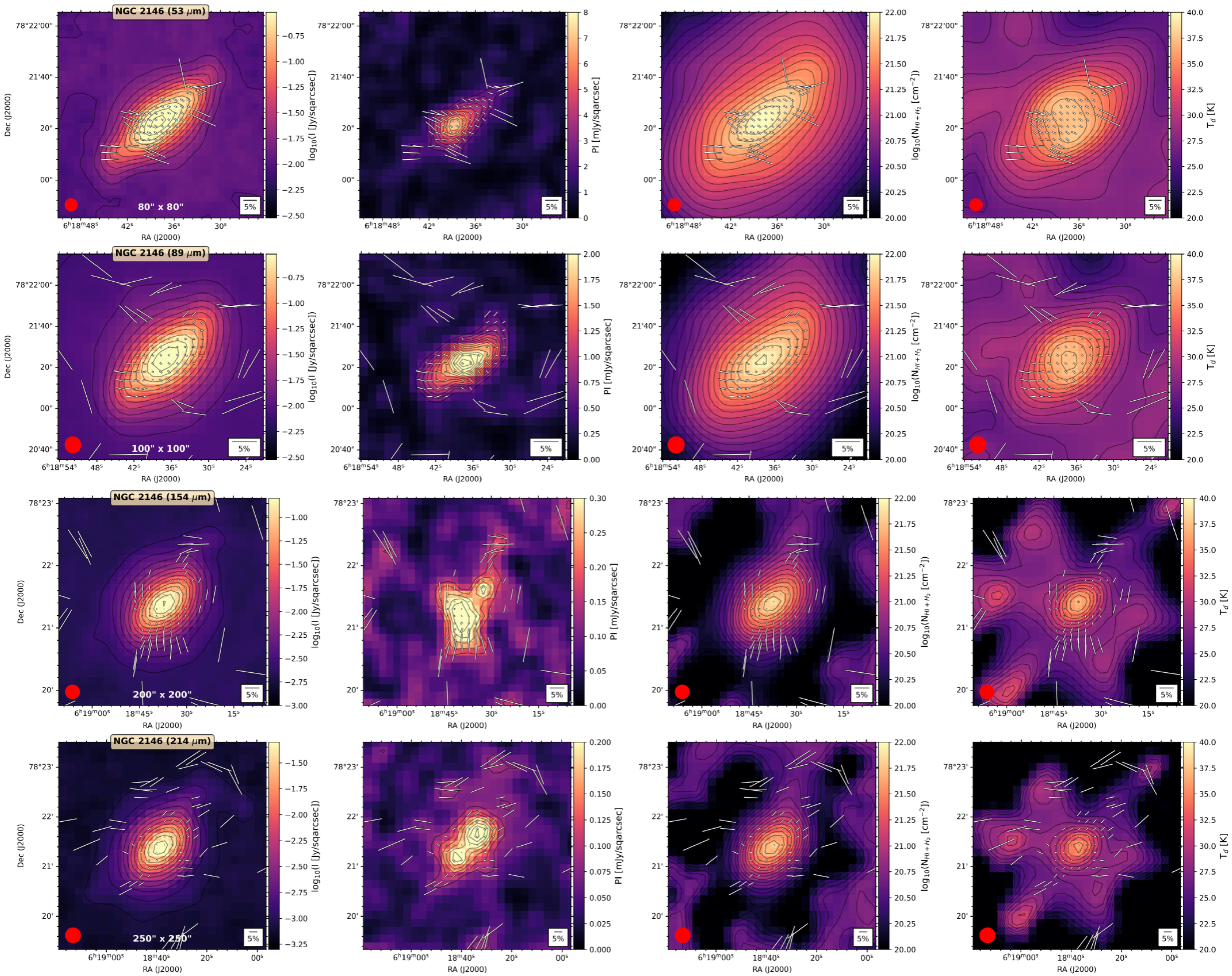}
\caption{NGC~2146 at $53$ (first row), $89$ (second row), $154$ (third row), and $214$ (Fourth row) \um\ (Group 3). 
\textit{First column:} Total intensity (colorscale) in logarithmic scale with contours starting at $8\sigma$ increasing in steps of $2^{n}\sigma$, with $n= 3, 3.5, 4, \dots$ the polarization fraction measurements (white lines) with a legend of $5$\% (bottom right) are shown. The polarization fraction measurements with $PI/\sigma_{PI}\ge3$, $P\le20$\%, and $I/\sigma_{I} \ge 20$ were selected. The beam size (red circle) is shown.
\textit{Second column:} Polarization map over the polarized intensity (colorscale) in linear scale with contours starting at $3 \sigma$ increasing in steps of $1\sigma$.
\textit{Third column:} Polarization map over the column density (colorscale) in logarithmic scale with contours starting at $\log_{10}{(N_{\rm HI+H_{2}} [{\rm cm}^{-2}]) = 20}$ increasing in steps of $\log_{10}{N_{\rm HI+H_{2}} [{\rm cm}^{-2}] = 0.1}$.
\textit{Fourth column:} Polarization map over the dust temperature (colorscale) in linear scale with contours starting at $20$ K increasing in steps of $1$ K.
\label{fig:figA9}}
\epsscale{2.}
\end{figure*}

\begin{figure*}[ht!]
\includegraphics[angle=0,scale=0.31]{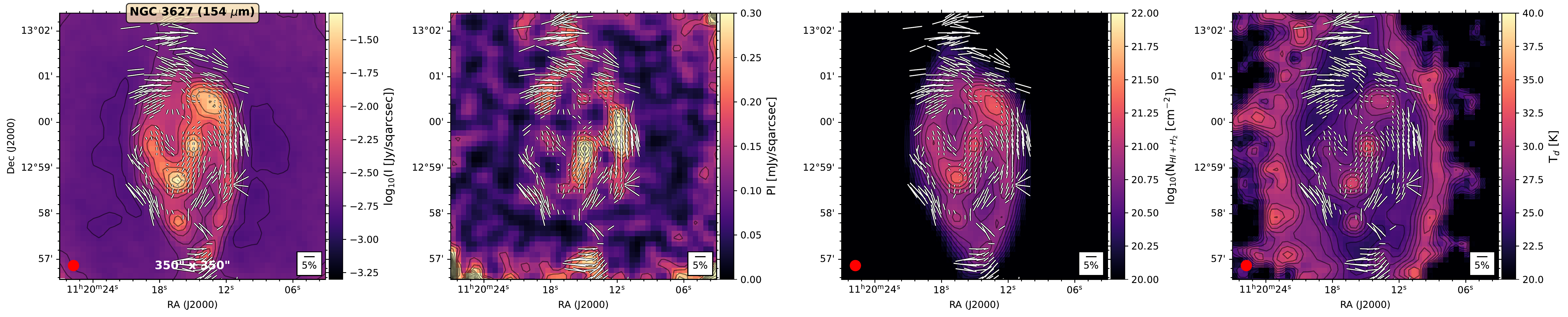}
\caption{NGC~3627 at $154$ \um\ (Group 3). 
\textit{First column:} Total intensity (colorscale) in logarithmic scale with contours starting at $8\sigma$ increasing in steps of $2^{n}\sigma$, with $n= 3, 3.5, 4, \dots$ the polarization fraction measurements (white lines) with a legend of $5$\% (bottom right) are shown. The polarization fraction measurements with $PI/\sigma_{PI}\ge3$, $P\le20$\%, and $I/\sigma_{I} \ge 40$ were selected. The beam size (red circle) is shown.
\textit{Second column:} Polarization map over the polarized intensity (colorscale) in linear scale with contours starting at $3 \sigma$ increasing in steps of $1\sigma$.
\textit{Third column:} Polarization map over the column density (colorscale) in logarithmic scale with contours starting at $\log_{10}{(N_{\rm HI+H_{2}} [{\rm cm}^{-2}]) = 20}$ increasing in steps of $\log_{10}{N_{\rm HI+H_{2}} [{\rm cm}^{-2}] = 0.1}$.
\textit{Fourth column:} Polarization map over the dust temperature (colorscale) in linear scale with contours starting at $20$ K increasing in steps of $1$ K.
\label{fig:figA10}}
\epsscale{2.}
\end{figure*}

\begin{figure*}[ht!]
\includegraphics[angle=0,scale=0.31]{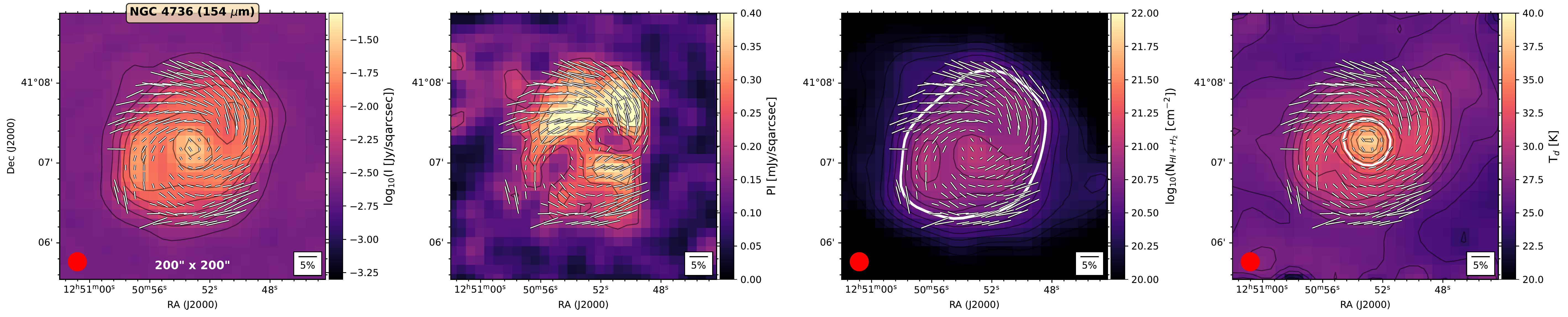}
\caption{NGC~4736 at $154$ \um\ (Group 2). 
\textit{First column:} Total intensity (colorscale) in logarithmic scale with contours starting at $8\sigma$ increasing in steps of $2^{n}\sigma$, with $n= 3, 3.5, 4, \dots$ the polarization fraction measurements (white lines) with a legend of $5$\% (bottom right) are shown. The polarization fraction measurements with $PI/\sigma_{PI}\ge3$, $P\le20$\%, and $I/\sigma_{I} \ge 20$ were selected. The beam size (red circle) is shown.
\textit{Second column:} Polarization map over the polarized intensity (colorscale) in linear scale with contours starting at $3 \sigma$ increasing in steps of $1\sigma$.
\textit{Third column:} Polarization map over the column density (colorscale) in logarithmic scale with contours starting at $\log_{10}{(N_{\rm HI+H_{2}} [{\rm cm}^{-2}]) = 20}$ increasing in steps of $\log_{10}{N_{\rm HI+H_{2}} [{\rm cm}^{-2}] = 0.1}$.
\textit{Fourth column:} Polarization map over the dust temperature (colorscale) in linear scale with contours starting at $20$ K increasing in steps of $1$ K.
The white contours in the $N_{\rm HI+H_{2}}$ maps show the cut-off column density and dust temperature values shown in Table \ref{tab:PI_fits} and described in Section \ref{subsec:PIPlots} and \ref{subsec:PTdPlots}, respectively.
\label{fig:figA11}}
\epsscale{2.}
\end{figure*}

\begin{figure*}[ht!]
\includegraphics[angle=0,scale=0.31]{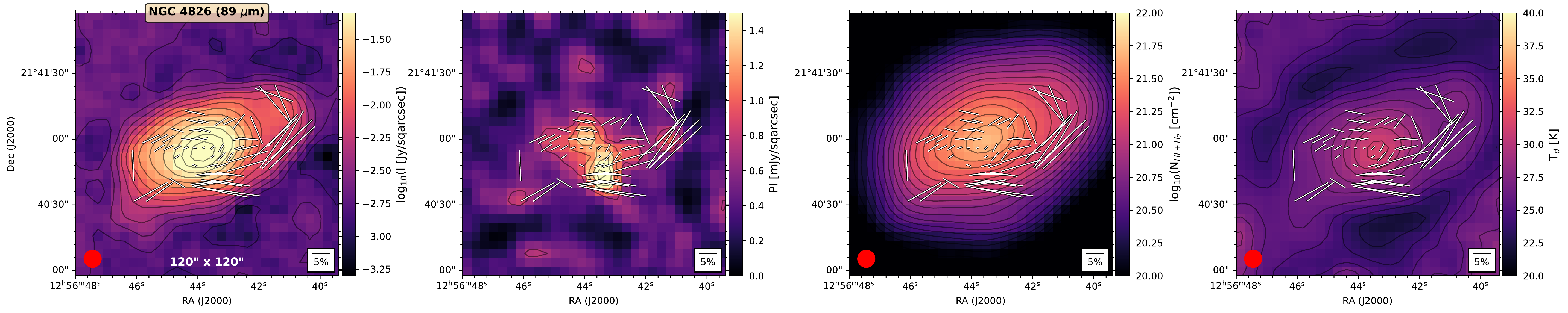}
\caption{NGC~4826 at $154$ \um\ (Group 3). 
\textit{First column:} Total intensity (colorscale) in logarithmic scale with contours starting at $8\sigma$ increasing in steps of $2^{n}\sigma$, with $n= 3, 3.5, 4, \dots$ the polarization fraction measurements (white lines) with a legend of $5$\% (bottom right) are shown. The polarization fraction measurements with $PI/\sigma_{PI}\ge3$, $P\le20$\%, and $I/\sigma_{I} \ge 20$ were selected. The beam size (red circle) is shown.
\textit{Second column:} Polarization map over the polarized intensity (colorscale) in linear scale with contours starting at $3 \sigma$ increasing in steps of $1\sigma$.
\textit{Third column:} Polarization map over the column density (colorscale) in logarithmic scale with contours starting at $\log_{10}{(N_{\rm HI+H_{2}} [{\rm cm}^{-2}]) = 20}$ increasing in steps of $\log_{10}{N_{\rm HI+H_{2}} [{\rm cm}^{-2}] = 0.1}$.
\textit{Fourth column:} Polarization map over the dust temperature (colorscale) in linear scale with contours starting at $20$ K increasing in steps of $1$ K.
\label{fig:figA12}}
\epsscale{2.}
\end{figure*}

\begin{figure*}[ht!]
\includegraphics[angle=0,scale=0.31]{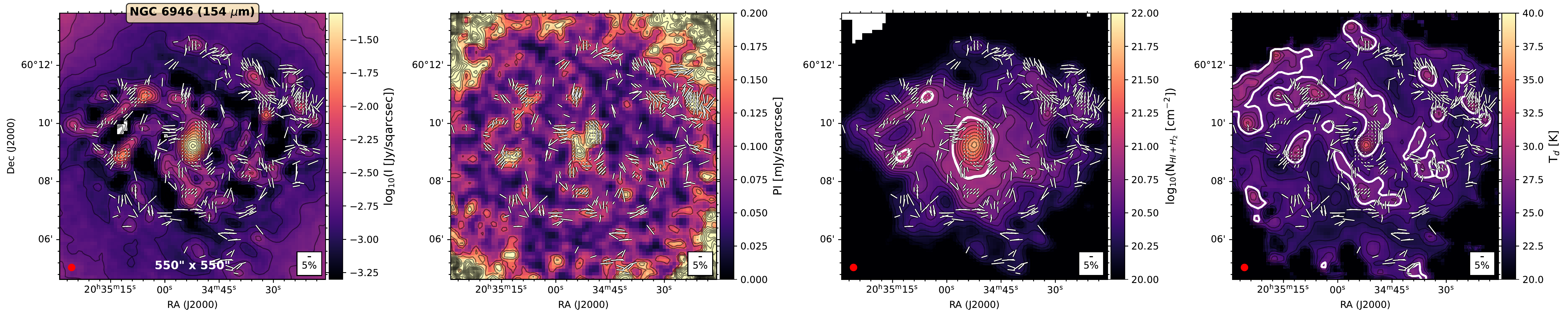}
\caption{NGC~6946 at $154$ \um\ (Group 2). 
\textit{First column:} Total intensity (colorscale) in logarithmic scale with contours starting at $8\sigma$ increasing in steps of $2^{n}\sigma$, with $n= 3, 3.5, 4, \dots$ the polarization fraction measurements (white lines) with a legend of $5$\% (bottom right) are shown. The polarization fraction measurements with $PI/\sigma_{PI}\ge3$, $P\le20$\%, and $I/\sigma_{I} \ge 20$ were selected. The beam size (red circle) is shown.
\textit{Second column:} Polarization map over the polarized intensity (colorscale) in linear scale with contours starting at $3 \sigma$ increasing in steps of $1\sigma$.
\textit{Third column:} Polarization map over the column density (colorscale) in logarithmic scale with contours starting at $\log_{10}{(N_{\rm HI+H_{2}} [{\rm cm}^{-2}]) = 20}$ increasing in steps of $\log_{10}{N_{\rm HI+H_{2}} [{\rm cm}^{-2}] = 0.1}$.
\textit{Fourth column:} Polarization map over the dust temperature (colorscale) in linear scale with contours starting at $20$ K increasing in steps of $1$ K.
The white contours in the $N_{\rm HI+H_{2}}$ and $T_{\rm d}$ maps show the cut-off column density and dust temperature values shown in Table \ref{tab:PI_fits} and described in Section \ref{subsec:PIPlots} and \ref{subsec:PTdPlots}, respectively.
\label{fig:figA13}}
\epsscale{2.}
\end{figure*}

\begin{figure*}[ht!]
\includegraphics[angle=0,scale=0.31]{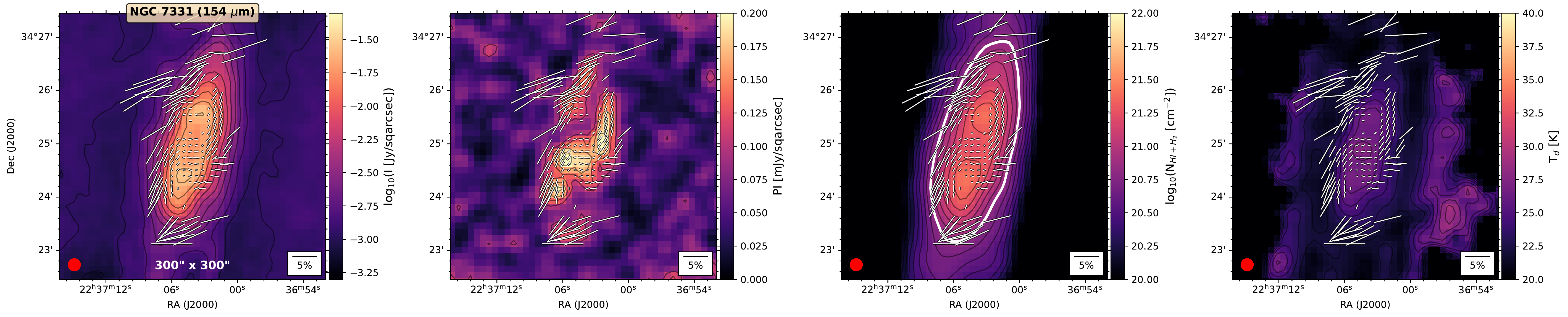}
\caption{NGC~7331 at $154$ \um\ (Group 2). 
\textit{First column:} Total intensity (colorscale) in logarithmic scale with contours starting at $8\sigma$ increasing in steps of $2^{n}\sigma$, with $n= 3, 3.5, 4, \dots$ the polarization fraction measurements (white lines) with a legend of $5$\% (bottom right) are shown. The polarization fraction measurements with $PI/\sigma_{PI}\ge3$, $P\le20$\%, and $I/\sigma_{I} \ge 45$ were selected. The beam size (red circle) is shown.
\textit{Second column:} Polarization map over the polarized intensity (colorscale) in linear scale with contours starting at $3 \sigma$ increasing in steps of $1\sigma$.
\textit{Third column:} Polarization map over the column density (colorscale) in logarithmic scale with contours starting at $\log_{10}{(N_{\rm HI+H_{2}} [{\rm cm}^{-2}]) = 20}$ increasing in steps of $\log_{10}{N_{\rm HI+H_{2}} [{\rm cm}^{-2}] = 0.1}$.
\textit{Fourth column:} Polarization map over the dust temperature (colorscale) in linear scale with contours starting at $20$ K increasing in steps of $1$ K.
The white contours in the $N_{\rm HI+H_{2}}$ maps show the cut-off column density and dust temperature values shown in Table \ref{tab:PI_fits} and described in Section \ref{subsec:PIPlots} and \ref{subsec:PTdPlots}, respectively.
\label{fig:figA14}}
\epsscale{2.}
\end{figure*}

This section shows the individual polarization maps of the HAWC+ observations for each galaxy and wavelength. In Figures~\ref{fig:figA1}-\ref{fig:figA14}, the first and second columns show the total and polarized surface brightness with overlaid polarization measurements, respectively. The length of each polarization measurement is proportional to the polarization fraction, where a legend of $5$\% polarization fraction is shown in each panel. The third and fourth columns show the column density, $N_{\rm HI+H_{2}}$, and dust temperatures, T$_{\rm d}$ within the same FOV and pixel scale as the total and polarized surface brightness observations. 

To compute the temperature and column density maps, we registered and binned $70-160$~\um\ \textit{Herschel} observations taken with PACS to the pixel scales of the HAWC+ observations. This approach ensures that images at all wavelengths have the same pixel scale and array dimensions. Then, for every pixel we fit an emissivity modified blackbody function 

\begin{equation}\label{eq:BB}
B(\tau,T) = \frac{2hc}{\lambda^{3}}\frac{1-e^{-\tau_{250}\left(\frac{\lambda}{\lambda_{250}}\right)^{\beta}}}{e^{\frac{hc}{\lambda k T}}-1} 
\end{equation}
\noindent
with a constant dust emissivity index $\beta=1.5$. A dust emissivity index of $1.5$ has been typically used \citep[][see also \citealt{dale2002,galametz2014}]{H1983}. A value of  $\beta = 1.60\pm0.38$ has been estimated to describe 65 local luminous IR galaxies \citep{Casey2012}. We derived the molecular hydrogen optical depth as $N_{\rm HI+H_{2}}= \tau_{250}/(k_{250}\mu m_{\rm H})$, with the dust opacity $k_{250} = 0.1$ cm$^{2}$ g$^{-1}$ at $250$ \um\, and the mean molecular weight per hydrogen atom $\mu=2.8$ \citep{H1983}. $\tau_{250}$ is a free parameter estimated at a reference wavelength of $250$ \um~using Eq. \ref{eq:BB}.

%%%%%%%%%%%%%%%%%%%
%%%% FIGURE A1 %%%%
%%%%%%%%%%%%%%%%%%%
%%%%%%%%%%%%%%

%%%%%%%%%%%%%%
%%%% FIGURE A2 %%%%
%%%%%%%%%%%%%%
%%%%%%%%%%%%%%

%%%%%%%%%%%%%%
%%%% FIGURE A3 %%%%
%%%%%%%%%%%%%%
%%%%%%%%%%%%%%

%%%%%%%%%%%%%%
%%%% FIGURE A4 %%%%
%%%%%%%%%%%%%%
%%%%%%%%%%%%%%

%%%%%%%%%%%%%%
%%%% FIGURE A5 %%%%
%%%%%%%%%%%%%%
%%%%%%%%%%%%%%

%%%%%%%%%%%%%%
%%%% FIGURE A6 %%%%
%%%%%%%%%%%%%%
%%%%%%%%%%%%%%

%%%%%%%%%%%%%%
%%%% FIGURE A7 %%%%
%%%%%%%%%%%%%%
%%%%%%%%%%%%%%

%%%%%%%%%%%%%%
%%%% FIGURE A8 %%%%
%%%%%%%%%%%%%%
%%%%%%%%%%%%%%

%%%%%%%%%%%%%%
%%%% FIGURE A9 %%%%
%%%%%%%%%%%%%%
%%%%%%%%%%%%%%

%%%%%%%%%%%%%%
%%%% FIGURE A10 %%%%
%%%%%%%%%%%%%%
%%%%%%%%%%%%%%

%%%%%%%%%%%%%%
%%%% FIGURE A11 %%%%
%%%%%%%%%%%%%%
%%%%%%%%%%%%%%

%%%%%%%%%%%%%%
%%%% FIGURE A12 %%%%
%%%%%%%%%%%%%%
%%%%%%%%%%%%%%

%%%%%%%%%%%%%%
%%%% FIGURE A13 %%%%
%%%%%%%%%%%%%%
%%%%%%%%%%%%%%

%%%%%%%%%%%%%%
%%%% FIGURE A14 %%%%
%%%%%%%%%%%%%%
%%%%%%%%%%%%%%

%%%%%%%%%%%%%%%%%%%%%
%%%% APPENDIX 3: TABLES %%%%%
%%%%%%%%%%%%%%%%%%%%%

\section{Tables}\label{app:A2}

This section contains the tabulated measurements from the main body of the manuscript.

%%%%%%%%%%%%%%%%%
%%%% TABLE 4 %%%%
%%%%%%%%%%%%%%%%%
\begin{deluxetable}{lcccccc}[h]
\tablecaption{Median and integrated polarization fraction of galaxies. \emph{Columns, from left to right:} 
a) Galaxy name, 
b) central wavelength of the band, 
c) median polarization fraction based on individual measurements from Eq. \ref{eq:Phist}, 
d) median polarization fraction based on the integrated Stokes $IQU$ of the full galaxy from Eq. \ref{eq:Pint}.
\label{tab:PPA} 
}
\tablewidth{0pt}
\tablehead{\colhead{Galaxy} & 	\colhead{Band}  & \colhead{$\langle P^{\rm hist} \rangle$}  & \colhead{$\langle P^{\rm int} \rangle$}  \\ 
  &	\colhead{(\um)}	&	\colhead{(\%)}	&		\colhead{(\%)} 	 \\
\colhead{(a)} & \colhead{(b)} & \colhead{(c)} & \colhead{(d)}} 
\startdata
Centaurus A	&	89	&	$	5.7	\pm	5.3	$	&	$	1.7	\pm	0.8	$	\\
Circinus	&	53	&	$	5.1	\pm	3.6	$	&	$	1.3	\pm	0.8	$	\\
	&	89	&	$	4.5	\pm	4.8	$	&	$	1.6	\pm	0.6	$			\\
	&	214	&	$	11.4\pm	5.8	$	&	$	8.4	\pm	0.8	$			\\
M51	        &	154	&	$	3.4	\pm	4.5	$	&	$	1.4	\pm	0.7	$	\\
M82	        &	53	&	$	2.5	\pm	1.1	$	&	$	1.2	\pm	0.2	$	\\
	        &	89	&	$	1.9	\pm	1.6	$	&	$	1.0	\pm	0.2	$	\\
	        &	154	&	$	1.6	\pm	1.6	$	&	$	1.0	\pm	0.3	$	\\
	        &	214	&	$	2.7	\pm	1.9	$	&	$	1.4	\pm	0.2	$	\\
M83	        &	154	&	$	3.9	\pm	4.0	$	&	$	0.7	\pm	0.3	$	\\
NGC 253	    &	89	&	$	1.5	\pm	3.1	$	&	$	0.8	\pm	0.6	$	\\
	    &	154	&	$	1.0	\pm	0.7	$	&	$	0.4	\pm	0.4	$		\\
NGC 1068	&	53	&	$	3.4	\pm	2.6	$	&	$	0.4	\pm	0.4	$	\\
	&	89	&	$	4.2	\pm	2.8	$	&	$	0.9	\pm	0.3	$			\\
NGC 1097	&	89	&	$	2.6	\pm	2.8	$	&	$	1.8	\pm	0.3	$	\\
	&	154	&	$	4.1	\pm	2.7	$	&	$	1.2	\pm	0.8	$			\\
NGC 2146	&	53	&	$	2.0	\pm	2.7	$	&	$	1.6	\pm	0.6	$	\\
	&	89	&	$	0.8	\pm	1.5	$	&	$	0.5	\pm	0.2	$			\\
	&	154	&	$	1.2	\pm	1.9	$	&	$	0.7	\pm	0.3	$			\\
	&	214	&	$	1.6	\pm	1.5	$	&	$	1.8	\pm	0.3	$			\\
NGC 3627	&	154	&	$	3.3	\pm	2.4	$	&	$	0.3	\pm	0.3	$	\\
NGC 4736	&	154	&	$	3.2	\pm	1.7	$	&	$	0.9	\pm	0.2	$	\\
NGC 4826	&	89	&	$	5.3	\pm	4.9	$	&	$	1.4	\pm	0.5	$	\\
NGC 6946	&	154	&	$	8.6	\pm	4.8	$	&	$	0.7	\pm	0.3	$	\\
NGC 7331	&	154	&	$	1.6	\pm	2.4	$	&	$	1.0	\pm	0.2	$	\\
\enddata
%\tablenotetext{{\dagger}}{}
%\tablenotetext{{\star}}{}
\end{deluxetable}

%%%%%%%%%%%%%
%%%% TABLE 6 %%%%
%%%%%%%%%%%%%
\begin{deluxetable*}{lcccccccccccc}[h]
\tablecaption{Power-law fitting parameters of the P-N$_{\rm HI+H2}$ and P-T$_{\rm d}$ relations. \emph{Columns, from left to right:} 
a) Galaxy name, 
b) central wavelength of the band, 
c) $P_{o}$ is the scale factor,
d) $\alpha_{1}$ is the power-law index before the cut-off column density, $N_{\rm HI+H2}^{\rm c}$, 
e) cut-off column density in units of cm$^{-2}$ in logarithmic scale, 
f) $\alpha_{2}$ is the power-law index after the cut-off column density, $N_{\rm HI+H2}^{\rm c}$, 
g) $P_{o}^{T_{\rm d}}$ is the scale factor,
h) $\alpha_{1}^{T_{\rm d}}$ is the power-law index before the cut-off dust temperature, $T_{\rm d}^{\rm c}$, 
i) cut-off dust temperature in units of K, 
j) $\alpha_{2}^{T_{\rm d}}$ is the power-law index after the cut-off dust temperature, $T_{\rm d}^{\rm c}$.
\label{tab:PI_fits} 
}
\tablewidth{0pt}
\tablehead{\colhead{Object}  & \colhead{Band} & \multicolumn{4}{c}{P-N$_{\rm HI+H2}$} & \multicolumn{4}{c}{P-T$_{\rm d}$} \\
& & \colhead{$P_{0}$} & \colhead{$\alpha_{1}$} & \colhead{$\log_{10}($N$_{\rm HI+H2}^{c}$)} & \colhead{$\alpha_{2}$} &        \colhead{$P_{0}^{T_{\rm d}}$} & \colhead{$\alpha_{1}^{T_{\rm d}}$} & \colhead{T$_{\rm d}^{\rm c}$} & \colhead{$\alpha_{2}^{T_{\rm d}}$} \\ 
  		& &  \colhead{(\%)} &	& \colhead{(cm$^{-2}$)}	& &\colhead{(\%)}	&	& \colhead{(K)}	&	 	 \\
\colhead{(a)} & \colhead{(b)} & \colhead{(c)} & \colhead{(d)} & \colhead{(e)} & \colhead{(f)} & \colhead{(g)} & \colhead{(h)} & \colhead{(i)} & \colhead{(j)} } 
\startdata
CentaurusA	&	89	&	$	1.5	^{+0.3	}_{	-0.3	}	$	&	$	-0.99	^{	+0.01	}_{	-0.01	}	$	&		-									&		-								&	$	0.6	^{	+0.1	}_{	-0.1	}	$	&	$	-11.32^{	+0.68	}_{	-0.61	}	$	&		-							&		-							\\
Circinus		&	53	&	$	3.0	^{+0.1	}_{	-0.1	}	$	&	$	-0.81	^{	+0.01	}_{	-0.01	}	$	&		-									&		-								&	$	1.2	^{	+0.5	}_{	-0.5	}	$	&	$	-6.83	^{	+3.37	}_{	-3.13	}	$	&		-							&		-							\\
			&	89	&	$	3.3	^{+0.2	}_{	-0.1	}	$	&	$	-0.93	^{	+0.03	}_{	-0.03	}	$	&		-									&		-								&	$	3.7	^{	+0.2	}_{	-0.2	}	$	&	$	-7.49	^{	+1.38	}_{	-1.65	}	$	&	$	28.5^{	+0.5	}_{	-0.4	}	$	&	$	-1.49	^{	+0.10	}_{	-0.10	}	$	\\
			&	214	&	$	2.4	^{+0.3	}_{	-0.2	}	$	&	$	-0.97	^{	+0.03	}_{	-0.03	}	$	&	$	21.69	^{	+0.05	}_{	-0.05	}	$	&	$	-0.34	^{	+0.16	}_{	-0.18	}	$	&	$	2.9	^{	+0.3	}_{	-0.3	}	$	&	$	-7.78	^{	+0.77	}_{	-0.65	}	$	&	$	27.4^{	+0.4	}_{	-0.4	}	$	&	$	-1.36	^{	+0.43	}_{	-0.42	}	$	\\
M51			&	154	&	$	7.7	^{+0.5	}_{	-0.5	}	$	&	$	-0.28^{	+0.21	}_{	-0.22	}	$	&	$	20.62	^{	+0.02	}_{	-0.02	}	$	&	$	-0.99	^{	+0.01	}_{	-0.02	}	$	&	$	0.4	^{	+0.1	}_{	-0.1	}	$	&	$	-11.78	^{	+0.22	}_{	-0.21	}	$	&		-							&		-							\\
M82			&	53	&	$	2.6	^{+0.2	}_{	-0.1	}	$	&	$	-0.86	^{	+0.16	}_{	-0.14	}	$	&	$	22.04	^{	+0.05	}_{	-0.06	}	$	&	$	-0.26	^{	+0.04	}_{	-0.03	}	$	&	$	2.0	^{	+0.1	}_{	-0.1	}	$	&	$	-7.24	^{	+1.28	}_{	-1.40	}	$	&	$	37.6^{	+0.4	}_{	-0.2	}	$	&	$	-0.01	^{	+0.01	}_{	-0.01	}	$	\\
			&	89	&	$	2.3	^{+0.2	}_{	-0.2	}	$	&	$	-0.65	^{	+0.09	}_{	-0.10	}	$	&	$	21.48	^{	+0.07	}_{	-0.10	}	$	&	$	-0.35	^{	+0.01	}_{	-0.01	}	$	&	$	1.0	^{	+0.1	}_{	-0.1	}	$	&	$	-4.81	^{	+0.37	}_{	-0.94	}	$	&	$	38.4^{	+0.2	}_{	-0.8	}	$	&	$	-0.00	^{	+0.01	}_{	-0.01	}	$	\\
			&	154	&	$	1.0	^{+0.1	}_{	-0.1	}	$	&	$	-0.99	^{	+0.01	}_{	-0.01	}	$	&	$	21.30	^{	+0.01	}_{	-0.01	}	$	&	$	-0.32	^{	+0.01	}_{	-0.01	}	$	&	$	0.8	^{	+0.1	}_{	-0.1	}	$	&	$	-5.74	^{	+0.20	}_{	-0.19	}	$	&	$	34.9^{	+0.1	}_{	-0.1	}	$	&	$	-3.89	^{	+0.10	}_{	-0.10	}	$	\\
			&	214	&	$	0.6	^{+0.1	}_{	-0.1	}	$	&	$	-0.78	^{	+0.05	}_{	-0.05	}	$	&	$	21.79	^{	+0.06	}_{	-0.07	}	$	&	$	-0.13	^{	+0.07	}_{	-0.07	}	$	&	$	0.5	^{	+0.1	}_{	-0.1	}	$	&	$	-6.17	^{	+0.37	}_{	-0.34	}	$	&	$	36.7^{	+0.7	}_{	-0.7	}	$	&	$	-0.44	^{	+0.44	}_{	-0.49	}	$	\\
M83			&	154	&	$	2.3	^{+0.3	}_{	-0.3	}	$	&	$	-0.96	^{	+0.05	}_{	-0.04	}	$	&	$	21.09	^{	+0.06	}_{	-0.07	}	$	&	$	-0.51	^{	+0.04	}_{	-0.04	}	$	&	$	1.2	^{	+0.1	}_{	-0.1	}	$	&	$	-4.35	^{	+0.26	}_{	-0.24	}	$	&	$	33.0^{	+0.5	}_{	-0.5	}	$	&	$	-2.78	^{	+0.31	}_{	-0.22	}	$	\\
NGC253		&	89	&	$	0.2	^{+0.1	}_{	-0.1	}	$	&	$	-1.00	^{	+0.01	}_{	-0.01	}	$	&		-									&		-								&	$	3.0	^{	+0.1	}_{	-0.1	}	$	&	$	-8.00	^{	+0.01	}_{	-0.01	}	$	&	$	28.0^{	+0.1	}_{	-0.1	}	$	&	$	-3.00	^{	+0.01	}_{	-0.01	}	$	\\
			&	154	&	$	0.4	^{+0.1	}_{	-0.1	}	$	&	$	-0.43	^{	+0.13	}_{	-0.12	}	$	&		-									&		-								&	$	0.5	^{	+0.1	}_{	-0.1	}	$	&	$	-0.95	^{	+0.41	}_{	-0.43	}	$	&		-							&		-							\\
NGC1068		&	53	&	$	4.0	^{+0.1	}_{	-0.1	}	$	&	$	-0.56	^{	+0.06	}_{	-0.11	}	$	&		-									&		-								&	$	0.3	^{	+0.1	}_{	-0.1	}	$	&	$	-11.08^{	+0.89	}_{	-0.80	}	$	&		-							&		-							\\
			&	89	&	$	0.9	^{+0.1	}_{	-0.1	}	$	&	$	-1.00	^{	+0.01	}_{	-0.01	}	$	&		-									&		-								&	$	0.9	^{	+0.1	}_{	-0.1	}	$	&	$	-9.16	^{	+0.48	}_{	-0.45	}	$	&	$	33.7^{	+0.3	}_{	-0.3	}	$	&	$	-1.16	^{	+0.69	}_{	-0.68	}	$	\\
NGC1097		&	89	&	$	5.3	^{+0.3	}_{	-0.7	}	$	&	$	-0.37	^{	+0.13	}_{	-0.12	}	$	&	$	21.33	^{	+0.04	}_{	-0.03	}	$	&	$	-0.96	^{	+0.07	}_{	-0.04	}	$	&	$	0.7	^{	+0.1	}_{	-0.1	}	$	&	$	-4.87	^{	+0.34	}_{	-0.37	}	$	&		-							&		-							\\
			&	154	&	$	5.9	^{+0.9	}_{	-0.1	}	$	&	$	-0.45	^{	+0.06	}_{	-0.05	}	$	&	$	21.28	^{	+0.03	}_{	-0.03	}	$	&	$	-0.88	^{	+0.13	}_{	-0.12	}	$	&	$	0.8	^{	+0.2	}_{	-0.2	}	$	&	$	-4.85	^{	+0.77	}_{	-0.78	}	$	&		-							&		-							\\
NGC2146		&	53	&	$	0.6	^{+0.1	}_{	-0.1	}	$	&	$	-0.93	^{	+0.09	}_{	-0.07	}	$	&		-									&		-								&	$	0.6	^{	+0.1	}_{	-0.2	}	$	&	$	-9.58	^{	+1.43	}_{	-1.48	}	$	&		-							&		-							\\
			&	89	&	$	0.3	^{+0.1	}_{	-0.1	}	$	&	$	-0.74	^{	+0.04	}_{	-0.04	}	$	&		-									&		-								&	$	0.3	^{	+0.1	}_{	-0.1	}	$	&	$	-5.11	^{	+0.61	}_{	-0.65	}	$	&		-							&		-							\\
			&	154	&	$	0.2	^{+0.1	}_{	-0.1	}	$	&	$	-0.99	^{	+0.02	}_{	-0.01	}	$	&		-									&		-								&	$	0.1	^{	+0.1	}_{	-0.1	}	$	&	$	-7.56	^{	+0.32	}_{	-0.30	}	$	&		-							&		-							\\
			&	214	&	$	0.3	^{+0.1	}_{	-0.1	}	$	&	$	-0.84	^{	+0.06	}_{	-0.06	}	$	&		-									&		-								&	$	0.2	^{	+0.1	}_{	-0.1	}	$	&	$	-6.13	^{	+0.43	}_{	-0.46	}	$	&		-							&		-							\\
NGC3627		&	154	&	$	0.8	^{+0.1	}_{	-0.1	}	$	&	$	-0.97	^{	+0.03	}_{	-0.03	}	$	&		-									&		-								&	$	1.2	^{	+0.2	}_{	-0.2	}	$	&	$	-7.24	^{	+0.66	}_{	-0.63	}	$	&	-	&	-	\\
NGC4736		&	154	&	$	4.4	^{+0.2	}_{	-0.2	}	$	&	$	-0.06	^{	+0.04	}_{	-0.04	}	$	&	$	20.61	^{	+0.01	}_{	-0.01	}	$	&	$	-0.98	^{	+0.03	}_{	-0.02	}	$	&	$	2.0	^{	+0.4	}_{	-0.4	}	$	&	$	-5.08	^{	+0.62	}_{	-0.61	}	$	&	-	&	-	\\
NGC4826		&	89	&	$	1.8	^{+0.2	}_{	-0.2	}	$	&	$	-0.96	^{	+0.05	}_{	-0.04	}	$	&		-									&		-								&	$	1.3	^{	+0.4	}_{	-0.4	}	$	&	$	-11.21^{	+1.04	}_{	-0.79	}	$	&		-							&		-							\\
NGC6946		&	154	&	$	3.1	^{+0.1	}_{	-0.1	}	$	&	$	-0.49	^{	+0.01	}_{	-0.01	}	$	&	$	20.91	^{	+0.01	}_{	-0.01	}	$	&	$	-1.00	^{	+0.01	}_{	-0.01	}	$	&	$	3.2	^{	+0.2	}_{	-0.2	}	$	&	$	-11.60^{	+0.42	}_{	-0.40	}	$	&	-	&	-	\\
NGC7331		&	154	&	$	6.0	^{+0.7	}_{	-0.3	}	$	&	$	-0.41	^{	+0.13	}_{	-0.09	}	$	&	$	20.81	^{	+0.04	}_{	-0.03	}	$	&	$	-0.99	^{	+0.02	}_{	-0.01	}	$	&	$	1.0	^{	+0.1	}_{	-0.1	}	$	&	$	-8.60	^{	+0.72	}_{	-0.73	}	$	&	-								&		-							\\
\enddata
\end{deluxetable*}

\section{Figures of the polarization fraction with column density and dust temperature}\label{App:A3}

This section shows the $P-N_{\rm HI+H_{2}}$ and $P-T_{\rm d}$ relations for all the galaxies associated with Groups 1, 2, and 3 described in Sections \ref{subsec:PIPlots} and \ref{subsec:PTdPlots}.

%%%%%%%%%%%%%%
%%%% FIGURE A3_1 %%%%
%%%%%%%%%%%%%%
\begin{figure*}[ht!]
\includegraphics[angle=0,width=\textwidth]{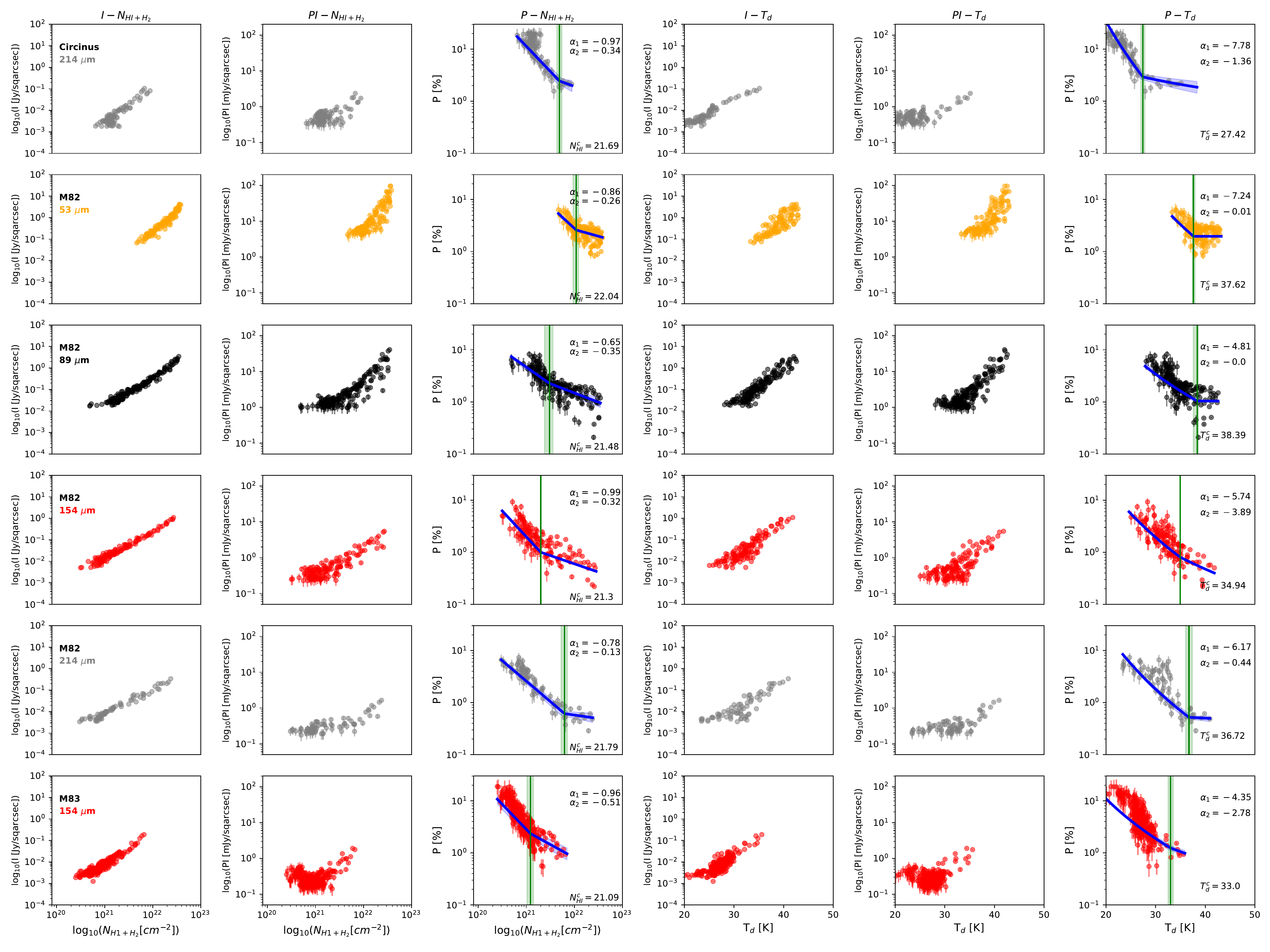}
\caption{Polarization fraction vs. $N_{\rm HI+H_{2}}$ and $T_{\rm d}$ for the galaxies associated with Group 1 shown in Figure \ref{fig:fig5} and \ref{fig:fig6}. The same polarization fraction measurements shown in Figure \ref{fig:fig1} are used.
\label{fig:A3_fig1}}
\epsscale{2.}
\end{figure*}
%%%%%%%%%%%%%%

%%%%%%%%%%%%%%
%%%% FIGURE A3_2 %%%%
%%%%%%%%%%%%%%
\begin{figure*}[ht!]
\includegraphics[angle=0,width=\textwidth]{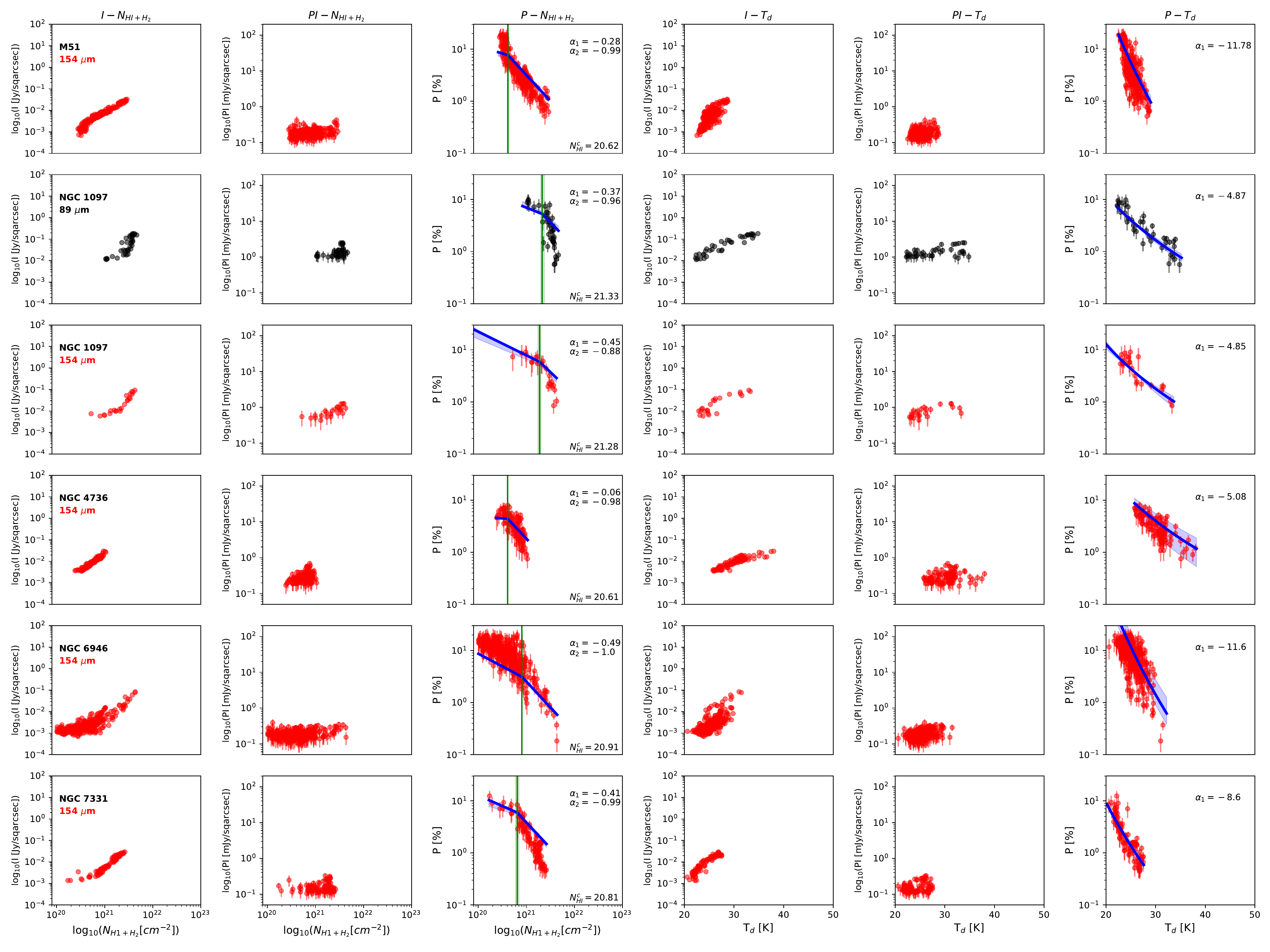}
\caption{Polarization fraction vs. $N_{\rm HI+H_{2}}$ and $T_{\rm d}$ for the galaxies associated with Group 2 shown in Figure \ref{fig:fig5} and \ref{fig:fig6}. The same polarization fraction measurements shown in Figure \ref{fig:fig1} are used.
\label{fig:A3_fig2}}
\epsscale{2.}
\end{figure*}
%%%%%%%%%%%%%%

%%%%%%%%%%%%%%
%%%% FIGURE A3_3 %%%%
%%%%%%%%%%%%%%
\begin{figure*}[ht!]
\includegraphics[angle=0,width=\textwidth]{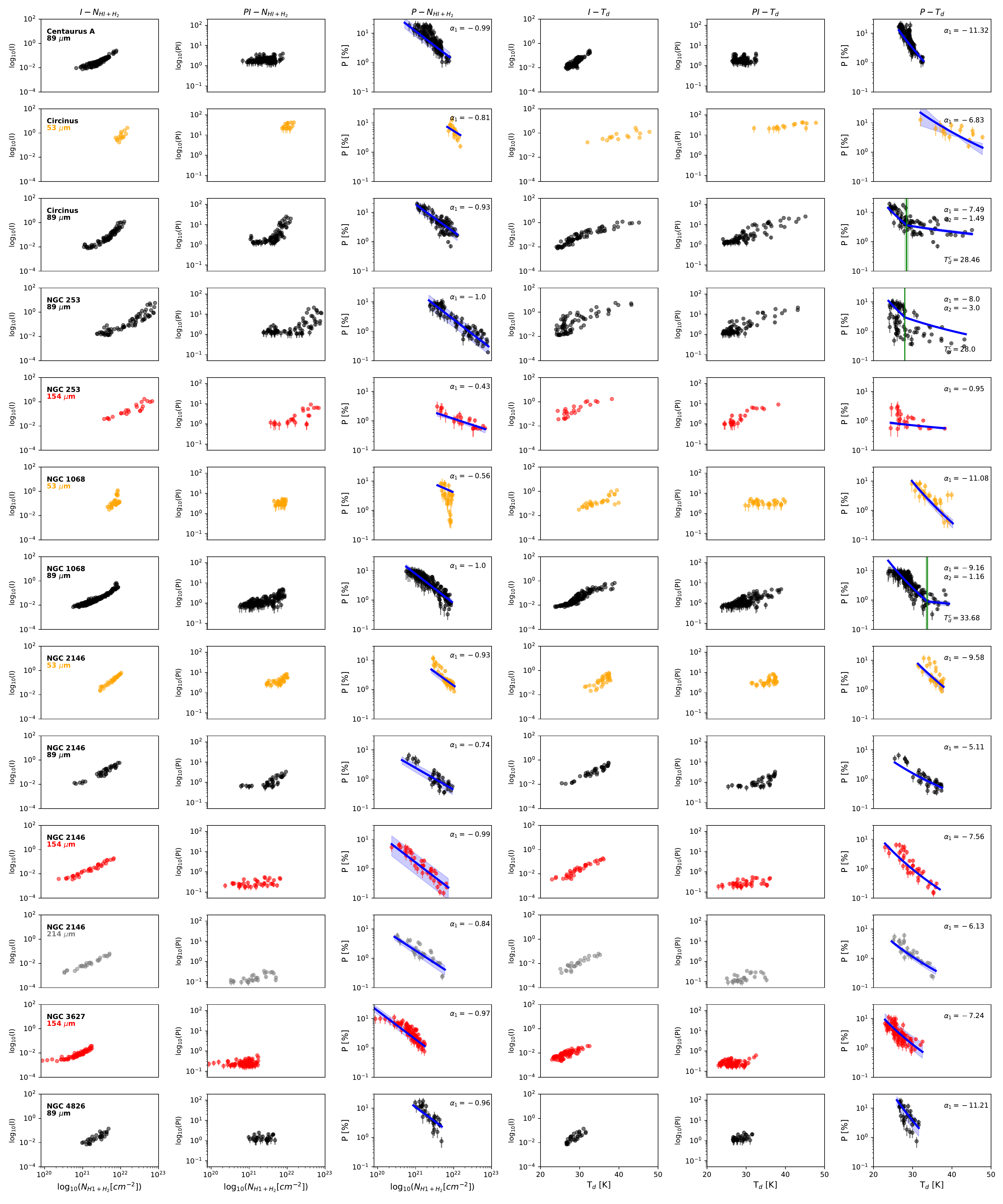}
\caption{Polarization fraction vs. $N_{\rm HI+H_{2}}$ and $T_{\rm d}$ for the galaxies associated with Group 3 shown in Figure \ref{fig:fig5} and \ref{fig:fig6}. The same polarization fraction measurements shown in Figure \ref{fig:fig1} are used.
\label{fig:A3_fig3}}
\epsscale{2.}
\end{figure*}
%%%%%%%%%%%%%%

\section{Depolarization effect due to the effect of $\gamma$ for intermediate inclinations}\label{App:A4}

Section \ref{subsec:DIS_Inc} shows that the polarization fraction may change across the galaxy disks for intermediate inclinations. The depolarization effect is caused by a change of the B-field orientation in the plane of the sky across the galaxy disk. 

Taking the first approach from Section \ref{subsec:DIS_Inc}, we have that the polarization fraction varies as

\begin{equation}
    P(\gamma,i) = \frac{p_{0}\cos^{2} \gamma \cos^{2} i}{1- p_{0}(\frac{\cos^{2} \gamma}{2} - \frac{1}{3})}
\end{equation}
\noindent
. Taking that the inclination is constant across the galaxy disk, then the ratio between the polarization fractions across the disk due to a change in $\gamma$ can be defined as

\begin{equation}\label{eq:DP}
    \Delta P_{1} = \frac{P_{\gamma 1}(\gamma  = 0)}{P_{\gamma 1}(\gamma')} = P_{\gamma 1,\rm max} \left[ \frac{1}{p_{o}} - \left(\frac{\cos^2 \gamma'}{2} - \frac{1}{3}\right) \right]
\end{equation}
\noindent
where $\gamma'$ is the B-field orientation with respect to the plane of the sky. 

Using the second approach from Section \ref{subsec:DIS_Inc}, the change in polarization fraction is 

\begin{equation}\label{eq:DP2}
    \Delta P_{2} = \frac{P_{\gamma 2}(\gamma  = 0)}{P_{\gamma 2}(\gamma')} = \frac{1}{\cos^{2}\gamma'}
\end{equation}
\noindent

We show the ratio of the polarization fraction ($\Delta P_{1}$ and $\Delta P_{2}$) as a function of $\Delta \gamma = \gamma' - \gamma = \gamma'$ ($\gamma = 0^{\circ}$) for three values of the maximum polarization efficiency, $p_{\gamma 1, \rm max} = 1$\%, $4.3$\%, and $20$\%. As an example, for a galaxy with inclination $i=40^{\circ}$ and assuming that $\gamma \sim i$, the regions of the galaxy along the axis of rotation in the plane of the sky can suffer a depolarization of up to a factor of $1.2$ to $5.1$ within the maximum polarization ranges of of $p_{\gamma 1, \rm max} = 1-20$\% for $\Delta P_{1}$, and up to a factor of $1.7$ for the pure geometrical approach ($\Delta P_{2}$).

%%%%%%%%%%%%%%
%%%% FIGURE A4_1 %%%%
%%%%%%%%%%%%%%
\begin{figure}[ht!]
\includegraphics[angle=0,width=\columnwidth]{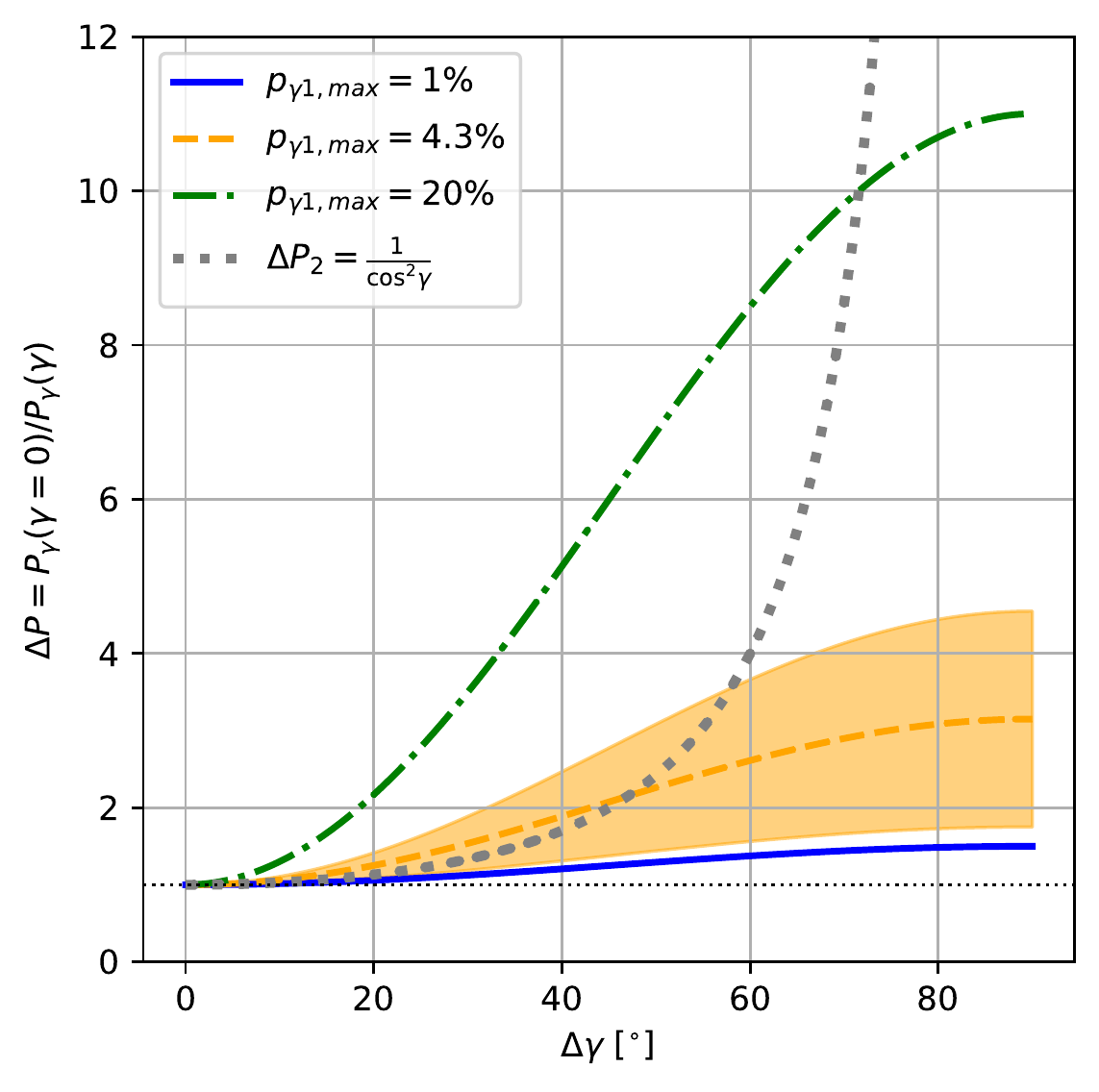}
\caption{Variation of the polarization fraction with $\gamma$ assuming a galaxy with constant inclination (Section \ref{subsec:DIS_Inc}). The ratio of the polarization fraction ($\Delta P_{1}$, Eq. \ref{eq:DP}) as a function of $\gamma$ for three values of the maximum polarization fraction, $p_{\gamma 1,max} = 1$\% (blue solid line), $4.3$\% (orange dashed line), and $20$\% (green dash-dotted line) are shown. The median polarization of $\langle P^{hist} \rangle = 4.3\pm2.8$\% for the galaxies with intermediate inclinations, $30-60^{\circ}$, are displayed as the maximum polarization fraction in orange with the uncertainty shown as an orange shadowed area. The ratio of the polarization fraction assuming a galaxy with constant inclination for $\Delta P_{2}$ (grey dotted line, Eq. \ref{eq:DP2}) is shown.
\label{fig:A4_fig1}}
\epsscale{2.}
\end{figure}
%%%%%%%%%%%%%%

\bibliography{references}



\end{document}